\documentclass[11pt]{article}

\usepackage{amssymb}
\usepackage{amsthm}

\usepackage[french]{babel}
\usepackage[T1]{fontenc}

\usepackage[latin1]{inputenc}

\usepackage{logique}

\newcommand{\NN}{\mathbb{N}}
\newcommand{\CC}{\mathbb{C}}

\newcommand{\nn}{\mbox{\sffamily n}}
\newcommand{\cp}{\mbox{\sffamily cp}}
\newcommand{\indi}{^{\mbox{\footnotesize int}}}
\newcommand{\inde}{^{\mbox{\footnotesize ent}}}
\newcommand{\ppt}{{\mbox{\small <}}}

\newcommand{\trl}{\triangleleft}

\newcommand{\forcec}{\;[\!]\hspace{-0.6em}-}

\newcommand{\sqle}{\sqsubseteq}
\newcommand{\et}{{\scriptstyle\land}}

\newtheorem{theorem}{Théorème}
\newtheorem{lemma}[theorem]{Lemme}
\newtheorem{corollary}[theorem]{Corollaire}
\newtheorem{proposition}[theorem]{Proposition}

\author{Jean-Louis Krivine\\
\footnotesize{krivine@pps.jussieu.fr}
}

\title{Algèbres de réalisabilité~:\\
un programme pour bien ordonner $\mathbb{R}$}
\date{\footnotesize {23 mars 2010}}

\begin{document}
\maketitle\noindent

\section*{Introduction}\noindent
Le principal problème, pour transformer les preuves mathématiques en programmes, est naturellement posé par les
\emph{axiomes}~: en effet, on sait depuis longtemps comment traiter une preuve en logique intuitionniste pure
(i.e. sans axiome), y compris au second ordre~\cite{curry,howard,girard}.\\
Le premier de ces axiomes est le \emph{tiers exclu}, et il paraissait insurmontable. La solution, tout à fait
surprenante, a été donnée par T. Griffin~\cite{griffin} en 1990 et c'est là une découverte essentielle pour la logique.
Dès ce moment, il était clair que tous les autres axiomes allaient suivre, en se plaçant dans un cadre adéquat.

\smallskip\noindent
La \emph{théorie de la réalisabilité classique} constitue un tel cadre~: elle est développée
dans~\cite{krivine2,krivine3}, où on traite les axiomes de l'\emph{Analyse} (arithmétique du second ordre
avec choix dépendant).\\
Dans~\cite{krivine5}, on commence à s'occuper de l'axiome du choix général, avec l'existence d'un ultrafiltre
sur~$\NN$~; l'outil principal est la notion de \emph{structure de réalisabilité}, dans laquelle les programmes
sont écrits en $\lbd$-calcul.\\
On la remplace ici par celle d'\emph{algèbre de réalisabilité}, plus simple, et beaucoup plus utilisable du point de
vue informatique. Il s'agit d'une variante de la notion usuelle d'\emph{algèbre combinatoire}. Le langage de programmation n'est donc plus le $\lbd$-calcul, mais un système convenable de combinateurs~; les $\lbd$-termes ne sont considérés que comme des notations ou abréviations, fort utiles au demeurant~: un $\lbd$-terme est infiniment plus lisible que sa compilation en combinateurs.

\smallskip\noindent
On montre ici comment transformer en programmes les preuves utilisant l'axiome du choix dépendant et~:\\
i)~l'existence d'un ultrafiltre non trival sur $\NN$~;\\
ii)~l'existence d'un bon ordre sur $\mathbb{R}$.\\
Bien entendu, (ii) implique (i) mais, la méthode utilisée pour (i) est intéressante, car elle donne des
programmes plus simples. Ce point est important, parce qu'un nouveau pro\-blème apparaît maintenant,
capital et fort difficile~: interpréter les programmes obtenus, c'est-à-dire expliquer leur comportement.
Un travail passionnant et de longue haleine.

\smallskip\noindent
Le cadre logique est donné par la \emph{logique classique du second ordre}, autrement dit,
le schéma de compréhension. Toutefois, comme on utilise une relation d'appartenance sur les individus,
il s'agit, en fait, d'une logique d'ordre 3 au moins. C'est d'ailleurs indispensable puisque, si
l'axiome du choix dépendant sur $\mathbb{R}$ est exprimable comme un schéma au second ordre, les
axiomes~(i) et~(ii) ne le sont pas.\\
En utilisant la mé\-thode exposée dans~\cite{krivine1}, on peut obtenir les mêmes résultats dans ZF\/.

\smallskip\noindent
Il me paraît clair que la technique utilisée ici permettra de traiter tous les axiomes ``na\-turels''
introduits en théorie des ensembles. C'est déjà fait pour l'\emph{hypothèse du continu}, qui fera l'objet
d'un prochain article. L'axiome du choix et l'hypothèse généralisée du continu dans ZF ne me semblent pas
soulever de problème sérieux, à part celui-ci~: il faudra se servir du \emph{forcing avec classes
propres} d'Easton~\cite{easton} à l'intérieur du modèle de réalisabilité, ce qui menace d'être très lourd.\\
Un problème ouvert fort intéressant est posé par les axiomes de grand cardinaux, comme l'exis\-tence d'un cardinal
mesurable, ou par l'axiome de détermination.

\smallskip\noindent
Mais le problème ouvert essentiel reste de comprendre ce que font les programmes obtenus, et ainsi d'arriver à
\emph{les exécuter}. Je crois que bien des surprises nous attendent là.\\
En effet, au fur et à mesure qu'on réalise les axiomes usuels des mathématiques, on est amené à
introduire des outils tout à fait standard et indispensables en programmation système~:
pour la loi de Peirce, ce sont les continuations (particulièrement utilisées pour les exceptions)~;
pour l'axiome du choix dépendant, c'est l'horloge et la numérotation des processus~; pour l'axiome
de l'ultrafiltre et le bon ordre sur $\mathbb{R}$, \c{c}a n'est rien de moins que la lecture et l'écriture
dans une mémoire globale, autrement dit \emph{l'affectation}.\\
On peut raisonnablement conjecturer que ces outils ne sont pas mobilisés pour rien, et donc que les programmes
fort complexes qu'on obtient par ce travail de formalisation accomplissent des tâches intéressantes. Lesquelles~?

\smallskip\noindent
{\small{\bfseries Remarque.}\\
Le problème de transformer en programmes les preuves utilisant certains axiomes doit être posé correctement
du point de vue informatique. Prenons comme exemple une preuve d'un théorème d'arithmétique, utilisant
un bon ordre de ${\cal P}(\NN)$~: en restreignant cette preuve à la classe des ensembles constructibles,
on la transforme aisément en une preuve du même théorème n'utilisant plus ce bon ordre. Il suffirait donc,
ensuite, de transformer cette nouvelle preuve en programme.\\
Mais l'extraction du programme aura été effectuée sur une preuve \emph{profondément différente de la preuve originale}. De plus, avec ce procédé, il est impossible d'associer un programme à l'axiome du bon ordre lui-même.
Du point de vue informatique, il y a là un grave défaut de \emph{modularité}~: au lieu d'avoir mis l'axiome
du bon ordre dans une \emph{bibliothèque de programmes}, on est obligé de recommencer le travail de programmation
à chaque nouvelle preuve.\\
La méthode exposée ici utilise seulement le $\lbd$-terme \emph{extrait de la preuve originale}, qui contient
donc une instruction pour l'axiome du bon ordre sur ${\cal P}(\NN)$, qui n'est pas encore implantée.
Par une compilation convenable, elle le transforme en un programme qui réalise le théorème considéré.\\
Comme corollaire de cette technique, on obtient un programme associé à l'axiome du bon ordre, que l'on peut
mettre en bibliothèque pour le réutiliser.}

\section*{Algèbres de réalisabilité}\noindent
Une \emph{algèbre de réalisabilité} est constituée par trois ensembles~:
$\LLbd$ (ensemble des \emph{termes}), $\PPi$ (ensemble des \emph{piles}),
$\LLbd\star\PPi$ (ensemble des \emph{processus}) avec les opérations sui\-vantes~:

\smallskip\noindent
$(\xi,\eta)\mapsto(\xi)\eta$ de $\LLbd^2$ dans $\LLbd$ (\emph{application})~;\\
$(\xi,\pi)\mapsto\xi\ps\pi$ de $\LLbd\fois\PPi$ dans $\PPi$ (\emph{empiler})~;\\
$(\xi,\pi)\mapsto\xi\star\pi$ de $\LLbd\fois\PPi$ dans $\LLbd\star\PPi$ (\emph{processus})~;\\
$\pi\mapsto\kk_\pi$ de $\PPi$ dans $\LLbd$ (\emph{continuation}).

\smallskip\noindent
On a, dans $\LLbd$, des éléments distingués \ $B,C,E,I,K,W,\ccc$, appelés \emph{combinateurs élémentaires} ou \emph{instructions}.

\smallskip\noindent
{\bfseries Notation.} Le terme $(\ldots(((\xi)\eta_1)\eta_2)\ldots)\eta_n$ sera aussi noté
$(\xi)\eta_1\eta_2\ldots\eta_n$ ou $\xi\eta_1\eta_2\ldots\eta_n$.\\
Par exemple~: \ $\xi\eta\zeta=(\xi)\eta\zeta=(\xi\eta)\zeta=((\xi)\eta)\zeta$. 

\smallskip\noindent
On définit une relation de préordre, notée $\succ$, sur $\LLbd\star\PPi$. C'est la plus petite relation
réflexive et transitive telle que, quels que soient $\xi,\eta,\zeta\in\LLbd$ et $\pi,\varpi\in\PPi$,
on ait~:

\smallskip\noindent
$(\xi)\eta\star\pi\succ\xi\star\eta\ps\pi$.\\
$I\star\xi\ps\pi\succ\xi\star\pi$.\\
$K\star\xi\ps\eta\ps\pi\succ\xi\star\pi$.\\
$E\star\xi\ps\eta\ps\pi\succ(\xi)\eta\star\pi$.\\
$W\star\xi\ps\eta\ps\pi\succ\xi\star\eta\ps\eta\ps\pi$.\\
$C\star\xi\ps\eta\ps\zeta\ps\pi\succ\xi\star\zeta\ps\eta\ps\pi$.\\
$B\star\xi\ps\eta\ps\zeta\ps\pi\succ(\xi)(\eta)\zeta\star\pi$.\\
$\ccc\star\xi\ps\pi\succ\xi\star\kk_\pi\ps\pi$.\\
$\kk_\pi\star\xi\ps\varpi\succ\xi\star\pi$.

\smallskip\noindent
On se donne enfin une partie $\bbot$ de $\LLbd\star\PPi$ qui est un segment terminal pour
ce préordre, c'est-à-dire que~: \ $\p\in\bbot$, $\p'\succ\p$ $\Fl$ $\p'\in\bbot$.\\
Autrement dit, on demande que $\bbot$ ait les propriétés suivantes~:

\smallskip\noindent
$(\xi)\eta\star\pi\notin\bbot\Fl\xi\star\eta\ps\pi\notin\bbot$.\\
$I\star\xi\ps\pi\notin\bbot\Fl\xi\star\pi\notin\bbot$.\\
$K\star\xi\ps\eta\ps\pi\notin\bbot\Fl\xi\star\pi\notin\bbot$.\\
$E\star\xi\ps\eta\ps\pi\notin\bbot\Fl(\xi)\eta\star\pi\notin\bbot$.\\
$W\star\xi\ps\eta\ps\pi\notin\bbot\Fl\xi\star\eta\ps\eta\ps\pi\notin\bbot$.\\
$C\star\xi\ps\eta\ps\zeta\ps\pi\notin\bbot\Fl\xi\star\zeta\ps\eta\ps\pi\notin\bbot$.\\
$B\star\xi\ps\eta\ps\zeta\ps\pi\notin\bbot\Fl(\xi)(\eta)\zeta\star\pi\notin\bbot$.\\
$\ccc\star\xi\ps\pi\notin\bbot\Fl\xi\star\kk_\pi\ps\pi\notin\bbot$.\\
$\kk_\pi\star\xi\ps\varpi\notin\bbot\Fl\xi\star\pi\notin\bbot$.

\subsection*{\cc-termes et $\lbd$-termes}\noindent
On appelle \ \emph{\cc-terme} \ un terme construit avec des variables, les combinateurs élémentaires
$B,C,E,I,K,W,\ccc$ et l'application (fonction binaire). Un \cc-terme est dit \emph{clos} s'il est sans va\-riable~;
il est alors aussi appelé \emph{quasi-preuve} et a une valeur dans $\LLbd$.

\smallskip\noindent
Etant donné un \cc-terme $t$ et une variable $x$, on définit le \cc-terme $\lbd x\,t$ par récurrence sur $t$~;
pour cela, on utilise le premier cas applicable dans la liste suivante~:

\smallskip\noindent
1.~$\lbd x\,t=(K)t$\label{def_lbd} si $t$ ne contient pas $x$.\\
2.~$\lbd x\,x=I$.\\
3.~$\lbd x\,tu=(C\lbd x(E)t)u$ si $u$ ne contient pas $x$.\\
4.~$\lbd x\,tx=(E)t$ si $t$ ne contient pas $x$.\\
5.~$\lbd x\,tx=(W)\lbd x(E)t$ (si $t$ contient $x$).\\
6.~$\lbd x(t)(u)v=\lbd x(B)tuv$ (si $uv$ contient $x$).

\smallskip\noindent
On voit facilement que cette réécriture se termine~: en effet, les règles 1~à~5 diminuent le nombre
d'atomes \emph{du terme initial} présents sous $\lbd x$, et la règle~6 ne peut être appliquée consécutivement qu'un nombre fini de fois.

\smallskip\noindent
Les \emph{$\lbd$-termes} sont définis de la façon habituelle.

\smallskip\noindent
Tout $\lbd$-terme $P$, comportant éventuellement $B,C,E,I,K,W,\ccc$, définit donc
un \cc-terme que nous notons~$|P|$. Si \ $P$ est un $\lbd$-terme clos, il a donc une valeur dans $\LLbd$.

\smallskip\noindent
{\small{\bfseries Remarque.} La notation $\lbd x$ est donc utilisée dans deux sens différents~:
dans les $\lbd$-termes, c'est un constituant de la syntaxe~; dans les \cc-termes, c'est une abréviation.
Dans cet article, sauf pour le théorème~\ref{subst}, c'est dans ce dernier sens que nous l'utiliserons
exclusivement.}

\begin{theorem}\label{subst}
Si $t=|P|$ et $u=|Q|$, alors $t[u/x]=|P[Q/x]|$.
\end{theorem}\noindent
Preuve par récurrence sur la longueur de $P$. C'est immédiat si $P$ est un atome, ou si $P=P_0P_1$.\\
Si $P=\lbd y\,P'$, alors $t=\lbd y\,t'$ avec $t=|P|,t'=|P'|$. On a $t'[u/x]=|P'[Q/x]|$ par hypothèse de récurrence.
Donc $|P[Q/x]|=|\lbd y\,P'[Q/x]|=\lbd y|P'[Q/x]|=\lbd y\,t'[u/x]$. Comme $t=\lbd y\,t'$, il reste à montrer
$\lbd y\,t'[u/x]=(\lbd y\,t')[u/x]$, ce qui est le lemme~\ref{subst1}.

\cqfd

\begin{lemma}\label{subst1}
Si $t,u$ sont des \cc-termes, on a $\lbd y\,t[u/x]=(\lbd y\,t)[u/x]$.
\end{lemma}\noindent
Preuve par récurrence sur le nombre de règles utilisées pour traduire $\lbd y\,t$. On considère la
première règle utilisée.\\
Si c'est la règle~1, $t$ ne contient pas $y$ et $\lbd y\,t=Kt$. Or, $u$ ne contient pas $y$ par hypothèse,
donc $t[u/x]$ ne le contient pas non plus. On a donc $\lbd y\,t[u/x]=Kt[u/x]$, d'où le résultat.\\
Si c'est l'une des autres règles, c'est trivial.

\cqfd

\smallskip\noindent
{\small{\bfseries Remarque.} Le théorème~\ref{subst} n'est pas utilisé dans la suite.}

\begin{theorem}\label{beta_red_gauche}
Si $t$ est un \cc-terme ne comportant que les variables $x_1,\ldots,x_n$, et si
$\xi_1,\ldots,\xi_n\in\LLbd$, alors $\lbd x_1\ldots\lbd x_n\,t\star\xi_1\ps\ldots\ps\xi_n\ps\pi
\succ t[\xi_1/x_1,\ldots,\xi_n/x_n]\star\pi$.
\end{theorem}\noindent
En raisonnant par récurrence sur $n$, on est ramené au cas où $n=1$, qui est donné par le
lemme~\ref{brg1}.

\begin{lemma}\label{brg1}
Si $t$ est un \cc-terme ne comportant que la variable $x$, et si $\xi\in\LLbd$, alors~:\\
$\lbd x\,t\star\xi\ps\pi\succ t[\xi/x]\star\pi$.
\end{lemma}\noindent
Preuve par récurrence sur le nombre de règles~1 à~6 utilisées pour traduire le terme $\lbd x\,t$.
On considère la première règle utilisée~.\\
Si c'est la règle~1, on a \ $\lbd x\,t\star\xi\ps\pi\equiv(K)t\star\xi\ps\pi\succ t\star\pi
\equiv t[\xi/x]\star\pi$ puisque $x$ n'est pas dans $t$.\\
Si c'est la règle~2, on a $t=x$ et \ $\lbd x\,t\star\xi\ps\pi\equiv I\star\xi\ps\pi\succ
\xi\star\pi\equiv t[\xi/x]\star\pi$.\\
Si c'est la règle~3, on a $t=uv$ \ et \ $\lbd x\,t\star\xi\ps\pi\equiv(C\lbd x(E)u)v\star\xi\ps\pi\succ
C\star\lbd x(E)u\ps v\ps\xi\ps\pi\\
\succ\lbd x(E)u\star\xi\ps v\ps\pi\succ(E)u[\xi/x]\star v\ps\pi$ (par hypothèse de récurrence)
$\succ E\star u[\xi/x]\ps v\ps\pi$\\
$\succ(u[\xi/x])v\star\pi\equiv t[\xi/x]\star\pi$ puisque $x$ n'est pas dans $v$.\\
Si c'est la règle~4, on a $t=ux$ \ et \ $\lbd x\,t\star\xi\ps\pi\equiv(E)u\star\xi\ps\pi\succ
E\star u\ps\xi\ps\pi\succ u\xi\star\pi\equiv t[\xi/x]\star\pi$ puisque $u$ ne contient pas $x$.\\
Si c'est la règle~5, on a $t=ux$ \ et \ $\lbd x\,t\star\xi\ps\pi\equiv(W)\lbd x(E)u\star\xi\ps\pi\succ
W\star\lbd x(E)u\ps\xi\ps\pi\\
\succ\lbd x(E)u\star\xi\ps\xi\ps\pi\succ(E)u[\xi/x]\star\xi\ps\pi$ (par hypothèse de récurrence)
$\succ E\star u[\xi/x]\ps\xi\ps\pi$\\
$\succ(u[\xi/x])\xi\star\pi\equiv t[\xi/x]\star\pi$.\\
Si c'est la règle~6, on a $t=(u)(v)w$ \ et \ $\lbd x\,t\star\xi\ps\pi\equiv
\lbd x(B)uvw\star\xi\ps\pi\succ$\\
$\succ(B)u[\xi/x]v[\xi/x]w[\xi/x]\star\pi$  (par hypothèse de récurrence)\\
$\succ B\star u[\xi/x]\ps v[\xi/x]\ps w[\xi/x]\ps\pi\succ
(u[\xi/x])(v[\xi/x])w[\xi/x]\star\pi\equiv t[\xi/x]\star\pi$.

\cqfd

\subsection*{Déduction naturelle}\noindent
Avant d'expliciter le langage formel que nous allons utiliser, il est bon de décrire informellement
les structures (modèles) que nous avons en vue. Ce sont des structures du second ordre, à deux types d'objets~: les {\em individus} appelés aussi {\em conditions} et les prédicats (d'arité diverses).
Comme il s'agit d'une description intuitive, on se limite aux modèles dits {\em pleins}.\\
Un tel modèle est constitué de~:\\
$\bullet$~~un ensemble infini $P$ (ensemble des individus ou conditions).\\
$\bullet$~~l'ensemble des prédicats d'arité $k$ est ${\cal P}(P^k)$ (modèle plein).\\
$\bullet$~~des fonctions de $P^k$ dans $P$.\\
En particulier, on a un individu $0$ et une fonction bijective $s:P\to(P\setminus\{0\})$.
Cela permet de définir l'ensemble des entiers $\ennl$ comme le plus petit ensemble contenant $0$ et clos par $s$.\\
On a aussi une condition notée $\1$ et une application notée \ $\et$ \ de $P^2$ dans $P$.\\
$\bullet$~~des relations (prédicats fixés) sur $P$. En particulier, on a la relation
d'égalité sur les individus et le sous-ensemble \ $\C$ des \emph{conditions non triviales}.\\
$\C[p\et q]$ se lit~:  ``$p$ et $q$ sont deux conditions \emph{compatibles}''.

\smallskip\noindent
On passe maintenant au langage formel, pour écrire des formules et des preuves concernant ces
structures. Il est cons\-titué par~:

\smallskip\noindent
$\bullet$~~des \emph{variables d'individu} ou \emph{variables de conditions} \ notées $x,y,\ldots$
ou $p,q,\ldots$\\
$\bullet$~~des \emph{variables de prédicat} ou \emph{variables du second ordre} \  $X,Y,\ldots$~;
\ chaque variable de prédicat a une arité qui est dans $\ennl$.\\
$\bullet$~~des \emph{symboles de fonction sur les individus} $f,g,\ldots$~; chacun d'eux a une arité
qui est dans $\ennl$.\\
On a, en particulier, un symbole de fonction d'arité $k$ pour chaque fonction récursive $f:\ennl^k\to\ennl$.
Ce symbole sera noté aussi $f$.\\
On a aussi un symbole de constante $\1$ (qui représente la plus grande condition) et un symbole
de fonction binaire $\et$ (qui représente la fonction $\inf$ sur les conditions).

\smallskip\noindent
Les \emph{termes} sont formés à la façon habituelle avec les variables et les symboles de fonction.

\smallskip\noindent
Les \emph{formules atomiques} sont de la forme $X(t_1,\ldots,t_n)$, où $X$ est une variable
de prédicat d'arité $n$, et $t_1,\ldots,t_n$ sont des termes.

\smallskip\noindent
Les {formules} sont construites comme d'habitude, à partir des formules atomiques, à l'aide
des seuls symboles logiques \ $\to,\pt$~:\\
$\bullet$~~chaque formule atomique est une formule~;\\
$\bullet$~~si $A,B$ sont des formules, alors $A\to B$ est une formule~;\\
$\bullet$~~si $A$ est une formule, alors $\pt x\,A$ et $\pt X\,A$ sont des formules.

\smallskip\noindent
{\bfseries Notations.} La formule $A_1\to(A_2\to(\ldots(A_n\to B)\ldots)$ sera notée
$A_1,A_2,\ldots,A_n\to B$.\\
Les symboles logiques usuels sont définis comme suit~:\\
($X$ est une variable de prédicat d'arité~$0$, appelée aussi \emph{variable propositionnelle})\\
$\bot\equiv\pt X\,X$~; $\neg A\equiv A\to\bot$~; $A\lor B\equiv(A\to\bot),(B\to\bot)\to\bot$~;
$A\land B\equiv(A,B\to\bot)\to\bot$~;\\
$\ex\;\mbox{\bfseries\sffamily y}\,F\equiv(\pt\mbox{\bfseries\sffamily y}(F\to\bot)\to\bot$
(où {\bfseries\sffamily y} est une variable d'individu ou de prédicat).\\
Plus généralement, on écrira \ $\ex\;\mbox{\bfseries\sffamily y}\{F_1,\ldots,F_k\}$ \ pour \
$\pt\;\mbox{\bfseries\sffamily y}(F_1,\ldots,F_k\to\bot)\to\bot$.\\
On pourra noter \ $\vec{F}$ \ une suite finie de formules \ $F_1,\ldots,F_k$~;\\
on écrira alors \ $\ex\;\mbox{\bfseries\sffamily y}\{\vec{F}\}$ \ et \
$\pt\;\mbox{\bfseries\sffamily y}(\vec{F}\to\bot)\to\bot$.

\smallskip\noindent
$x=y$ est la formule $\pt Z(Zx\to Zy)$, où $Z$ est une variable de prédicat unaire.

\smallskip\noindent
Les règles de la déduction naturelle sont les suivantes (les $A_i$ sont des formules, les $x_i$
sont des variables de \cc-terme, $t,u$ sont des \cc-termes)~:

\smallskip\noindent
1.~$x_1:A_1,\ldots,x_n:A_n\vdash x_i:A_i$.\\
2.~$x_1:A_1,\ldots,x_n:A_n\vdash t:A\to B$, \ \ $x_1:A_1,\ldots,x_n:A_n\vdash u:A$ \ \ $\Fl$\\ \hspace*{\fill}$x_1:A_1,\ldots,x_n:A_n\vdash tu:B$.\\
3.~$x_1:A_1,\ldots,x_n:A_n,x:A\vdash t:B$ \ \ $\Fl$ \ \
$x_1:A_1,\ldots,x_n:A_n\vdash\lbd x\,t:A\to B$.\\
4.~$x_1:A_1,\ldots,x_n:A_n\vdash t:A$ \ \ $\Fl$ \ \
$x_1:A_1,\ldots,x_n:A_n\vdash t:\pt${\bfseries\sffamily x}$\,A$ \ \
quelle que soit la variable {\bfseries\sffamily x} (individu ou prédicat) qui n'apparaît pas dans
$A_1,\ldots,A_n$.\\
5.~$x_1:A_1,\ldots,x_n:A_n\vdash t:\pt x\,A$ \ \ $\Fl$ \ \
$x_1:A_1,\ldots,x_n:A_n\vdash t:A[\tau/x]$ \ \ où $x$ est une variable d'individu
et $\tau$ un terme.\\
6.~$x_1:A_1,\ldots,x_n:A_n\vdash t:\pt X\,A$ \ \ $\Fl$ \ \
$x_1:A_1,\ldots,x_n:A_n\vdash t:A[F/Xy_1\ldots y_k]$ \ \
où $X$ est une variable de prédicat d'arité $k$ \ et $F$ une formule quelconque.

\smallskip\noindent
{\small{\bfseries Remarque.}\\
Dans la notation $A[F/Xy_1\ldots y_k]$, les variables \ $y_1,\ldots,y_k$ sont liées.
Une notation plus usuelle est $A[\lbd\!y_1\ldots\lbd\!y_k\,F/X]$. Je ne l'emploie pas, car
cela introduit un troisième usage de $\lbd$.}

\subsection*{Réalisabilité}\noindent
Etant donnée une algèbre de réalisabilité ${\cal A}=(\LLbd,\PPi,\LLbd\star\PPi)$,
un \emph{${\cal A}$-modèle} ${\cal M}$ est l'ensemble des données suivantes~:\\
$\bullet$~~Un ensemble infini $P$ qui est le domaine de variation des variables d'individu.\\
$\bullet$~~Le domaine de variation des variables de prédicat d'arité $k$ est ${\cal P}(\PPi)^{P^k}$.\\
$\bullet$~~A chaque symbole de fonction $f$ d'arité $k$, on associe une fonction de $P^k$ dans $P$,
notée $\ov{f}$ ou même $f$ s'il n'y a pas d'ambigüité.\\
En particulier, on a donc un élément distingué $0$ de $P$ et une fonction $s:P\to P$
(interprétation du symbole $s$). On suppose que $s$ est une bijection de $P$ sur $P\setminus\{0\}$.
On peut alors confondre $s^n0\in P$ avec l'entier $n$. On a donc $\ennl\subset P$.\\
Chaque fonction récursive $f:\ennl^k\to\ennl$ est, par hypothèse, un symbole de fonction. Bien entendu,
on suppose que son interprétation $\ov{f}:P^k\to P$ prend les mêmesvaleurs que $f$ sur $\ennl^k$.\\
Enfin, on a aussi une condition $\1\in P$ et une fonction binaire $\et$ de $P^2$ dans $P$.

\smallskip\noindent
Un \emph{terme clos} (resp. une \emph{formule close}) \emph{à paramètres dans le modèle ${\cal M}$}
est, par définition, un terme (resp. une formule) où on a remplacé les occurrences libres de chaque
variable par un \emph{paramètre}, c'est-à-dire un objet du même type du modèle ${\cal M}$~:
une condition pour une variable d'individu, une application de $P^k$ dans ${\cal P}(\PPi)$ pour
une variable de prédicat $k$-aire.\\
Chaque terme clos $t$, à paramètres dans ${\cal M}$ a une valeur $\ov{t}\in P$.

\smallskip\noindent
Une {\em interprétation} $\cal I$ est une application qui associe un individu (condition)
à chaque va\-riable d'individu et un paramètre d'arité $k$ à chaque variable du second
ordre d'arité $k$.\\
${\cal I}[x\lf p]$ (resp. ${\cal I}[X\lf{\cal X}]$) est l'interprétation obtenue en
changeant, dans ${\cal I}$ la valeur de la variable $x$ (resp. $X$) et en lui donnant
la valeur $p\in P$ (resp. ${\cal X}\in{\cal P}(\PPi)^{P^k}$).\\
Pour toute formule $F$ (resp. terme $t$), on désigne par $F^{\cal I}$ (resp. $t^{\cal I})$
la formule close (resp. le terme clos) avec paramètres obtenue en remplaçant chaque variable
libre par la valeur donnée par $\cal I$.

\smallskip\noindent
Pour chaque formule close $F^{\cal I}$ à paramètres dans ${\cal M}$, on définit deux
valeurs de vérité~:\\
$\|F^{\cal I}\|\subset\PPi$ et $|F^{\cal I}|\subset\LLbd$.\\
$|F^{\cal I}|$ est défini par~: \ $\xi\in|F^{\cal I}|$ \ $\Dbfl$ \
$(\pt\pi\in\|F^{\cal I}\|)\,\xi\star\pi\in\bbot$.\\
$\|F^{\cal I}\|$ est définie par récurrence sur $F$~:\\
$\bullet$~~$F$ est atomique~: alors $F^{\cal I}$ est de la forme ${\cal X}(t_1,\ldots,t_k)$ où
${\cal X}:P^k\to{\cal P}(\PPi)$ et les $t_i$ sont des termes clos à paramètres dans ${\cal M}$.
On pose \ $\|{\cal X}(t_1,\ldots,t_k)\|={\cal X}(\ov{t}_1,\ldots,\ov{t}_k)$.\\
$\bullet$~~$F\equiv A\to B$~: on pose \
$\|F^{\cal I}\|=\{\xi\ps\pi~;\;\xi\in|A^{\cal I}|,\pi\in\|B^{\cal I}\|\}$.\\
$\bullet$~~$F\equiv\pt x\,A$~: on pose \ $\|F^{\cal I}\|=\bigcup\{\|A^{{\cal I}[x\lf p]}\|~;\;p\in P\}$.\\
$\bullet$~~$F\equiv\pt X\,A$~: on pose \
$\|F^{\cal I}\|=\bigcup\{\|A^{{\cal I}[X\lf{\cal X}]}\|~;\;{\cal X}\in{\cal P}(\PPi)^{P^k}\}$
si $X$ est une variable de prédicat $k$-aire.

\smallskip\noindent
{\bfseries Notation.} \ On écrira \ $\xi\force F$ \ pour \ $\xi\in|F|$.

\begin{theorem}[Lemme d'adéquation]\label{adequat}\ \\
Si \ $x_1:A_1,\ldots,x_k:A_k\vdash t:A$ \ et si \
$\xi_1\force A_1^{\cal I},\ldots,\xi_k\force A_k^{\cal I}$, où ${\cal I}$ est une interprétation,
alors \ $t[\xi_1/x_1,\ldots,\xi_k/x_k]\force A^{\cal I}$.\\
En particulier, si $A$ est close et si \ $\vdash t:A$, alors $t\force A$.
\end{theorem}\noindent
Preuve par récurrence sur la longueur de la démonstration de \ $x_1:A_1,\ldots,x_n:A_n\vdash t:A$.\\
On considère la dernière règle utilisée.

\smallskip\noindent
1. On a $t=x_i,A\equiv A_i$. Or, on a supposé $\xi_i\force A_i^{\cal I}$ et c'est le résultat cherché.

\smallskip\noindent
2. On a $t=uv$ et on a déjà obtenu \ $x_1:A_1,\ldots,x_k:A_k\vdash u:B\to A$ \ et \
$x_1:A_1,\ldots,x_k:A_k\vdash v:B$.
Etant donnée $\pi\in\|A^{\cal I}\|$, on doit montrer \
$(uv)[\xi_1/x_1,\ldots,\xi_k/x_k]\star\pi\in\bbot$.\\
Par hypothèse sur $\bbot$, il suffit de montrer \
$u[\xi_1/x_1,\ldots,\xi_k/x_k]\star v[\xi_1/x_1,\ldots,\xi_k/x_k]\ps\pi\in\bbot$.\\
Par hypothèse de récurrence, on a \ $v[\xi_1/x_1,\ldots,\xi_k/x_k]\force B^{\cal I}$ et par suite~:\\
$v[\xi_1/x_1,\ldots,\xi_k/x_k]\ps\pi\in\|B^{\cal I}\to A^{\cal I}\|$.\\
Or, par hypothèse de récurrence, on a aussi $u[\xi_1/x_1,\ldots,\xi_k/x_k]\force B^{\cal I}\to A^{\cal I}$,
d'où le résultat.

\smallskip\noindent
3. On a $A=B\to C$, $t=\lbd x\,u$. On doit montrer~:\\
$\lbd x\,u[\xi_1/x_1,\ldots,\xi_k/x_k]\force B^{\cal I}\to C^{\cal I}$
et on considère donc $\xi\force B^{\cal I}$, $\pi\in\| C^{\cal I}\|$. On est ramené à
montrer $\lbd x\,u[\xi_1/x_1,\ldots,\xi_k/x_k]\star\xi\ps\pi\in\bbot$. Pour cela,
par hypothèse sur $\bbot$ et d'après le lemme~\ref{brg1}, il suffit de montrer
$u[\xi/x,\xi_1/x_1,\ldots,\xi_k/x_k]\star\pi\in\bbot$.\\
Cela résulte de l'hypothèse de récurrence appliquée à \
$x_1:A_1,\ldots,x_n:A_n,x:B\vdash u:C$.

\smallskip\noindent
4. On a $A\equiv\pt X\,B$, $X$ n'étant pas libre dans $A_1,\ldots,A_n$.
On doit montrer~:\\
$t[\xi_1/x_1,\ldots,\xi_k/x_k]\force(\pt X\,B)^{\cal I}$, c'est-à-dire
$t[\xi_1/x_1,\ldots,\xi_k/x_k]\force B^{\cal J}$ avec
${\cal J}={\cal I}[X\leftarrow{\cal X}]$.
Or on a, par hypothèse, $\xi_i\force A_i^{\cal I}$  donc $\xi_i\force A_i^{\cal J}$~:
en effet, comme $X$ n'est pas libre dans $A_i$, on a $\|A_i^{\cal I}\|=\|A_i^{\cal J}\|$.
L\ 'hypothèse de récurrence donne alors le résultat.

\smallskip\noindent
6. On a $A=B[F/Xy_1\ldots y_n]$ et on doit montrer~:\\
$t[\xi_1/x_1,\ldots,\xi_k/x_k]\force B[F/Xy_1\ldots y_n]^{\cal I}$
avec l'hypothèse $t[\xi_1/x_1,\ldots,\xi_k/x_k]\force(\pt X\,B)^{\cal I}$.\\
Cela découle du lemme~\ref{BF/Xy}.

\cqfd

\begin{lemma}\label{BF/Xy}
$\|B[F/Xy_1\ldots y_n]^{\cal I}\|=\|B^{{\cal I}[X\lf{\cal X}]}\|$ où
${\cal X}:P^n\to{\cal P}(\PPi)$ est défini par~:\\
${\cal X}(p_1,\ldots,p_n)=\| F^{{\cal I}[y_1\lf p_1,\ldots,y_n\lf p_n]}\|$.
\end{lemma}\noindent
Preuve par récurrence sur $B$. C'est trivial si $X$ n'est pas libre dans $B$.
Le seul cas intéressant de la récurrence est $B=\pt Y\,C$, et on a donc $Y\ne X$. On a alors~:\\
$\|B[F/Xy_1\ldots y_n]^{\cal I}\|=\|(\pt Y\,C[F/Xy_1\ldots y_n])^{\cal I}\|=
\bigcup_{\cal Y}\|C[F/Xy_1\ldots y_n]^{{\cal I}[Y\leftarrow{\cal Y}]}\|$.\\
Par hypothèse de récurrence, cela donne
$\bigcup_{\cal Y}\| C^{{\cal I}[Y\leftarrow{\cal Y}][X\leftarrow{\cal X}]}\|$, soit
$\bigcup_{\cal Y}\| C^{{\cal I}[X\leftarrow{\cal X}][Y\leftarrow{\cal Y}]}\|$
c'est-à-dire $\|(\pt Y\,C)^{{\cal I}[X\leftarrow{\cal X}]}\|$.

\cqfd

\begin{lemma}\label{continuation}
Soient ${\cal X,Y}\subset\PPi$ des valeurs de vérité. Si $\pi\in{\cal X}$, \ alors \
$\kk_\pi\force{\cal X}\to{\cal Y}$.
\end{lemma}\noindent
Soient $\xi\force{\cal X}$ et $\rho\in{\cal Y}$~; on doit montrer $\kk_\pi\star\xi\ps\rho\in\bbot$, soit
$\xi\star\pi\in\bbot$, ce qui est clair.

\cqfd

\begin{proposition}[Loi de Peirce]
$\ccc\force\pt X\pt Y(((X\to Y)\to X)\to X)$.
\end{proposition}\noindent
On doit montrer que \ $\ccc\force(({\cal X}\to{\cal Y})\to{\cal X})\to{\cal X}$.
Soient donc \ $\xi\force({\cal X}\to{\cal Y})\to{\cal X}$ et $\pi\in\|{\cal X}\|$~;
on doit montrer que $\ccc\star\xi\ps\pi\in\bbot$, ou encore $\xi\star\kk_\pi\ps\pi\in\bbot$.
D'après l'hypothèse sur $\xi$ et $\pi$, il suffit de montrer que $\kk_\pi\force{\cal X}\to{\cal Y}$,
ce qui résulte du lemme~\ref{continuation}.

\cqfd

\begin{proposition}
i) Si $t\force A\to B$, \ alors \ $\pt u(u\force A\Fl tu\force B)$.\\
ii) Si $\pt u(u\force A\Fl tu\force B)$, \ alors \ $(E)t\force A\to B$.
\end{proposition}\noindent
i) D'après $tu\star\pi\succ t\star u\ps\pi$.\\
ii) D'après $(E)t\star u\ps\pi\succ tu\star\pi$.

\cqfd

\subsubsection*{Symboles de prédicat}\noindent
On utilisera dans la suite des \emph{formules étendues} utilisant des
\emph{symboles (ou constantes) de prédicat sur les individus} \ {\sf R,S},\ldots \
Chacun d'eux a une arité, qui est dans~$\ennl$.\\
On a, en particulier, un symbole de prédicat unaire $\C$ (pour représenter l'ensemble
des conditions non triviales).\\
On ajoute aux règles de construction des formules, les règles~:

\smallskip\noindent
$\bullet$~~Si $F$ est une formule, {\sf R} une constante de prédicat $n$-aire
et $t_1,\ldots,t_n$ sont des termes, alors \ ${\sf R}(t_1,\ldots,t_n)\to F$ \ et \
${\sf R}(t_1,\ldots,t_n)\mapsto F$ sont des formules.\\
$\bullet$~~$\top$ est une formule atomique.

\smallskip\noindent
Dans la définition d'un ${\cal A}$-modèle ${\cal M}$, on ajoute la clause~:

\smallskip\noindent
$\bullet$~~A chaque symbole de relation {\sf R} d'arité $n$, on associe une application, notée
$\ov{\sf R}_{\cal M}$ ou $\ov{\sf R}$, de $P^n$ dans ${\cal P}(\LLbd)$.
On écrira aussi \ $|{\sf R}(p_1,\ldots,p_n)|$, \ au lieu de \ $\ov{\sf R}(p_1,\ldots,p_n)$, pour
$p_1,\ldots,p_n\in P$.\\
En particulier, on a une application $\ov{\sf C}:P\to{\cal P}(\LLbd)$, que l'on notera $|\C[p]|$.

\smallskip\noindent
On définit comme suit la valeur de vérité dans ${\cal M}$ d'une formule étendue~:

\smallskip\noindent
$\|\top\|=\vide$.\\
$\|({\sf R}(t_1,\ldots,t_n)\to F)^{\cal I}\|=
\{t\ps\pi;\;t\in|{\sf R}(t_1^{\cal I},\ldots,t_n^{\cal I})|,\pi\in\|F^{\cal I}\|\}$.\\
$\|({\sf R}(t_1,\ldots,t_n)\mapsto F)^{\cal I}\|=\|F^{\cal I}\|$ si
$I\in|{\sf R}(t_1^{\cal I},\ldots,t_n^{\cal I})|$~;\\
$\|({\sf R}(t_1,\ldots,t_n)\mapsto F)^{\cal I}\|=\vide$ sinon.

\begin{proposition}\label{R_simple}\ \\
i)~$\lbd x(x)I\force\pt X\pt x_1\ldots\pt x_n[({\sf R}(x_1,\ldots,x_n)\to X)\to
({\sf R}(x_1,\ldots,x_n)\mapsto X)]$.\\
ii)~Si on a \ $|{\sf R}(p_1,\ldots,p_n)|\ne\vide$ $\Fl$ $I\in|{\sf R}(p_1,\ldots,p_n)|$ \
quels que soient $p_1,\ldots,p_n\in P$, alors~:\\
$K\force\pt X\pt x_1\ldots\pt x_n[({\sf R}(x_1,\ldots,x_n)\mapsto X)\to
({\sf R}(x_1,\ldots,x_n)\to X)]$.
\end{proposition}\noindent
Trivial.

\cqfd

\smallskip\noindent
{\small{\bfseries Remarque.} D'après la proposition~\ref{R_simple}, on voit que, si l'application
$\ov{\sf R}:P^n\to{\cal P}(\LLbd)$ ne prend que les valeurs $\{I\}$ et $\vide$, on peut
remplacer \ ${\sf R}(t_1,\ldots,t_n)\to F$ \ par \ ${\sf R}(t_1,\ldots,t_n)\mapsto F$.}

\smallskip\noindent
On définit le prédicat binaire $\simeq$ en posant $|p\simeq q|=\{I\}$ si $p=q$ \ et \
$|p\simeq q|=\vide$ si $p\ne q$.\\
D'après la remarque ci-dessus, on peut remplacer \ $p\simeq q\to F$ par $p\simeq q\mapsto F$.
La proposition~\ref{egal_simple} montre qu'on peut aussi remplacer \ $p=q\to F$ par $p\simeq q\mapsto F$.

\smallskip\noindent
{\bfseries Notations.} On écrira \ $p=q\mapsto F$ \ au lieu de \ $p\simeq q\mapsto F$. On a donc~:\\
$\|p=q\mapsto F\|=\|F\|$ \ si $p=q$~; \ $\|p=q\mapsto F\|=\vide$ \ si $p\ne q$.\\
On écrira \ $p\ne q$ \ pour \ $p=q\mapsto\bot$. On a donc~:\\
$\|p\ne q\|=\PPi$ si $p=q$ et \ $\|p\ne q\|=\vide$ si $p\ne q$.

\smallskip\noindent
L'utilisation de $p=q\mapsto F$ \ au lieu de \ $p=q\to F$, et de $p\ne q$ au lieu de $p=q\to\bot$,
simplifie beaucoup le calcul de la valeur de vérité d'une formule comportant le symbole $=$.

\begin{proposition}\label{egal_simple}\ \\
i)~$\lbd x\,xI\force\pt X\pt x\pt y((x=y\to X)\to(x=y\mapsto X))$~;\\
ii)~$\lbd x\lbd y\,yx\force\pt X\pt x\pt y((x=y\mapsto X),x=y\to X)$.
\end{proposition}\noindent
i)~Soient $a,b\in P\,,\;{\cal X}\subset\PPi,\xi\force a=b\to{\cal X}$
et $\pi\in\|a=b\mapsto{\cal X}\|$.\\
On a donc $a=b$, d'où $I\force a=b$, donc $\xi\star I\ps\pi\in\bbot$, d'où $\lbd x\,xI\star\xi\ps\pi\in\bbot$.\\
ii)~Soient maintenant $\eta\force(a=b\mapsto{\cal X}),\;\zeta\force a=b$ et $\rho\in\|{\cal X}\|$.\\
On montre que \ $\lbd x\lbd y\,yx\star\eta\ps\zeta\ps\rho\in\bbot$
autrement dit \ $\zeta\star\eta\ps\rho\in\bbot$.\\
Si $a=b$, alors $\eta\force{\cal X}$, $\zeta\force\pt Y(Y\to Y)$.
On a $\eta\ps\rho\in\|{\cal X}\to{\cal X}\|$, donc $\zeta\star\eta\ps\rho\in\bbot$.\\
Si $a\ne b$, alors $\zeta\force\top\to\bot$, donc $\zeta\star\eta\ps\rho\in\bbot$.\\
Dans les deux cas, on a le résultat voulu.

\cqfd

\smallskip\noindent
{\small{\bfseries Remarque.}\\
Soient $R$ une partie de $P^k$ et $1_R:P^k\to\{0,1\}$ sa fonction caractéristique, définie par~:\\
$1_R(p_1,\ldots,p_n)=1$ (resp. $=0$) si $(p_1,\ldots,p_n)\in R$ (resp. $(p_1,\ldots,p_n)\notin R$).\\
On étend le prédicat $R$ au modèle ${\cal M}$ en posant~:\\
$|R(p_1,\ldots,p_n)|=\{I\}$ (resp. $=\vide$) si $(p_1,\ldots,p_n)\in R$ (resp. $(p_1,\ldots,p_n)\notin R$).\\
D'après les propositions~\ref{R_simple} et~\ref{egal_simple}, on voit que $R(x_1,\ldots,x_n)$ et
$1_R(x_1,\ldots,x_n)=1$ sont interchangeables. Plus précisément, on a~:\\
$I\force\pt X\pt x_1\ldots\pt x_n((R(x_1,\ldots,x_n)\mapsto X)\dbfl(1_R(x_1,\ldots,x_n)=1\mapsto X))$.}

\smallskip\noindent
Pour chaque formule $A[x_1,\ldots,x_k]$, on peut définir le symbole de prédicat $k$-aire $N_A$, en posant
$|N_A(p_1,\ldots,p_k)|=\{\kk_\pi;\;\pi\in\|A[p_1,\ldots,p_k]\|\}$. La proposition~\ref{N_A_negA} montre que
$N_A$ et $\neg A$ sont interchangeables~; cela peut simplifier les calculs de valeurs de vérité.

\begin{proposition}\label{N_A_negA}\ \\
i)~~$I\force\pt x_1\ldots\pt x_k(N_A(x_1,\ldots,x_k)\to\neg A(x_1,\ldots,x_k))$~;\\
ii)~~$\ccc\force\pt x_1\ldots\pt x_k((N_A(x_1,\ldots,x_k)\to\bot)\to A(x_1,\ldots,x_k))$.
\end{proposition}\noindent
i)~~Soient $p_1,\ldots,p_k\in P$, $\pi\in\|A(p_1,\ldots,p_k)\|$, $\xi\force A(p_1,\ldots,p_k)$ et $\rho\in\PPi$.
On doit montrer~:\\
$I\star\kk_\pi\ps\xi\ps\rho\in\bbot$, soit $\xi\star\pi\in\bbot$, ce qui est évident.\\
ii)~~Soient $\eta\force N_A(p_1,\ldots,p_k)\to\bot$ \ et \ $\pi\in\|A(p_1,\ldots,p_k)\|$. On doit montrer~:\\
$\ccc\star\eta\ps\pi\in\bbot$, soit $\eta\star\kk_\pi\ps\pi\in\bbot$, \ ce qui est clair, puisque \
$\kk_\pi\in|N_A(p_1,\ldots,p_k)|$.

\cqfd

\subsubsection*{Combinateur de point fixe}\noindent

\begin{theorem}\label{bien_fonde}
On pose \ $\Y=AA$ \ avec $A=\lbd a\lbd f(f)(a)af$. On a alors \ $\Y\star\xi\ps\pi\succ\xi\star\Y\xi\ps\pi$.\\
Soit $f:P^2\to P$ telle que $f(x,y)=1$ soit une relation bien fondée sur $P$. On alors~:\\
i)~~$\Y\force\pt X\{\pt x[\pt y(f(y,x)=1\mapsto Xy)\to Xx]\to\pt x\,Xx\}$.\\
ii)~~$\Y\force\pt X_1\ldots\pt X_k\\
\hspace*{\fill}\{\pt x[\pt y(X_1y,\ldots,X_ky\to f(y,x)\ne1),X_1x,\ldots,X_kx\to\bot]
\to\pt x(X_1x,\ldots,X_kx\to\bot)\}$.
\end{theorem}\noindent
La propriété \ $\Y\star\xi\ps\pi\succ\xi\star\Y\xi\ps\pi$ \ est immédiate, d'après
le théorème~\ref{beta_red_gauche}.\\
i)~~On fixe ${\cal X}:P\to{\cal P}(\PPi)$, $p\in P$ et \
$\xi\force\pt x[\pt y(f(y,x)=1\mapsto{\cal X}y)\to{\cal X}x]$. On montre, par induction sur
la relation bien fondée $f(x,y)=1$, que $\Y\star\xi\ps\pi\in\bbot$ pour tout $\pi\in{\cal X}p$.\\
Soit donc $\pi\in{\cal X}p$~; d'après (i), on a \ $\Y\star\xi\ps\pi\succ\xi\star\Y\xi\ps\pi$ \ et
il suffit donc de montrer que \ $\xi\star\Y\xi\ps\pi\in\bbot$.
Par hypothèse, on a $\xi\force\pt y(f(y,p)=1\mapsto{\cal X}y)\to{\cal X}p$~; il suffit donc
de montrer que $\Y\xi\force f(q,p)=1\mapsto{\cal X}q$ pour tout $q\in P$.
C'est évident si $f(q,p)\ne1$, par définition de \ $\mapsto$.\\
Si $f(q,p)=1$, on doit montrer $\Y\xi\force{\cal X}q$, soit $\Y\star\xi\ps\rho\in\bbot$ pour tout
$\rho\in{\cal X}q$. Or, cela découle de l'hypothèse d'induction.

\smallskip\noindent
ii)~~La preuve est presque identique~: on fixe ${\cal X}_1,\ldots,{\cal X}_k:P\to{\cal P}(\PPi)$, $p\in P$ et\\
$\xi\force\pt x[\pt y({\cal X}_1y,\ldots,{\cal X}_ky\to f(y,x)\ne1),{\cal X}_1x,\ldots,{\cal X}_kx\to\bot]$.
On montre, par induction sur la relation bien fondée $f(x,y)=1$, que $\Y\star\xi\ps\pi\in\bbot$
pour tout $\pi\in\|{\cal X}_1p,\ldots,{\cal X}_kp\to\bot\|$.\\
Comme précédemment, on est ramené à montrer que~:\\
$\Y\xi\force{\cal X}_1q,\ldots,{\cal X}_kq\to f(q,p)\ne1$ pour tout $q\in P$~; c'est évident si $f(q,p)\ne1$.\\
Si $f(q,p)=1$, on doit montrer $\Y\xi\force{\cal X}_1q,\ldots,{\cal X}_kq\to\bot$, ou encore~:\\
$\Y\star\xi\ps\rho\in\bbot$ pour tout $\rho\in\|{\cal X}_1q,\ldots,{\cal X}_kq\to\bot\|$.
Or, cela découle de l'hypothèse d'induction.

\cqfd

\subsection*{Entiers, mise en mémoire et fonctions récursives}\noindent
On a un symbole de constante $0$ et un symbole de fonction unaire $s$ qui est interprété,
dans le modèle ${\cal M}$ par une fonction bijective $s:P\to(P\setminus\{0\})$.\\
Rappelons que nous avons identifié $s^n0$ avec l'entier $n$ et qu'on suppose donc que $\ennl\subset P$.

\smallskip\noindent
On désigne par int$(x)$ la formule $\pt X(\pt y(Xy\to Xsy),X0\to Xx)$.

\smallskip\noindent
Soit $u=(u_n)_{n\in\NN}$ une suite d'éléments de $\LLbd$.
On définit le symbole de prédicat unaire $e_u$ en posant~: \ $|e_u(s^n0)|=\{u_n\}$~; \
$|e_u(p)|=\vide$ si $p\notin\ennl$.

\begin{theorem}\label{memoire_gen}
Soient $T_u,S_u\in\LLbd$ tels que l'on ait \ $S_u\force(\top\to\bot),\top\to\bot$ \ et~:\\
$T_u\star\phi\ps\nu\ps\pi\succ\nu\star S_u\ps\phi\ps u_0\ps\pi$~; \
$S_u\star\psi\ps u_n\ps\pi\succ\psi\star u_{n+1}\ps\pi$\\
quels que soient $\nu,\phi,\psi\in\LLbd$ et $\pi\in\PPi$. Alors~:\\
$T_u\force\pt X\pt x[(e_u(x)\to X),$ int$(x)\to X]$.\\
$T_u$ est appelé opérateur de mise en mémoire.
\end{theorem}\noindent
Soient \ $p\in P$, $\phi\force e_u(p)\to X$, \ $\nu\force$int$(p)$ \ et \ $\pi\in\|X\|$.
On doit montrer \ $T_u\star\phi\ps\nu\ps\pi\in\bbot$ \ autrement dit \ $\nu\star S_u\ps\phi\ps u_0\ps\pi\in\bbot$.

\smallskip\noindent
$\bullet$~~Si $p\notin\ennl$, on définit le prédicat unaire $Y$ en posant~:\\
$Y(q)\equiv\top$ si $q\in\ennl$~; \ $Y(q)\equiv\top\to\bot$ si $q\notin\ennl$.\\
On a donc, évidemment, $\phi\force Y(0)$ et $u_0\ps\pi\in\|Y(p)\|$.\\
Or, par hypothèse sur $\nu$, on a $\nu\force\pt y(Yy\to Ysy),Y0\to Yp$. Il suffit donc de montrer que~:\\
$S_u\force\pt y(Yy\to Ysy)$, soit \ $S_u\force Y(q)\to Y(sq)$ pour tout $q\in P$.\\
C'est évident si $q\in\ennl$, puisqu'alors $\|Y(sq)\|=\vide$.\\
Si $q\notin\ennl$, on doit montrer \	
$S_u\force(\top\to\bot),\top\to\bot$, ce qui résulte de l'hypothèse.

\smallskip\noindent
$\bullet$~~Si $p\in\ennl$, on a $p=s^p0$~; on définit le prédicat unaire $Y$ en posant~:\\
$\|Ys^i0\|=\{u_{p-i}\ps\pi\}$ pour $0\le i\le p$ et
$\|Yq\|=\vide$ si $q\notin\{s^i0;\;0\le i\le p\}$.\\
Par hypothèse sur $\nu,\phi,\pi$, on a~:\\
$\nu\force\pt y(Yy\to Ysy),Y0\to Ys^p0$~; \ $\phi\force Y0$~; \ $u_0\ps\pi\in\|Ys^p0\|$.\\
Il suffit donc de montrer que \ $S_u\force\pt y(Yy\to Ysy)$, soit $S_u\force Yq\to Ysq$ pour tout $q\in P$.\\
C'est évident si $q\notin\{s^i0;\;0\le i<p\}$, car alors $\|Ysq\|=\vide$.
Si $q=s^i0$ avec $i<p$, soit $\xi\force Yq$~;\\
on doit montrer \ $S_u\star\xi\ps u_{p-i-1}\ps\pi\in\bbot$. Or, on a \
$S_u\star\xi\ps u_{p-i-1}\ps\pi\succ\xi\star u_{p-i}\ps\pi$ \ qui est dans~$\bbot$, par hypothèse sur $\xi$.

\cqfd

\smallskip\noindent
{\bfseries Notation.} On définit les $\cc$-termes clos \ $\ul{0}=\lbd x\lbd y\,y$~; \
$\sig=\lbd n\lbd f\lbd x(f)(n)fx$~; et, pour chaque $n\in\ennl$, on pose \ $\ul{n}=(\sig)^n\ul{0}$.
On définit le symbole de prédicat
unaire ent(x) en posant~:\\
$|$ent$(n)|=\{\ul{n}\}$ si $n\in\ennl$~;\\
$|$ent$(p)|=\vide$ si $p\notin\ennl$.\\
Autrement dit, ent(x) est le prédicat $e_u(x)$ lorsque la suite $u$ est $(\ul{n})_{n\in\ennl}$.

\begin{theorem}\label{mm1}\ \\
On pose \ $T=\lbd f\lbd n(n)Sf\ul{0}$, \ avec \ $S=\lbd g\lbd x(g)(\sigma)x$. On a alors~:\\
i)~~$T\force\pt X\pt x(($ent$(x)\to X),$ int$(x)\to X)$.\\
ii)~~$I\,\force\pt x(($ent$(x)\to$int$(x))$.
\end{theorem}\noindent
$T$ est donc un opérateur de mise en mémoire.

\smallskip\noindent
i)~~On a immédiatement, d'après le théorème~\ref{beta_red_gauche}~:\\
$T\star\phi\ps\nu\ps\pi\succ\nu\star S\ps\phi\ps\ul{0}\ps\pi$~; \
$S\star\psi\ps(\sig)^n\ul{0}\ps\pi\succ\psi\star(\sig)^{n+1}\ul{0}\ps\pi$\\
quels que soient $\nu,\phi,\psi\in\LLbd$ et $\pi\in\PPi$.\\
On vérifie que \ $S\force(\top\to\bot),\top\to\bot$~: en effet, si $\xi\force\top\to\bot$, alors
$S\star\xi\ps\eta\ps\pi\succ\xi\star\sig\eta\ps\pi\in\bbot$ quels que soient $\eta\in\LLbd$ et $\pi\in\PPi$
(d'après le théorème~\ref{beta_red_gauche}).\\
Le résultat est alors immédiat, d'après le théorème~\ref{memoire_gen}.

\smallskip\noindent
ii)~~On doit montrer \ $I\force$ ent$(p)\to$ int$(p)$ pour tout $p\in P$. On peut supposer
$p\in\ennl$ (sinon ent$(p)=\vide$ et le résultat est trivial).
On doit alors montrer~:\\
$I\star\sig^p\ul{0}\ps\rho\in\bbot$ sachant que $\rho\in\|$int$(s^p0)\|$.\\
Il existe donc un prédicat unaire $X:P\to{\cal P}(\PPi)$, $\phi\force\pt y(Xy\to Xsy)$,
$\omega\force X0$ et $\pi\in\|Xs^p0\|$ tels que $\rho=\phi\ps\omega\ps\pi$.
On doit montrer \ $(\sig)^p\ul{0}\star\phi\ps\omega\ps\pi\in\bbot$. On montre, en fait, par récurrence
sur~$p$, que  \ $(\sig)^p\ul{0}\star\phi\ps\omega\ps\pi\in\bbot$ \ \emph{pour tout $\pi\in\|Xs^p0\|$}.\\
Pour $p=0$, soit $\pi\in\|X0\|$~; on doit montrer $\ul{0}\star\phi\ps\omega\ps\pi\in\bbot$,
soit $\omega\star\pi\in\bbot$, ce qui est évident puisque $\omega\force X0$.\\
Pour passer de $p$ à $p+1$, soit $\pi\in\|Xs^{p+1}0\|$. On a~:\\
$\sig^{p+1}\ul{0}\star\phi\ps\omega\ps\pi\equiv
(\sig)(\sig)^p\ul{0}\star\phi\ps\omega\ps\pi\succ\sig\star\sig^p\ul{0}\ps\phi\ps\omega\ps\pi
\succ\phi\star(\sig^p\ul{0})\phi\omega\ps\pi$.\\
Or, par hypothèse de récurrence, on a $\sig^p\ul{0}\star\phi\ps\omega\ps\rho\in\bbot$ pout tout
$\rho\in\|Xs^p0\|$. Il en résulte que $(\sig^p\ul{0})\phi\omega\force Xs^p0$. Comme
$\phi\force Xs^p0\to Xs^{p+1}0$, on a bien \ $\phi\star(\sig^p\ul{0})\phi\omega\ps\pi\in\bbot$.

\cqfd

\smallskip\noindent
Le théorème~\ref{mm1} montre qu'on peut utiliser le prédicat ent$(x)$ au lieu de int$(x)$,
ce qui simplifie beaucoup les calculs. En particulier, on définit le \emph{quantificateur universel
restreint aux entiers} $\pt x\indi$ en posant \ $\pt x{\indi}F\equiv\pt x($int$(x)\to F)$.\\
On peut donc le remplacer par \emph{le quantificateur universel restreint à ent$(x)$} défini par~:\\
$\pt x\inde\,F\equiv\pt x($ent$(x)\to F)$. On a alors \
$\|\pt x\inde\,F\|=\{\ul{n}\ps\pi;\;n\in\ennl,\pi\in\|F[s^n0/x]\|\}$.\\
La valeur de vérité de la formule \ $\pt x\inde\,F$ \ est donc beaucoup plus simple que celle de la
formule \ $\pt x^{\indi}F$.

\begin{theorem}\label{red_tte_f}
Soit $\phi:\ennl\to\ennl$ une fonction récursive. Il existe un $\lbd$-terme clos $\theta$ tel que,
si $m\in\ennl$, $n=\phi(m)$ et $f$ est une $\lbd$-variable, alors $\theta\ul{m}f$ se réduit à
$f\ul{n}$ par réduction de tête faible.
\end{theorem}\noindent
Il s'agit d'une variante du théorème de représentation des fonctions récursives par des
$\lbd$-termes. Elle est démontrée dans~\cite{krivine3}.

\cqfd

\begin{theorem}
Soit $\phi:\ennl^k\to\ennl$ une fonction récursive. On définit, dans ${\cal M}$, un symbole de fonction
$f$ en posant $f(s^{m_1}0,\ldots,s^{m_k}0)=s^n0$ avec $n=\phi(m_1,\ldots,m_k)$~; on prolonge $f$
de façon arbitraire sur $P^k\setminus\ennl^k$. Alors, il existe une quasi-preuve \ $\theta$ telle que~:\\
$\theta\force\pt x_1\ldots\pt x_k[$int$(x_1),\ldots,\,$int$(x_k)\to$int$(f(x_1,\ldots,x_k))]$.
\end{theorem}\noindent
Pour simplifier, on suppose $k=1$.
D'après le théorème~\ref{mm1}, il suffit de trouver une quasi-preuve $\theta$ telle que \
$\theta\force\pt x[e(x),(e(f(x))\to\bot)\to\bot]$. Autrement dit~:\\
$\theta\force e(p),(e(f(p))\to\bot)\to\bot$ pour tout $p\in P$.\\
On peut supposer que $p=s^m0$ (sinon, $e(p)=\vide$ et le résultat est trivial).\\
On a donc \ $e(p)=\{\ul{m}\}$~; on doit donc avoir $\theta\star\ul{m}\ps\xi\ps\pi\in\bbot$ pour tout
$\pi\in\PPi$ et $\xi\force e(s^n0)\to\bot$, avec $n=\phi(m)$.\\
On prend le $\lbd$-terme \ $\theta$ donné par le théorème~\ref{red_tte_f}. D'après ce théorème, on a~:\\
$\theta\star\ul{m}\ps\xi\ps\pi\succ\xi\star\ul{n}\ps\pi$, qui est dans $\bbot$, par hypothèse sur $\xi$.

\cqfd

\smallskip\noindent
{\bfseries Remarque.} On a ainsi réalisé par des quasi-preuves, tous les axiomes de l'arithmétique
du second ordre, avec un symbole de fonction pour chaque fonction récursive.

\subsection*{Algèbres standard}\noindent
Une algèbre de réalisabilité ${\cal A}$ est dite \emph{standard} si
son ensemble de termes $\Lbd$ et son ensemble de piles $\Pi$ sont définis comme suit~:\\
On a un ensemble dénombrable $\Pi_0$ qui est l'ensemble des \emph{constantes de pile}.\\
Les termes et les piles de ${\cal A}$ sont les suites finies d'éléments de l'ensemble~:\\
\centerline{$\Pi_0\cup\{B,C,E,I,K,W,\ccc,\vsig,\chi,\chi',\kk,(,),[,],\ps\}$}

\noindent
qui sont obtenus par les règles suivantes~:

\smallskip\noindent
$\bullet$~~$B,C,E,I,K,W,\ccc,\vsig,\chi,\chi'$ sont des termes~;\\
$\bullet$~~chaque élément de $\Pi_0$ est une pile~;\\
$\bullet$~~si $\xi,\eta$ sont des termes, alors $(\xi)\eta$ est un terme~;\\
$\bullet$~~si $\xi$ est un terme et $\pi$ une pile, alors $\xi\ps\pi$ est une pile~;\\
$\bullet$~~si $\pi$ est une pile, alors $\kk[\pi]$ est un terme.

\smallskip\noindent
Un terme de la forme $\kk[\pi]$ est appelé \emph{continuation}. Il sera noté aussi $\kk_\pi$.

\smallskip\noindent
L'ensemble des processus de l'algèbre ${\cal A}$ est $\Lbd\fois\Pi$.\\
Si $\xi\in\Lbd$ et $\pi\in\Pi$, le couple $(\xi,\pi)$ est noté $\xi\star\pi$.

\smallskip\noindent
Une pile est donc de la forme \ $\pi=\xi_1\ps\ldots\ps\xi_n\ps\pi_0$, où $\xi_1,\ldots,\xi_n\in\Lbd$
et $\pi_0\in\Pi_0$ ($\pi_0$ est une constante de pile). Etant donné un terme $\tau$, on pose
$\pi^\tau=\xi_1\ps\ldots\ps\xi_n\ps\tau\ps\pi_0$.

\smallskip\noindent
On fixe une bijection récursive de $\Pi$ sur $\ennl$, notée \ $\pi\mapsto\nn_\pi$.

\smallskip\noindent
On définit une relation de préordre, notée $\succ$, sur $\Lbd\star\Pi$. C'est la plus petite relation
réflexive et transitive telle que, quels que soient $\xi,\eta,\zeta\in\Lbd$ et $\pi,\varpi\in\Pi$,
on ait~:

\smallskip\noindent
$(\xi)\eta\star\pi\succ\xi\star\eta\ps\pi$.\\
$I\star\xi\ps\pi\succ\xi\star\pi$.\\
$K\star\xi\ps\eta\ps\pi\succ\xi\star\pi$.\\
$E\star\xi\ps\eta\ps\pi\succ(\xi)\eta\star\pi$.\\
$W\star\xi\ps\eta\ps\pi\succ\xi\star\eta\ps\eta\ps\pi$.\\
$C\star\xi\ps\eta\ps\zeta\ps\pi\succ\xi\star\zeta\ps\eta\ps\pi$.\\
$B\star\xi\ps\eta\ps\zeta\ps\pi\succ(\xi)(\eta)\zeta\star\pi$.\\
$\ccc\star\xi\ps\pi\succ\xi\star\kk_\pi\ps\pi$.\\
$\kk_\pi\star\xi\ps\varpi\succ\xi\star\pi$.\\
$\vsig\star\xi\ps\pi\succ\xi\star\ul{\nn}_\pi\ps\pi$.\\
$\chi\star\xi\ps\pi^\tau\succ\xi\star\tau\ps\pi$.\\
$\chi'\star\xi\ps\tau\ps\pi\succ\xi\star\pi^\tau$.

\smallskip\noindent
On se donne enfin une partie $\bbot$ de $\Lbd\star\Pi$ qui est un segment terminal pour
ce préordre, c'est-à-dire que~: \ $\p\in\bbot$, $\p'\succ\p$ $\Fl$ $\p'\in\bbot$.\\
Autrement dit, on demande que $\bbot$ ait les propriétés suivantes~:

\smallskip\noindent
$(\xi)\eta\star\pi\notin\bbot\Fl\xi\star\eta\ps\pi\notin\bbot$.\\
$I\star\xi\ps\pi\notin\bbot\Fl\xi\star\pi\notin\bbot$.\\
$K\star\xi\ps\eta\ps\pi\notin\bbot\Fl\xi\star\pi\notin\bbot$.\\
$E\star\xi\ps\eta\ps\pi\notin\bbot\Fl(\xi)\eta\star\pi\notin\bbot$.\\
$W\star\xi\ps\eta\ps\pi\notin\bbot\Fl\xi\star\eta\ps\eta\ps\pi\notin\bbot$.\\
$C\star\xi\ps\eta\ps\zeta\ps\pi\notin\bbot\Fl\xi\star\zeta\ps\eta\ps\pi\notin\bbot$.\\
$B\star\xi\ps\eta\ps\zeta\ps\pi\notin\bbot\Fl(\xi)(\eta)\zeta\star\pi\notin\bbot$.\\
$\ccc\star\xi\ps\pi\notin\bbot\Fl\xi\star\kk_\pi\ps\pi\notin\bbot$.\\
$\kk_\pi\star\xi\ps\varpi\notin\bbot\Fl\xi\star\pi\notin\bbot$.\\
$\vsig\star\xi\ps\pi\notin\bbot\Fl\xi\star \ul{\nn}_\pi\ps\pi\notin\bbot$.\\
$\chi\star\xi\ps\pi^\tau\notin\bbot\Fl\xi\star\tau\ps\pi\notin\bbot$.\\
$\chi'\star\xi\ps\tau\ps\pi\notin\bbot\Fl\xi\star\pi^\tau\notin\bbot$.

\smallskip\noindent
{\bfseries Remarque.} Le seul élément non fixé dans une algèbre de réalisabilité standard
est donc l'ensemble de processus $\bbot$.

\subsubsection*{L'axiome du choix sur les individus (ACI)}\noindent
Soient ${\cal A}$ une algèbre de réalisabilité standard et ${\cal M}$ un ${\cal A}$-modèle, dont
l'ensemble d'indi\-vidus est noté $P$. On a alors~:

\begin{theorem}[ACI]\label{ACI}
Pour chaque formule close \ $\pt x_1\ldots\pt x_m\pt y\,F$ \ avec paramètres, il existe une fonction
$f:P^{m+1}\to P$ telle que l'on ait~:\\
i)~~$\vsig\force\pt x_1\ldots\pt x_m(\pt x($int$(x)\to F[f(x_1,\ldots,x_m,x)/y])\to\pt y\,F)$.\\
ii)~~$\vsig\force\pt x_1\ldots\pt x_m(\pt x($ent$(x)\to F[f(x_1,\ldots,x_m,x)/y])\to\pt y\,F)$.
\end{theorem}\noindent
Pour $p_1,\ldots,p_m,k\in P$, on définit $f(p_1,\ldots,p_m,k)$ de façon arbitraire si $k\notin\ennl$.\\
Si $k\in\ennl$, on a $k=\nn_{\pi_k}$ pour une pile $\pi_k\in\Pi$ et une seule. 
On définit la fonction $f(p_1,\ldots,p_m,k)$ au moyen de l'axiome du choix, de façon que, s'il existe
$q\in P$ tel que \ $\pi_k\in\|F[p_1,\ldots,p_m,q]\|$, on ait $\pi_k\in\|F[p_1,\ldots,p_m,f(p_1,\ldots,p_m,k)]\|$.

\smallskip\noindent
i)~On doit montrer \ $\vsig\force\pt x($int$(x)\to F[p_1,\ldots,p_m,f(p_1,\ldots,p_m,x)])\to F[p_1,\ldots,p_m,q]$,
quels que soient $p_1,\ldots,p_m,q\in P$.\\
Soient donc \ $\xi\force\pt x($int$(x)\to F[p_1,\ldots,p_n,f(p_1,\ldots,p_n,x)])$ et
$\pi\in\|F[p_1,\ldots,p_m,q]\|$~; on doit montrer
$\vsig\star\xi\ps\pi\in\bbot$, soit \ $\xi\star\ul{\nn}_\pi\ps\pi\in\bbot$. Or, on a~:\\
$\xi\force$int$(\nn_\pi)\to F[p_1,\ldots,p_m,f(p_1,\ldots,p_m,\nn_\pi)]$ par hypothèse sur $\xi$~;\\
$\ul{\nn}_\pi\force$int$(\nn_\pi)$ d'après le théorème~\ref{adequat}~;\\
$\pi\in\|F[p_1,\ldots,p_m,f(p_1,\ldots,p_m,\nn_\pi)]\|$ par hypothèse sur $\pi$ et par définition de $f$.

\smallskip\noindent
ii)~La preuve est la même~; on observe simplement que \ $\ul{\nn}_\pi\in|$ent$(\nn_\pi)|$.

\cqfd

\subsection*{Modèles génériques}\noindent
A partir d'une algèbre de réalisabilité \emph{standard} ${\cal A}$ et d'un ${\cal A}$-modèle ${\cal M}$,
on construit une nouvelle algèbre de réalisabilité ${\cal B}$ et un ${\cal B}$-modèle ${\cal N}$,
qui est dit \emph{générique} sur ${\cal M}$.
Nous définirons ensuite le \emph{forcing}, qui est une transformation syntaxique sur les formules~;
c'est l'outil essentiel pour calculer les valeurs de vérité dans le modèle générique~${\cal N}$.

\smallskip\noindent
On considère donc une algèbre de réalisabilité standard ${\cal A}$ et
un ${\cal A}$-modèle ${\cal M}$ dont l'ensem\-ble d'individus est $P$.\\
On a un prédicat unaire $\C:P\to{\cal P}(\LLbd)$, une fonction binaire $\et:P^2\to P$
et un individu distingué $\1\in P$. On suppose que les données $\{\C,\et,\1\}$ constituent
une \emph{structure de forcing dans ${\cal M}$}, ce qui veut dire qu'on a la propriété suivante~:

\smallskip\noindent
Il existe six quasi-preuves $\alpha_0,\alpha_1,\alpha_2,\beta_0,\beta_1,\beta_2$ telles que~:

\smallskip\noindent
$\tau\in|\C[(p\et q)\et r]|$ $\Fl$ $\alpha_0\tau\in|\C[p\et(q\et r)]|$~;\\
$\tau\in|\C[p]|$ $\Fl$ $\alpha_1\tau\in|\C[p\et\1]|$~;\\
$\tau\in|\C[p\et q]|$ $\Fl$ $\alpha_2\tau\in|\C[q]|$~;\\
$\tau\in|\C[p]|$ $\Fl$ $\beta_0\tau\in|\C[p\et p]|$~;\\
$\tau\in|\C[p\et q]|$ $\Fl$ $\beta_1\tau\in|\C[q\et p]|$~;\\
$\tau\in|\C[((p\et q)\et r)\et s]|$ $\Fl$ $\beta_2\tau\in|\C[(p\et(q\et r))\et s]|$.

\smallskip\noindent
Nous appellerons \emph{$\C$-expression} une suite de symboles de la forme $\gamma=(\delta_0)(\delta_1)\ldots(\delta_k)$ où chaque $\delta_i$ est l'une des quasi-preuves
$\alpha_0,\alpha_1,\alpha_2,\beta_0,\beta_1,\beta_2$.\\
Une telle expression n'est pas un $\cc$-terme, mais $\gamma\tau$ en est un, pour tout $\cc$-terme $\tau$~;\\
le terme \ $\gamma\tau=(\delta_0)(\delta_1)\ldots(\delta_k)\tau$ sera aussi écrit $(\gamma)\tau$.

\smallskip\noindent
{\bfseries Notation.}
Un \emph{$\et$-terme} est un terme écrit avec les variables $p_1,\ldots,p_k$, la constante~$\1$ et le symbole
de fonction binaire $\et$. Soient $t(p_1,\ldots,p_k),u(p_1,\ldots,p_k)$ deux $\et$-termes. La notation~:\\
\centerline{$\gamma::t(p_1,\ldots,p_k)\Fl u(p_1,\ldots,p_k)$}
signifie que $\gamma$ est une $\C$-expression telle que \
$\tau\in|\C[t(p_1,\ldots,p_k)]|$ $\Fl$ $(\gamma)\tau\in|\C[u(p_1,\ldots,p_k)]|$.

\smallskip\noindent
Avec cette notation, les hypothèses ci-dessus s'écrivent donc~:

\smallskip\noindent
$\alpha_0::(p\et q)\et r\Fl p\et(q\et r)$~; \ $\alpha_1::p\Fl p\et\1$~; \
$\alpha_2::p\et q\Fl q$~;\\
$\beta_0::p\Fl p\et p$~; $\beta_1::p\et q\Fl q\et p$~; 
$\beta_2::((p\et q)\et r)\et s\Fl(p\et(q\et r))\et s$.

\begin{lemma}\label{et_calc}
Il existe des \ $\C$-expressions $\beta'_0,\beta'_1,\beta'_2,\beta_3,\beta'_3$ telles que~:\\
$\beta'_0::p\et q\Fl (p\et q)\et q$~; \ $\beta'_1::(p\et q)\et r\Fl (q\et p)\et r$~; \
$\beta'_2::p\et(q\et r)\Fl (p\et q)\et r$~;\\
$\beta_3::p\et(q\et r)\Fl p\et(r\et q)$~; $\beta'_3::(p\et(q\et r))\et s\Fl(p\et(r\et q))\et s$.
\end{lemma}\noindent
On écrit la suite des transformations, avec la $\C$-expression qui l'exécute~:

\smallskip\noindent
$\bullet$~~$\beta'_0=(\beta_1)(\alpha_2)(\alpha_0)(\beta_0)$.\\
$p\et q;\;\beta_0\;;\;(p\et q)\et(p\et q)\;;\;\alpha_0\;;p\et(q\et(p\et q))\;;\;\alpha_2\;;
q\et(p\et q)\;;\;\beta_1\;; \; (p\et q)\et q$.

\smallskip\noindent
$\bullet$~~$\beta'_2=(\beta_1)(\alpha_0)(\beta_1)(\alpha_0)(\beta_1)$.\\
$p\et(q\et r)\;;\beta_1\;;\;(q\et r)\et p\;;\;\alpha_0\;;\;q\et(r\et p)\;;\;\beta_1\;;\;
(r\et p)\et q\;;\;\alpha_0\;;\;r\et(p\et q)\;;\;\beta_1\;;\;(p\et q)\et r$.

\smallskip\noindent
$\bullet$~~$\beta'_1=(\alpha_2)(\alpha_0)(\beta_2)(\beta_1)(\alpha_0)(\alpha_2)(\beta_1)(\beta'_2)
(\beta'_0)(\beta_1)$.\\
$(p\et q)\et r\;;\;\beta_1\;;\;r\et(p\et q)\;;\;\beta'_0\;\;(r\et(p\et q))\et(p\et q)\;;\beta'_2\;;
((r\et(p\et q))\et p)\et q\;;\;\beta_1\;;\;q\et((r\et(p\et q))\et p)\;;\\
\alpha_2\;;\;(r\et(p\et q))\et p\;;\;\alpha_0\;;\;r\et((p\et q)\et p)\;;\beta_1\;;\;
((p\et q)\et p)\et r\;;\;\beta_2\;;\;(p\et(q\et p))\et r\;;\;\alpha_0\;;\;p\et((q\et p)\et r)\;;\;\\
\alpha_2\;;\;(q\et p)\et r$.

\smallskip\noindent
$\bullet$~~$\beta_3=(\beta_1)(\beta'_1)(\beta_1)$.\\
$p\et(q\et r)\;;\;\beta_1\;;\;(q\et r)\et p\;;\;\beta'_1\;;\;(r\et q)\et p\;;\;\beta_1\;;\;
p\et(r\et q)$.

\smallskip\noindent
$\bullet$~~$\beta'_3=(\beta'_1)(\beta'_2)(\beta'_1)(\alpha_0)(\beta'_1)$.\\
$(p\et(q\et r))\et s\;;\;\beta'_1\;;\;((q\et r)\et p)\et s\;;\;\alpha_0\;;\;(q\et r)\et(p\et s)\;;\;
\beta'_1\;;\;(r\et q)\et(p\et s)\;;\;\beta'_2\;;\;((r\et q)\et p)\et s\;;\;\beta'_1\;;\;\\
(p\et(r\et q))\et s$.

\cqfd

\begin{theorem}\label{gamma_t_tu}
Soient $t,u$ deux $\et$-termes tels que toute variable de $u$ apparaisse dans $t$. Alors, il existe une \
$\C$-expression $\gamma$  telle que $\gamma::t\Fl t\et u$.
\end{theorem}\noindent

\begin{lemma}\label{gamma_t_tp}
Soient $t$ un $\et$-terme et $p$ une variable de $t$. Alors, il existe une \
$\C$-expression \ $\gamma$ \ telle que \ $\gamma::t\Fl t\et p$.
\end{lemma}\noindent
On raisonne par récurrence sur le nombre de symboles de $t$ qui se trouvent après la dernière occurrence de $p$.
Si ce nombre est $0$, on a $t=p$ ou $t=u\et p$. On a alors $\gamma=\beta_0$ ou $\beta'_0$
(lemme~\ref{et_calc}).\\
Sinon, on a $t=u\et v$~; si la dernière occurrence de $p$ est dans $u$, l'hypothèse de récurrence donne \
$\gamma'::v\et u\Fl(v\et u)\et p$. On a alors \ $\gamma=(\beta'_1)(\gamma')(\beta_1)$.\\
Si la dernière occurrence de $p$ est dans $v$, on a $v=v_0\et v_1$. Si cette occurrence est dans $v_0$,  
l'hypothèse de récurrence donne \ $\gamma'::u\et(v_1\et v_0)\Fl(u\et(v_1\et v_0))\et p$. On pose \ $\gamma=(\beta'_3)(\gamma')(\beta_3)$ (lemme~\ref{et_calc}).\\
Si cette occurrence est dans $v_1$, l'hypothèse de récurrence donne\\
$\gamma'::(u\et v_0)\et v_1\Fl((u\et v_0)\et v_1)\et p$. On pose alors $\gamma=(\beta_2)(\gamma')(\beta'_2)$.

\cqfd

\smallskip\noindent
On montre le théorème~\ref{gamma_t_tu} par récurrence sur la longueur de $u$.\\
Si $u=\1$, on a $\gamma=\alpha_1$~; si $u$ est une variable, on applique le lemme~\ref{gamma_t_tp}.\\
Si $u=v\et w$, l'hypothèse de récurrence donne \ $\gamma'::t\Fl t\et v$ \ et aussi \
$\gamma''::t\et v\Fl (t\et v)\et w$. On pose alors \ $\gamma=(\alpha_0)(\gamma'')(\gamma')$.

\cqfd

\begin{corollary}\label{gamma_t_u}
Soient $t,u$ deux $\et$-termes tels que toute variable de $u$ apparaisse dans $t$. Alors, il existe une \
$\C$-expression $\gamma$ telle que $\gamma::t\Fl u$.
\end{corollary}\noindent
D'après le théorème~\ref{gamma_t_tu}, on a \ $\gamma'::t\Fl t\et u$. On peut donc poser
$\gamma=(\alpha_2)(\gamma')$.

\cqfd

\begin{corollary}\label{CpqCqp}
Il existe des \ $\C$-expressions $\gamma_0,\gamma_I,\gamma_K,\gamma_E,\gamma_W,\gamma_C,\gamma_B,
\gamma_{\sf cc},\gamma_{\sf k}$ telles que~:\\
$\gamma_I::p\et q\Fl q$~; \ $\gamma_K::\1\et(p\et(q\et r))\Fl p\et r$~; \
$\gamma_E::\1\et(p\et(q\et r))\Fl (p\et q)\et r$~;\\
$\gamma_W::\1\et(p\et(q\et r))\Fl p\et(q\et(q\et r))$~; \
$\gamma_C::\1\et(p\et(q\et(r\et s)))\Fl p\et(r\et(q\et s))$~;\\
$\gamma_B::\1\et(p\et(q\et(r\et s)))\Fl (p\et(q\et r))\et s$~; \
$\gamma_{\sf cc}::\1\et(p\et q)\Fl p\et(q\et q)$~;\\
$\gamma_{\sf k}::p\et(q\et r)\Fl q\et p$.
\end{corollary}\noindent

\begin{lemma}\label{A_gamma}
Pour chaque \ $\C$-expression $\gamma$, on pose \ $\ov{\gamma}=\lbd x(\chi)\lbd y(\chi' x)(\gamma)y$.\\
On a alors \ $\ov{\gamma}\star\xi\ps\pi^\tau\succ\xi\star\pi^{\gamma\tau}$.
\end{lemma}\noindent
Immédiat, d'après le théorème~\ref{beta_red_gauche}. On aurait pu aussi prendre \
$\ov{\gamma}=(\chi)\lbd x\lbd y(\chi'y)(\gamma)x$.

\cqfd

\begin{proposition}\label{gamma::tu}
Si on a \ $\gamma::t(p_1,\ldots,p_k)]\Fl u(p_1,\ldots,p_k)$, alors~:\\
$(\ov{\gamma}\star\xi\ps\pi,t(p_1,\ldots,p_k))\succ(\xi\star\pi,u(p_1,\ldots,p_k))$.
\end{proposition}\noindent
Supposons $(\ov{\gamma}\star\xi\ps\pi,t(p_1,\ldots,p_k))\notin\bbbot$. Il existe donc
$\tau\in\C[t(p_1,\ldots,p_k)]$ tel que~:\\
$\ov{\gamma}\star\xi\ps\pi^\tau\notin\bbot$. On a donc $\xi\star\pi^{\gamma\tau}\notin\bbot$
et $\gamma\tau\in\C[u(p_1,\ldots,p_k)]$. Par suite~:\\
$(\xi\star\pi,u(p_1,\ldots,p_k))\notin\bbbot$.

\cqfd

\subsubsection*{L'algèbre ${\cal B}$}\noindent
On définit une algèbre de réalisabilité ${\cal B}$ dont l'ensemble des termes est
$\LLbd=\Lbd\fois P$, l'ensemble des piles est \ $\PPi=\Pi\fois P$ et l'ensemble des processus
est $\LLbd=(\Lbd\star\Pi)\fois P$.\\
L'ensemble de processus $\bbot_{\cal B}$ de cette algèbre sera noté $\bbbot$. Il est défini
comme suit~:\\
$(\xi\star\pi,p)\in\bbbot$ \ $\Dbfl$ \ $(\pt\tau\in\C[p])\,\xi\star\pi^\tau\in\bbot$.

\smallskip\noindent
Pour $(\xi,p)\in\LLbd$ et $(\pi,q)\in\PPi$), on pose~:

\smallskip\noindent
$(\xi,p)\star(\pi,q)=(\xi\star\pi,p\et q)$~;\\
$(\xi,p)\ps(\pi,q)=(\xi\ps\pi,p\et q)$.

\smallskip\noindent
Pour $(\xi,p),(\eta,q)\in\LLbd$, on pose~:

\smallskip\noindent
$(\xi,p)(\eta,q)=(\ov{\alpha}_0\xi\eta,p\et q)$.

\begin{lemma}
On a \ $(\xi,p)(\eta,q)\star(\pi,r)\notin\bbbot$ \ $\Fl$ \ $(\xi,p)\star(\eta,q)\ps(\pi,r)\notin\bbbot$.
\end{lemma}\noindent
Par hypothèse, on a \ $(\ov{\alpha}_0\xi\eta\star\pi,(p\et q)\et r)\notin\bbbot$~; il existe donc
$\tau\in\C[(p\et q)\et r]$ tel que~:\\
$\ov{\alpha}_0\xi\eta\star\pi^\tau\notin\bbot$. D'après le lemme~\ref{A_gamma}, on a
$\xi\star\eta\ps\pi^{\alpha_0\tau}\notin\bbot$~; comme $\alpha_0\tau\in\C[p\et(q\et r)]$,
on a \ $(\xi\star\eta\ps\pi,p\et(q\et r))\notin\bbot$ et donc
\ $(\xi,p)\star(\eta,q)\ps(\pi,r)\notin\bbbot$.

\cqfd

\smallskip\noindent
On définit les combinateurs élémentaires {\bfseries B, C, E, I, K, W, {\sffamily cc}} de l'algèbre ${\cal B}$
en posant~:

\smallskip\noindent
{\bf B} $=(B^*,\1)$~; {\bf C} $=(C^*,\1)$~; {\bf E} $=(E^*,\1)$~;
{\bf I} $=(I^*,\1)$~; {\bf K} $=(K^*,\1)$~; {\bf W} $=(W^*,\1)$~; {\bf\sffamily cc} $=(\ccc^*,\1)$\\
avec $B^*=\lbd x\lbd y\lbd z(\ov{\gamma}_B)(\ov{\alpha}_0x)(\ov{\alpha}_0)yz$~; $C^*=\ov{\gamma}_CC$~;
$E^*=\lbd x\lbd y(\ov{\gamma}_E)(\ov{\alpha}_0)xy$~; $I^*=\ov{\gamma}_II$~;\\
$K^*=\ov{\gamma}_KK$~; $W^*=\ov{\gamma}_WW$~; \
$\ccc^*=(\chi)\lbd x\lbd y(\ccc)\lbd k((\chi'y)(\gamma_{\sf cc})x)(\chi)\lbd x\lbd y(k)(\chi'y)(\gamma_{\sf k})x$.

\smallskip\noindent
On pose \ {\bf\sffamily k}$_{(\pi,p)}=(\kk^*_\pi,p)$ \ avec \
$\kk^*_\pi=(\chi)\lbd x\lbd y(\kk_\pi)(\chi'y)(\gamma_{\sf k})x$.

\begin{theorem}\label{combi_compo}
Quels que soient $\tilde{\xi},\tilde{\eta},\tilde{\zeta}\in\LLbd$ et $\tilde{\pi},\tilde{\varpi}\in\PPi$,
on a~:\\
$\mathbf{I}\star\tilde{\xi}\ps\tilde{\pi}\notin\bbbot$ \ $\Fl$ \
$\tilde{\xi}\star\tilde{\pi}\notin\bbbot$~;\\
$\mathbf{K}\star\tilde{\xi}\ps\tilde{\eta}\ps\tilde{\pi}\notin\bbbot$ \ $\Fl$ \ $\tilde{\xi}\star\tilde{\pi}\notin\bbbot$~;\\
$\mathbf{E}\star\tilde{\xi}\ps\tilde{\eta}\ps\tilde{\pi}\notin\bbbot$ \ $\Fl$ \ $(\tilde{\xi})\tilde{\eta}\star\tilde{\pi}\notin\bbbot$~;\\
$\mathbf{W}\star\tilde{\xi}\ps\tilde{\eta}\ps\tilde{\pi}\notin\bbbot$ \ $\Fl$ \ $\tilde{\xi}\star\tilde{\eta}\ps\tilde{\eta}\ps\tilde{\pi}\notin\bbbot$.\\
$\mathbf{B}\star\tilde{\xi}\ps\tilde{\eta}\ps\tilde{\zeta}\ps\tilde{\pi}\notin\bbbot$ \ $\Fl$ \ $(\tilde{\xi})(\tilde{\eta})\tilde{\zeta}\star\tilde{\pi}\notin\bbbot$~;\\
$\mathbf{C}\star\tilde{\xi}\ps\tilde{\eta}\ps\tilde{\zeta}\ps\tilde{\pi}\notin\bbbot$ \ $\Fl$ \ $\tilde{\xi}\star\tilde{\zeta}\ps\tilde{\eta}\ps\tilde{\pi}\notin\bbbot$.\\
{\bf\sffamily cc} $\star\,\tilde{\xi}\ps\tilde{\pi}\notin\bbbot$ \ $\Fl$ \
$\tilde{\xi}\;\star$ {\bf\sffamily k}$_{\tilde{\pi}}\ps\tilde{\pi}\notin\bbbot$.\\
{\bf\sffamily k}$_{\tilde{\pi}}\star\tilde{\xi}\ps\tilde{\varpi}\notin\bbbot$ \ $\Fl$ \
$\tilde{\xi}\star\tilde{\pi}\notin\bbbot$.
\end{theorem}\noindent
A titre d'exemples, on fait la démonstration pour {\bfseries W, B, {\sffamily k}}$_{\tilde{\pi}}$,
{\bf\sffamily cc}.\\
On pose $\tilde{\xi}=(\xi,p),\tilde{\eta}=(\eta,q),\tilde{\zeta}=(\zeta,r),
\tilde{\pi}=(\pi,s),\tilde{\varpi}=(\varpi,q)$.

\smallskip\noindent
Supposons \ $\mathbf{W}\star\tilde{\xi}\ps\tilde{\eta}\ps\tilde{\pi}\notin\bbbot$, donc
$(\ov{\gamma}_WW\star\xi\ps\eta\ps\pi,\1\et(p\et(q\et s)))\notin\bbbot$.\\
Il existe donc $\tau\in\C[\1\et(p\et(q\et s))]$ \ tel que \
$\ov{\gamma}_WW\star\xi\ps\eta\ps\pi^\tau\notin\bbot$.\\
Comme $\ov{\gamma}_WW\star\xi\ps\eta\ps\pi^\tau\succ\xi\star\eta\ps\eta\ps\pi^{\gamma_W\tau}$,
on a \ $\xi\star\eta\ps\eta\ps\pi^{\gamma_W\tau}\notin\bbot$.\\
Mais\ $\gamma_W\tau\in\C[p\et(q\et(q\et s))]$ et il en résulte que l'on a
$\tilde{\xi}\star\tilde{\eta}\ps\tilde{\eta}\ps\tilde{\pi}\notin\bbbot$.

\smallskip\noindent
Supposons \ $\mathbf{B}\star\tilde{\xi}\ps\tilde{\eta}\ps\tilde{\zeta}\ps\tilde{\pi}\notin\bbbot$, soit \
$(B^*\star\xi\ps\eta\ps\zeta\ps\pi,\1\et(p\et(q\et(r\et s))))\notin\bbbot$.\\
Il existe donc $\tau\in\C[\1\et(p\et(q\et(r\et s)))]$ \ tel que \
$B^*\star\xi\ps\eta\ps\zeta\ps\pi^\tau\notin\bbot$.\\
Or, on a \
$B^*\star\xi\ps\eta\ps\zeta\ps\pi^\tau\succ(\ov{\gamma}_B)(\ov{\alpha}_0\xi)(\ov{\alpha}_0)\eta\zeta\star\pi^\tau$
(d'après le théorème~\ref{beta_red_gauche})\\
$\succ(\ov{\alpha}_0\xi)(\ov{\alpha}_0)\eta\zeta\star\pi^{\gamma_B\tau}$ (d'après le lemme~\ref{A_gamma}).
Donc, on a \ $(\ov{\alpha}_0\xi)(\ov{\alpha}_0)\eta\zeta\star\pi^{\gamma_B\tau}\notin\bbot$.\\
Mais \ $\gamma_B\tau\in\C[(p\et(q\et r))\et s]$ et on a donc~:\\
$((\ov{\alpha}_0\xi)(\ov{\alpha}_0)\eta\zeta\star\pi,(p\et(q\et r))\et s)\notin\bbbot$,
autrement dit $(\tilde{\xi})(\tilde{\eta})\tilde{\zeta}\star\tilde{\pi}\notin\bbbot$.

\smallskip\noindent
Supposons \ {\bf\sffamily k}$_{\tilde{\pi}}\star\tilde{\xi}\ps\tilde{\varpi}\notin\bbbot$, \ soit \
$(\kk^*_\pi\star\xi\ps\varpi,s\et(p\et q))\notin\bbbot$. Il existe donc $\tau\in\C[s\et(p\et q)]$
tel que \ $\kk^*_\pi\star\xi\ps\varpi^\tau\notin\bbot$. Or, on a
$\kk^*_\pi\star\xi\ps\varpi^\tau\succ\lbd x\lbd y(\kk_\pi)(\chi'y)(\gamma_{\sf k})x\star\tau\ps\xi\ps\varpi\succ
(\kk_\pi)(\chi'\xi)(\gamma_{\sf k})\tau\star\varpi$ (d'après le théorème~\ref{beta_red_gauche})
$\succ(\chi'\xi)(\gamma_{\sf k})\tau\star\pi\succ\chi'\star\xi\ps\gamma_{\sf k}\tau\ps\pi\succ
\xi\star\pi^{\gamma_{\sf k}\tau}$.\\
On a donc \ $\xi\star\pi^{\gamma_{\sf k}\tau}\notin\bbot$~; mais, comme \ $\gamma_{\sf k}\tau\in\C[p\et s]$,
on a bien \ $\tilde{\xi}\star\tilde{\pi}\notin\bbbot$.

\smallskip\noindent
Supposons {\bf\sffamily cc} $\star\,\tilde{\xi}\ps\tilde{\pi}\notin\bbbot$, \ soit \
$(\ccc^*\star\xi\ps\pi,\1\et(p\et s))\notin\bbbot$. Il existe donc $\tau\in\C[\1\et(p\et s)]$
tel que \ $\ccc^*\star\xi\ps\pi^\tau\notin\bbot$. Or, on a~:\\
$\ccc^*\star\xi\ps\pi^\tau\succ
\lbd x\lbd y(\ccc)\lbd k((\chi'y)(\gamma_{\sf cc})x)(\chi)\lbd x\lbd y(k)(\chi'y)(\gamma_{\sf k})x\star
\tau\ps\xi\ps\pi\\
\succ(\ccc)\lbd k((\chi'\xi)(\gamma_{\sf cc})\tau)(\chi)\lbd x\lbd y(k)(\chi'y)(\gamma_{\sf k})x\star\pi\\
\succ((\chi'\xi)(\gamma_{\sf cc})\tau)(\chi)\lbd x\lbd y(\kk_\pi)(\chi'y)(\gamma_{\sf k})x\star\pi
\succ\chi'\star\xi\ps\gamma_{\sf cc}\tau\ps(\chi)\lbd x\lbd y(\kk_\pi)(\chi'y)(\gamma_{\sf k})x\ps\pi\\
\succ\xi\star(\chi)\lbd x\lbd y(\kk_\pi)(\chi'y)(\gamma_{\sf k})x\ps\pi^{\gamma_{\sf cc}\tau}
\equiv\xi\star\kk^*_\pi\ps\pi^{\gamma_{\sf cc}\tau}$.\\
Il en résulte que \ $\xi\star\kk^*_\pi\ps\pi^{\gamma_{\sf cc}\tau}\notin\bbot$. Or, on a
$\gamma_{\sf cc}\tau\in\C[p\et(s\et s)]$ et il en résulte que l'on a \
$(\xi,p)\star(\kk^*_\pi,s)\ps(\pi,s)\notin\bbbot$, \ c'est-à-dire \
$\tilde{\xi}\star$ {\bf\sffamily k}$_{\tilde{\pi}}\ps\tilde{\pi}\notin\bbbot$.

\cqfd

\smallskip\noindent
On a ainsi défini complètement l'algèbre de réalisabilité ${\cal B}$.

\smallskip\noindent
Pour chaque $\cc$-terme clos $t$ (quasi-preuve), désignons par $t_{\cal B}$ sa valeur dans l'algèbre
${\cal B}$ (sa valeur dans l'algèbre standard ${\cal A}$ est $t$ lui-même). On pose
$t_{\cal B}=(t^*,\1_t)$, où $t^*$ est une quasi-preuve et $\1_t$ une condition écrite avec $\1$,
$\et$ et les parenthèses, qui sont définis comme suit par récurrence sur $t$~:

\smallskip\noindent
$\bullet$~~Si $t$ est un combinateur élémentaire $B,C,E,I,K,W,{\sf cc}$, alors $t^*$ a déjà été défini~;
$\1_t=\1$.\\
$\bullet$~~$(tu)^*=\ov{\alpha}_0t^*u^*$~; $\1_{tu}=\1_t\et\1_u$.

\subsubsection*{Le modèle ${\cal N}$}\noindent
Le ${\cal B}$-modèle ${\cal N}$ a le même ensemble d'individus $P$ et les mêmes fonctions que ${\cal M}$.\\
Par définition, les prédicats d'arité~$k$ de ${\cal N}$ sont les applications de $P^k$ dans ${\cal P}(\PPi)$.
Mais comme $\PPi=\Pi\fois P$, ils s'identifient aux applications de $P^{k+1}$ dans ${\cal P}(\Pi)$,
c'est-à-dire aux prédicats d'arité~$k+1$ du modèle ${\cal M}$.\\
Chaque constante de prédicat \ {\sf R}, d'arité $k$, est interprétée dans le modèle ${\cal M}$,
par une application ${\sf R}_{\cal M}:P^k\to{\cal P}(\Lbd)$.
Dans le modèle ${\cal N}$, cette constante de prédicat est interprétée par l'application \
${\sf R}_{\cal N}:P^k\to{\cal P}(\LLbd)$, où \
${\sf R}_{\cal N}(p_1,\ldots,p_k)={\sf R}_{\cal M}(p_1,\ldots,p_k)\fois\{\1\}$.

\smallskip\noindent
Pour chaque formule close $F$ à paramètres dans ${\cal N}$, sa valeur de vérité, qui est une partie
de~$\PPi$, sera notée $\vv F\vv$. On écrira \ $(\xi,p)\fforce F$ pour exprimer que $(\xi,p)\in\LLbd$
réalise $F$, autrement dit \
$(\pt\pi\in\Pi)(\pt q\in P)((\pi,q)\in\vv F\vv)\Fl(\xi,p)\star(\pi,q)\in\bbbot)$.

\begin{theorem}\label{adequat_B}\ \\
Si on a \ $\vdash t:A$ en logique classique du second ordre, où $A$ est une formule close, alors\\
$t_{\cal B}=(t^*,\1_t)\fforce A$.
\end{theorem}\noindent 
Application immédiate du théorème~\ref{adequat} (lemme d'adéquation) dans le ${\cal B}$-modèle ${\cal N}$.

\cqfd

\begin{proposition}\label{1_to_p}\ \\
i) Si \ $(\xi,\1)\fforce F$, alors \ $(\ov{\gamma}\xi,p)\fforce F$ pour tout $p\in P$, avec \
$\gamma::p\et q\Fl\1\et q$.\\
ii) Soient $\xi,\eta\in\Lbd$ tels que $\xi\star\pi\succ\eta\star\pi$ pour toute $\pi\in\Pi$. On a alors~:\\
$(\xi\star\pi,p)\succ(\eta\star\pi,p)$ quels que soient $\pi\in\Pi$ et $p\in P$~;\\
$(\eta,p)\fforce F$ \ $\Fl$ \ $(\xi,p)\fforce F$ pour toute formule close $F$.
\end{proposition}\noindent
i)~~On doit montrer que, pour tout $(\pi,q)\in\vv F\vv$, on a \
$(\ov{\gamma}\xi,p)\star(\pi,q)\in\bbbot$, soit~:\\
$(\ov{\gamma}\xi\star\pi,p\et q)\in\bbbot$. Soit donc \ $\tau\in\C[p\et q]$, d'où $\gamma\tau\in\C[\1\et q]$.\\
Comme on a, par hypothèse, $(\xi\star\pi,\1\et q)\in\bbbot$, on en déduit \
$\xi\star\pi^{\gamma\tau}\in\bbot$ et donc $\ov{\gamma}\xi\star\pi^\tau\in\bbot$.\\
ii)~~Supposons $(\xi\star\pi,p)\notin\bbbot$~; il existe donc $\tau\in\C[p]$ tel que
$\xi\star\pi^\tau\notin\bbot$. On a donc $\eta\star\pi^\tau\notin\bbot$, d'où
$(\eta\star\pi,p)\notin\bbbot$.\\
Soit $(\pi,q)\in\vv F\vv$~; on a $(\eta,p)\star(\pi,q)\in\bbbot$, soit
$(\eta\star\pi,p\et q)\in\bbot$. D'après ce qu'on vient de voir, on a donc
$(\xi\star\pi,p\et q)\in\bbot$, donc $(\xi,p)\star(\pi,q)\in\bbbot$.

\cqfd

\subsubsection*{Les entiers du modèle ${\cal N}$}\noindent
Rappelons qu'on a posé~:\\
$\sig=\lbd n\lbd f\lbd x(f)(n)fx$, \ $\ul{0}=\lbd x\lbd y\,y$ \ et \
$\ul{n}=(\sig)^n\ul{0}$ \ pour tout entier $n$.\\  
On a donc \ $\sig_{\cal B}=(\sig^*,\1_\sig)$ \ et \
$\ul{n}_{\cal B}=((\sig)^n\ul{0})_{\cal B}=(\ul{n}^*,\1_{\ul{n}})$.\\
Donc $\ul{0}_{\cal B}=(KI)_{\cal B}=(K^*,\1)(I^*,\1)$ \ et \
$\ul{n+1}_{\cal B}=\sig_{\cal B}\ul{n}_{\cal B}=(\sig^*,\1_\sig)(\ul{n}^*,\1_{\ul{n}})$.\\
Les définitions par récurrence de $\ul{n}^*,\1_{\ul{n}}$ sont donc les suivantes~:\\
$\ul{0}^*=\ov{\alpha}_0K^*I^*$~; $(\ul{n+1})^*=\ov{\alpha}_0\sig^*\ul{n}^*$~;\\
$\1_{\ul{0}}=\1\et\1$~; $\1_{\ul{n+1}}=\1_\sig\et\1_{\ul{n}}$.

\smallskip\noindent
On peut définir le prédicat unaire \ ent$(x)$ \ dans le modèle ${\cal N}$ de deux façons distinctes~:

\smallskip\noindent
i)~~A partir du prédicat \ ent$(x)$ \ du modèle ${\cal M}$, en posant~:\\
$|$ent$(s^n0)|=\{(\ul{n},\1)\}$~; $|$ent$(p)|=\vide$ si $p\notin\ennl$.\\
ii)~~Directement par la définition de ent(x) dans le modèle ${\cal N}$~; nous notons ce prédicat
ent$_{\cal N}(x)$. On a donc~:\\
|ent$_{\cal N}(s^n0)|=\ul{n}_{\cal B}$~; |ent$_{\cal N}(p)|=\vide$ si $p\notin\ennl$.\\
D'après le théorème~\ref{memoire_gen}, appliqué dans le modèle ${\cal N}$, on sait que les
prédicats int$(x)$ et ent$_{\cal N}(x)$ sont interchangeables. Le théorème~\ref{entMN} montre que
les prédicats int$(x)$ et ent$(x)$ sont aussi interchangeables. On a ainsi trois prédicats
qui définissent les entiers dans le modèle ${\cal N}$~; c'est ent$(x)$ que nous utiliserons
le plus souvent dans la suite. En particulier, on remplacera le quantificateur $\pt x\indi$
par $\pt x\inde$.

\begin{theorem}\label{entMN}\ \\
Il existe deux quasi-preuves $T,J$ telles que~:\\
i)~~$(T,\1)\fforce\pt X\pt x(($ent$(x)\to X),$ int$(x)\to X)$.\\
ii)~~$(J,\1)\fforce\pt x($ent$(x)\to$int$(x))$.
\end{theorem}\noindent
i)~~On applique le théorème~\ref{memoire_gen} à la suite $u:\ennl\to\LLbd$ définie par $u_n=(\ul{n},\1)$.\\
On cherche deux quasi-preuves $T,S$ telles que~:\\
$(S,\1)\star(\psi,p)\ps(\ul{n},\1)\ps(\pi,r)\succ(\psi,p)\star(\ul{n+1},\1)\ps(\pi,r)$~; \
$(S,\1)\fforce\top\to\bot,\top\to\bot$.\\
$(T,\1)\star(\phi,p)\ps(\nu,q)\ps(\pi,r)\succ(\nu,q)\star(S,\1)\ps(\phi,p)\ps(\ul{0},\1)\ps(\pi,r)$.

\smallskip\noindent
D'après le théorème~\ref{memoire_gen}, on aura alors le résultat cherché~:\\
$(T,\1)\fforce\pt X\pt x(($ent$(x)\to X),$ int$(x)\to X)$.

\smallskip\noindent
On pose $S=\lbd f\lbd x(\ov{\gamma}f)(\sig)x$, avec $\gamma::\1\et(p\et(q\et r))\Fl p\et(q\et r)$.

\smallskip\noindent
On a alors \ $(S,\1)\star(\psi,p)\ps(\nu,q)\ps(\pi,r)\equiv
(S\star\psi\ps\nu\ps\pi,\1\et(p\et(q\et r)))\succ\\
(\ov{\gamma}\psi\star\sig\nu\ps\pi,\1\et(p\et(q\et r)))$
(théorème~\ref{beta_red_gauche} et proposition~\ref{1_to_p}(ii))\\
$\succ(\psi\star\sig\nu\ps\pi,p\et(q\et r))$ (proposition~\ref{gamma::tu})
$\equiv(\psi,p)\star(\sig\nu,q)\ps(\pi,r)$.\\
Supposons d'abord que $(\psi,p)\fforce\top\to\bot$~; on a alors 
$(\psi,p)\star(\sig\nu,q)\ps(\pi,r)\in\bbbot$ et donc~:\\
$(S,\1)\star(\psi,p)\ps(\nu,q)\ps(\pi,r)\in\bbbot$. Cela montre que
$(S,\1)\fforce\top\to\bot,\top\to\bot$.\\
Par ailleurs, en posant $\nu=\ul{n}$, d'où $\sig\nu=\ul{n+1}$, et $q=\1$, on a bien montré que~:\\
$(S,\1)\star(\psi,p)\ps(\ul{n},\1)\ps(\pi,r)\succ(\psi,p)\star(\ul{n+1},\1)\ps(\pi,r)$.

\smallskip\noindent
On pose ensuite $T=\lbd f\lbd x(\ov{\gamma}'x)Sf\ul{0}$, avec \
$\gamma'::\1\et(p\et(q\et r))]\Fl q\et(\1\et(p\et(\1\et r)))$.

\smallskip\noindent
On a alors \ $(T,\1)\star(\phi,p)\ps(\nu,q)\ps(\pi,r)\equiv
(T\star\phi\ps\nu\ps\pi,\1\et(p\et(q\et r)))\succ\\
(\ov{\gamma}'\nu\star S\ps\phi\ps\ul{0}\ps\pi,\1\et(p\et(q\et r)))$
(théorème~\ref{beta_red_gauche} et proposition~\ref{1_to_p}(ii))\\
$\succ(\nu\star S\ps\phi\ps\ul{0}\ps\pi,q\et(\1\et(p\et(\1\et r))))$
(proposition~\ref{gamma::tu})\\
$\equiv(\nu,q)\star(S,\1)\ps(\phi,p)\ps(\ul{0},\1)\ps(\pi,r)$ \
ce qui est le résultat cherché.

\smallskip\noindent
ii)~~On cherche une quasi-preuve $J$ telle que $(J,\1)\fforce\pt x($ent$(x)\to$int$(x))$. Il suffit d'avoir~:\\
$(J,\1)\fforce$ent$(s^n0)\to$int$(s^n0)$ pour tout $n\in\ennl$, puisque |ent$(p)|=\vide$
si $p\notin\ennl$.\\
Soit $(\pi,q)\in\vv$int$(n)\vv$~;
on doit avoir \ $(J,\1)\star(\ul{n},\1)\ps(\pi,q)\in\bbbot$, \ soit \
$(J\star\ul{n}\ps\pi,\1\et(\1\et q))\in\bbbot$.\\
Or, on a \ $(\ul{n}^*,\1_{\ul{n}})=((\sig)^n\ul{0})_{\cal B}\fforce$int$(s^n0)$
(théorème~\ref{adequat}, appliqué dans ${\cal B}$) et donc~:\\
$(\ul{n}^*,\1_{\ul{n}})\star(\pi,q)\in\bbbot$ \ ou encore \
$(\ul{n}^*\star\pi,\1_{\ul{n}}\et q)\in\bbbot$.

\smallskip\noindent
Soit donc $\tau\in\C[\1\et(\1\et q)]$~; on a alors \
$(\gamma)^n(\gamma_0)\tau\in\C[\1_n\et q]$\\
où $\gamma_0$ et $\gamma$ sont deux $\C$-expressions telles que~:\\
$\gamma_0::\1\et(\1\et q)\Fl(\1\et\1)\et q$~; \ $\gamma::p\et q\Fl(\1_\sig\et p)\et q$.\\
En effet, on a vu que $\1_{\ul{0}}=\1\et\1$ et $\1_{\ul{n+1}}=\1_\sig\et\1_{\ul{n}}$.
Il en résulte que, si $\tau\in\C[\1\et(\1\et q)]$, alors
$(\gamma_0)\tau\in\C[\1_{\ul{0}}\et q]$, \ d'où \ $(\gamma)^n(\gamma_0)\tau\in\C[\1_n\et q]$.
On a donc \ $\ul{n}^*\star\pi^{(\gamma)^n(\gamma_0)\tau}\in\bbot$.\\
On construit ci-dessous deux quasi-preuves $g,j$ telles que, pour tout $n\in\ennl$, on ait~:\\
a)~~$g\star\ul{n}\ps\xi\ps\pi^\tau\succ\xi\star\pi^{(\gamma)^n(\gamma_0)\tau}$~;\\
b)~~$j\star\ul{n}\ps\xi\ps\pi\succ\xi\star\ul{n}^*\ps\pi$.\\
En posant $J=\lbd x(gx)(j)x$, on a \ $J\star\ul{n}\ps\pi^\tau\succ\ul{n}^*\star\pi^{(\gamma)^n(\gamma_0)\tau}\in\bbot$,
ce qui est le résultat voulu.

\smallskip\noindent
a)~~On pose $g=\lbd k\lbd x(\ov{\gamma}_0)(k)\ov{\gamma}x$~; on a, d'après le
théorème~\ref{beta_red_gauche}~:\\
$g\star\ul{n}\ps\xi\ps\pi^\tau\succ\ov{\gamma}_0\star(\ul{n})\ov{\gamma}\xi\ps\pi^\tau\succ
(\ul{n})\ov{\gamma}\xi\star\pi^{(\gamma_0)\tau}$.\\
Il suffit donc de montrer que \
$(\ul{n})\ov{\gamma}\xi\star\pi^\tau\succ\xi\star\pi^{(\gamma)^n\tau}$ \
ce qu'on fait par récurrence sur $n$.\\
Si $n=0$, on a immédiatement \
$\ul{0}\star\ov{\gamma}\ps\xi\ps\pi^\tau\succ\xi\star\pi^\tau$ \ puisque $\ul{0}=\lbd x\lbd y\,y$.\\
Pour passer de $n$ à $n+1$, on a \ $(\ul{n+1})\ov{\gamma}\xi\star\pi^\tau\equiv
(\sig\ul{n})\ov{\gamma}\xi\star\pi^\tau\succ\sig\star\ul{n}\ps\ov{\gamma}\ps\xi\ps\pi^\tau\\
\succ\ov{\gamma}\star(\ul{n})\ov{\gamma}\xi\ps\pi^\tau\succ
(\ul{n})\ov{\gamma}\xi\star\pi^{(\gamma)\tau}\succ
\xi\star\pi^{(\gamma)^{n+1}\tau}$ \ par hypothèse de récurrence.

\smallskip\noindent
b)~~On pose $\beta=\ov{\alpha}_0\sig^*$, $U=\lbd g\lbd y(g)(\beta)y$
et $j=\lbd k\lbd f(k)Uf\ul{0}^*$.\\
On a donc \ $j\star\ul{n}\ps\xi\ps\pi\succ\ul{n}U\xi\star\ul{0}^*\ps\pi$. On montre,
par récurrence sur $n$, que~:\\
$\ul{n}U\xi\star\ul{k}^*\ps\pi\succ\xi\star(\ul{n+k})^*\ps\pi$ pour tout entier $k$,
ce qui donne le résultat voulu avec $k=0$.\\
Pour $n=0$, on a \ $\ul{0}U\xi\star\ul{k}^*\ps\pi\succ\xi\star\ul{k}^*\ps\pi$
puisque $\ul{0}=\lbd x\lbd y\,y$.\\
Pour passer de $n$ à $n+1$, on a \ $(\ul{n+1})\star U\ps\xi\ps\ul{k}^*\ps\pi\equiv
\sig\ul{n}\star U\ps\xi\ps\ul{k}^*\ps\pi\succ U\star\ul{n}U\xi\ps\ul{k}^*\ps\pi$\\
(puisque $\sig=\lbd n\lbd f\lbd x(f)(n)fx$) $\succ\ul{n}U\xi\star\beta\ul{k}^*\ps\pi
\equiv\ul{n}U\xi\star(\ul{k+1})^*\ps\pi\succ\xi\star(\ul{n+k+1})^*\ps\pi$\\
par hypothèse de récurrence.

\cqfd

\subsection*{Forcing}\noindent
Il s'agit d'évaluer les valeurs de vérité des formules dans le ${\cal B}$-modèle générique ${\cal N}$.\\
Pour chaque variable de prédicat $X$ d'arité $k$, on ajoute au langage une nouvelle variable de prédicat,
notée $X^+$, d'arité $k+1$. Dans le ${\cal A}$-modèle ${\cal M}$, on utilisera
les variables $X$ et $X^+$~; dans le ${\cal B}$-modèle ${\cal N}$, seulement les variables $X$.

\smallskip\noindent
A chaque paramètre du second ordre d'arité $k$ du modèle ${\cal N}$, soit \
${\cal X}:P^k\to{\cal P}(\PPi)$, on associe un paramètre du second ordre d'arité $k+1$, soit \
${\cal X}^+:P^{k+1}\to{\cal P}(\Pi)$ du modèle ${\cal M}$. Il est défini de façon évidente,
puisque $\PPi=\Pi\fois P$~; on pose~:\\
${\cal X}^+(p,p_1,\ldots,p_k)=\{\pi\in\Pi;\;(\pi,p)\in{\cal X}(p_1,\ldots,p_k)\}$

\smallskip\noindent
Pour toute formule $F$ \emph{écrite sans les variables $X^+$}, avec paramètres dans le modèle ${\cal N}$,
on définit, par récurrence sur $F$, une formule notée $p\forcec F$ (lire~``~$p$ force $F$~''),
avec paramètres dans le modèle ${\cal A}$, \emph{écrite avec les variables $X^+$}
et une variable libre $p$ de condition~:

\smallskip\noindent
Si $F$ est atomique de la forme \ $X(t_1,\ldots,t_k)$, alors $p\forcec F$ est \
$\pt q(\C[p\et q]\to X^+(q,t_1,\ldots,t_k))$.\\
Si $F$ est atomique de la forme \ ${\cal X}(t_1,\ldots,t_k)$, alors $p\forcec F$ est \
$\pt q(\C[p\et q]\to{\cal X}^+(q,t_1,\ldots,t_k))$.\\
Si $F\equiv(A\to B)$ où $A,B$ sont des formules, alors \ $p\forcec F$ est \
$\pt q(q\forcec A\to p\et q\forcec B)$.\\
Si $F\equiv({\sf R}(t_1,\ldots,t_k)\to B)$, où ${\sf R}$ est une constante de prédicat, alors~:\\
$p\forcec F$ \ est \ $({\sf R}(t_1,\ldots,t_k)\to p\forcec B)$.\\
Si $F\equiv(t_1=t_2\mapsto B)$, \ alors \ $p\forcec F$ \ est \ $(t_1=t_2\mapsto p\forcec B)$.\\
Si $F\equiv\pt x\,A$, \ alors \ $p\forcec F$ \ est \ $\pt x(p\forcec A)$.\\
Si $F\equiv\pt X\,A$, \ alors \ $p\forcec F$ \ est \ $\pt X^+(p\forcec A)$.

\smallskip\noindent
On a donc, en particulier~:\\
Si $F\equiv\pt x\inde\,A\,$, \ alors \ $p\forcec F$ \ est \ $\pt x\inde(p\forcec A)$.

\begin{lemma}\label{forcecXX+}
Soient $F$ une formule dont les variables libres sont parmi $X_1,\ldots,X_k$ et \
${\cal X}_1,\ldots,{\cal X}_k$ des paramètres du second ordre du modèle ${\cal N}$, d'arités correspondantes.
On a alors~:\\
$(p\forcec F)[{\cal X}^+_1/X^+_1,\ldots,{\cal X}^+_k/X^+_k]\equiv
(p\forcec F[{\cal X}_1/X_1,\ldots,{\cal X}_k/X_k])$.
\end{lemma}\noindent
Immédiat, par récurrence sur $F$.

\cqfd

\begin{theorem}\label{forcec-fforce}\ \\
Pour chaque formule close $F$ à paramètres dans le modèle ${\cal N}$, il existe deux quasi-preuves $\chi_F,\chi'_F$, qui ne dépendent que de la structure propositionnelle de $F$,
telles que l'on ait~:\\
$\xi\force(p\forcec F)$ \ $\Fl$ \ $(\chi_F\xi,p)\fforce F$~;\\
$(\xi,p)\fforce F$ \ $\Fl$ \ $\chi'_F\xi\force(p\forcec F)$\\
quels que soient $\xi\in\Lbd$ et $p\in P$.
\end{theorem}\noindent
La \emph{structure propositionnelle} de $F$ est le type simple construit avec un seul atome $O$ et
$\to$, obtenu à partir de $F$ en supprimant tous les quantificateurs, tous les symboles $\mapsto$
avec leur hypothèse, et en identifiant toutes les formules atomiques avec $O$.\\
Par exemple, la structure propositionnelle de la formule~:\\
$\pt X(\pt x(\pt y(f(x,y)=0\mapsto Xy)\to Xx)\to\pt x\,Xx)$
est $(O\to O)\to O$.

\smallskip\noindent
Preuve par récurrence sur la longueur de $F$.\\
$\bullet$~~Si $F$ est atomique, on a $F\equiv{\cal X}(t_1,\ldots,t_k)$~; on montre que $\chi_F=\chi$
et $\chi'_F=\chi'$. On a, en effet~:\\
$\|p\forcec F\|=\|\pt q(\C[p\et q]\to{\cal X}^+(q,t_1,\ldots,t_k)\|=
\bigcup_q\{\tau\ps\pi;\;\tau\in\C[p\et q],(\pi,q)\in\vv{\cal X}(t_1,\ldots,t_k)\vv\}$.\\
En effet, par définition de ${\cal X}^+$, on a \
$\pi\in\|{\cal X}^+(q,t_1,\ldots,t_k)\|$ $\Dbfl$ $(\pi,q)\in\vv{\cal X}(t_1,\ldots,t_k)\vv$.\\
On a donc~:\\
$(*)$\hspace{2em}$\xi\force(p\forcec F)$ \ $\Dbfl$ \
$(\pt q\in P)(\pt\tau\in\C[p\et q])(\pt\pi\in\Pi)
((\pi,q)\in\vv{\cal X}(t_1,\ldots,t_k)\vv\Fl\xi\star\tau\ps\pi\in\bbot)$.

\smallskip\noindent
Par ailleurs, on a \ $(\xi,p)\fforce F$ $\Dbfl$ 
$(\pt q\in P)(\pt\pi\in\Pi)
((\pi,q)\in\vv F\vv\Fl(\xi,p)\star(\pi,q)\in\bbbot)$\\
$\Dbfl$ \ $(\pt q\in P)(\pt\pi\in\Pi)
((\pi,q)\in\vv F\vv\Fl(\xi\star\pi,p\et q)\in\bbbot)$ \ d'où enfin, par définition de $\bbbot$~:

\smallskip\noindent
$(**)$\hspace{2em}$(\xi,p)\fforce F$ \ $\Dbfl$ \ $(\pt q\in P)(\pt\tau\in\C[p\et q])(\pt\pi\in\Pi)
((\pi,q)\in\vv F\vv\Fl\xi\star\pi^\tau\in\bbot)$.

\smallskip\noindent
Supposons que \ $\xi\force(p\forcec F)$. Comme \ $\chi\xi\star\pi^\tau\succ\xi\star\tau\ps\pi$, on a
d'après $(*)$~:\\
$(\pt q\in P)(\pt\tau\in\C[p\et q])(\pt\pi\in\Pi)
((\pi,q)\in\vv{\cal X}(t_1,\ldots,t_k)\vv\Fl\chi\xi\star\tau\ps\pi\in\bbot)$\\
et donc \ $(\chi\xi,p)\fforce F$ \ d'après $(**)$.\\
Inversement, supposons que  \ $(\xi,p)\fforce F$. En appliquant $(**)$ et le fait que
$\chi'\xi\star\tau\ps\pi\succ\xi\star\pi^\tau$, on obtient \
$(\pt q\in P)(\pt\tau\in\C[p\et q])(\pt\pi\in\Pi)
((\pi,q)\in\vv F\vv\Fl\chi'\xi\star\tau\ps\pi\in\bbot)$\\
et donc \ $\chi'\xi\force(p\forcec F)$ \ d'après $(*)$.

\smallskip\noindent
$\bullet$~~Si $F\equiv\pt X\,A$, alors $p\forcec F\equiv\pt X^+(p\forcec A)$.
On a donc \ $\xi\force(p\forcec F)\equiv\pt X^+(\xi\force(p\forcec A))$.\\
Par ailleurs, on a \
$(\xi,p)\fforce F\equiv\pt X((\xi,p)\fforce A)$.\\
Soient ${\cal X}:P^k\to{\cal P}(\PPi)$ un paramètre du second ordre du modèle ${\cal N}$, de même arité
que $X$, et ${\cal X}^+$ le paramètre correspondant du modèle ${\cal M}$.\\
Si \ $\xi\force(p\forcec F)$, alors on a $(\xi\force(p\forcec A))[{\cal X}^+/X^+]$, donc 
$\xi\force(p\forcec A[{\cal X}/X])$, d'après le lemme~\ref{forcecXX+}.\\
Par hypothèse de récurrence, on a \ $(\chi_A\xi,p)\fforce A[{\cal X}/X]$. Comme ${\cal X}$ est arbitraire,
on en déduit \ $(\chi_A\xi,p)\fforce\pt X\,A$.\\
Inversement, si on a \ $(\xi,p)\fforce F$, alors \ $(\xi,p)\fforce A[{\cal X}/X]$ pour tout ${\cal X}$.\\
Par hypothèse de récurrence, on a \ $\chi'_A\xi\force(p\forcec A[{\cal X}/X])$, d'où \
$\chi'_A\xi\force(p\forcec A)[{\cal X}^+/X^+])$, d'après le lemme~\ref{forcecXX+}.
Comme \ ${\cal X}^+$ est arbitraire, on en déduit \ $\chi'_A\xi\force\pt X^+(p\forcec A)$, c'est-à-dire \
$\chi'_A\xi\force(p\forcec\pt X\,A)$.

\smallskip\noindent
$\bullet$~~Si $F\equiv\pt x\,A$, alors $p\forcec F\equiv\pt x(p\forcec A)$. Donc
$\xi\force p\forcec F\equiv\pt x(\xi\force(p\forcec A))$.\\
Par ailleurs, \ $(\xi,p)\fforce F\equiv\pt x((\xi,p)\fforce A)$.\\
Le résultat est immédiat, d'après l'hypothèse de récurrence.

\smallskip\noindent
$\bullet$~~Si $F\equiv(t_1=t_2\mapsto A)$, alors
$p\forcec F\equiv t_1=t_2\mapsto p\forcec A$. Donc~:\\
$\xi\force(p\forcec F)\equiv(t_1=t_2\mapsto\xi\force(p\forcec A))$.\\
Par ailleurs, \ $(\xi,p)\fforce F\equiv(t_1=t_2\mapsto(\xi,p)\fforce A)$.\\
Le résultat est immédiat, d'après l'hypothèse de récurrence.

\smallskip\noindent
$\bullet$~~Si $F\equiv A\to B$, on a \ $p\forcec F\equiv\pt q(q\forcec A\to p\et q\forcec B)$ et donc~:\\
$(*)$\hspace{2em}
$\xi\force(p\forcec F)$ $\Fl$ $\pt\eta\pt q(\eta\force(q\forcec A)\to\xi\eta\force(p\et q\forcec B))$.\\
Supposons \ $\xi\force(p\forcec F)$ \ et posons \ $\chi_F=\lbd x\lbd y(\ov{\gamma}_0)(\chi_B)(x)(\chi'_A)y$.\\
On doit montrer \ $(\chi_F\xi,p)\fforce A\to B$~; soient donc $(\eta,q)\fforce A$ et $(\pi,r)\in\vv B\vv$.\\
On doit montrer \ $(\chi_F\xi,p)\star(\eta,q)\ps(\pi,r)\in\bbbot$ \ soit \
$(\chi_F\xi\star\eta\ps\pi,p\et(q\et r))\in\bbbot$.\\
Soit donc \ $\tau\in\C[p\et(q\et r)]$~; on doit montrer \ $\chi_F\xi\star\eta\ps\pi^\tau\in\bbot$ \
ou encore \ $\chi_F\star\xi\ps\eta\ps\pi^\tau\in\bbot$.

\smallskip\noindent
D'après l'hypothèse de récurrence appliquée à $(\eta,q)\fforce A$, on a \ $\chi'_A\eta\force(q\forcec A)$.\\
D'après $(*)$, on a donc \ $(\xi)(\chi'_A)\eta\force(p\et q\forcec B)$.\\
En appliquant de nouveau l'hypothèse de récurrence, on en déduit~:\\
$((\chi_B)(\xi)(\chi'_A)\eta,p\et q)\fforce B$. Mais, comme $(\pi,r)\in\vv B\vv$, on a alors~:\\
$((\chi_B)(\xi)(\chi'_A)\eta,p\et q)\star(\pi,r)\in\bbbot$, \ soit \
$((\chi_B)(\xi)(\chi'_A)\eta\star\pi,(p\et q)\et r)\in\bbbot$.\\
Comme \ $\tau\in\C[p\et(q\et r)]$, on a \ $\gamma_0\tau\in\C[(p\et q)\et r]$ \ et donc \
$(\chi_B)(\xi)(\chi'_A)\eta\star\pi^{\gamma_0\tau}\in\bbot$.\\
Mais, par définition de $\chi_F$, on a, d'après le théorème~\ref{beta_red_gauche}~:\\
$\chi_F\star\xi\ps\eta\ps\pi^\tau\succ(\chi_B)(\xi)(\chi'_A)\eta\star\pi^{\gamma_0\tau}$ \
ce qui donne le résultat voulu~: \ $\chi_F\star\xi\ps\eta\ps\pi^\tau\in\bbot$.

\smallskip\noindent
Supposons maintenant \ $(\xi,p)\fforce A\to B$~; on pose \
$\chi'_F=\lbd x\lbd y(\chi'_B)(\ov{\alpha}_0 x)(\chi_A)y$.\\
On doit montrer \ $\chi'_F\xi\force(p\forcec A\to B)$ \ c'est-à-dire \
$\pt q(\chi'_F\xi\force(q\forcec A\to p\et q\forcec B))$.\\
Soient donc \ $\eta\force q\forcec A$ et $\pi\in\|p\et q\forcec B\|$~; on doit montrer \
$\chi'_F\xi\star\eta\ps\pi\in\bbot$.\\
Par hypothèse de récurrence, on a \ $(\chi_A\eta,q)\fforce A$, donc
$(\xi,p)(\chi_A\eta,q)\fforce B$ \ ou encore, par définition de l'algèbre ${\cal B}$~: \
$((\ov{\alpha}_0\xi)(\chi_A)\eta,p\et q)\fforce B$.\\
En appliquant encore l'hypothèse de récurrence, on a \
$(\chi'_B)(\ov{\alpha}_0\xi)(\chi_A)\eta\force(p\et q\forcec B)$ \ et donc~:\\
$(\chi'_B)(\ov{\alpha}_0\xi)(\chi_A)\eta\star\pi\in\bbot$. Mais on a~:\\ $\chi'_F\xi\star\eta\ps\pi\succ\chi'_F\star\xi\ps\eta\ps\pi\succ
(\chi'_B)(\ov{\alpha}_0\xi)(\chi_A)\eta\star\pi$ \ d'après le théorème~\ref{beta_red_gauche}~;
d'où le résultat voulu.

\cqfd

\smallskip\noindent
Une formule $F$ est dite \emph{du premier ordre} si elle est obtenue par les règles suivantes~:\\
$\bullet$~~$\bot$ est du premier ordre.\\
$\bullet$~~Si $A,B$ sont du premier ordre, alors $A\to B$ est du premier ordre.\\
$\bullet$~~Si $B$ est du premier ordre, {\sf R} est un symbole de prédicat et $t_1,\ldots,t_k$ sont des termes avec
paramètres, alors ${\sf R}(t_1,\ldots,t_k)\to B$, $t_1=t_2\mapsto B$ sont du premier ordre.\\
$\bullet$~~Si $A$ est du premier ordre, $\pt x\,A$ est du premier ordre ($x$ est une variable d'individu).

\smallskip\noindent
{\small{\bfseries Remarques.}\\
i)~Si \ $A$ \ est une formule du premier ordre, il en est de même de \ $\pt x\inde\,A$.\\
ii)~Cette notion sera étendue plus loin (voir proposition~\ref{n_eps_p_b}).}

\begin{theorem}\label{premier_ordre}
Soit $F$ une formule close du premier ordre. Il existe deux quasi-preuves $\delta_F,\delta'_F$, qui ne dépendent que
de la structure propositionnelle de $F$, telles que l'on ait~:\\
$\xi\force(\C[p]\to F)$ \ $\Fl$ \ $(\delta_F\xi,p)\fforce F$~;\\
$(\xi,p)\fforce F$ \ $\Fl$ \ $\delta'_F\xi\force(\C[p]\to F)$\\
quels que soient $\xi\in\Lbd$ et $p\in P$.
\end{theorem}\noindent
On raisonne par récurrence sur la construction de $F$ suivant les règles ci-dessus.

\smallskip\noindent
$\bullet$~~Si $F$ est $\bot$, on pose~:\\
$\delta_\bot=\lbd x(\chi)\lbd y(x)(\alpha)y$ \ avec $\alpha::p\et q\Fl p$ .\\
$\delta'_\bot=\lbd x\lbd y(\chi'x)(\alpha')y$ \ avec $\alpha'::p\Fl p\et\1$ .

\smallskip\noindent
En effet, supposons \ $\xi\force\C[p]\to\bot$ et montrons \ $(\delta_\bot\xi,p)(\pi,q)\in\bbbot$,
soit \ $(\delta_\bot\xi\star\pi,p\et q)\in\bbbot$. Soit donc \ $\tau\in\C[p\et q]$, donc
$\alpha\tau\in\C[p]$, d'où \ $\xi\star\alpha\tau\ps\pi\in\bbot$, par hypothèse sur $\xi$, ce qui donne \
$\delta_\bot\xi\star\pi^\tau\in\bbot$.

\smallskip\noindent
Inversement, si \ $(\xi,p)\fforce\bot$, on a \ $(\xi,p)\star(\pi,\1)\equiv(\xi\star\pi,p\et\1)\in\bbbot$ \
pour toute $\pi\in\Pi$.\\
Or, si $\tau\in\C[p]$, on a \ $\alpha'\tau\in\C[p\et\1]$, donc \ $\xi\star\pi^{\alpha'\tau}\in\bbot$,
donc \ $\delta'_\bot\xi\star\tau\ps\pi\in\bbot$.\\
Donc $\delta'_\bot\xi\force\C[p]\to\bot$.

\smallskip\noindent
$\bullet$~~Si $F$ est $A\to B$, on pose~:\\
$\delta_{A\to B}=\lbd x\lbd y(\chi)\lbd z((\chi')(\delta_B)\lbd d((x)(\alpha)z)(\delta'_Ay)(\beta)z)(\gamma)z$ \ avec\\
$\alpha::p\et(q\et r)\Fl p$; \ $\beta::p\et(q\et r)\Fl q$~; \ $\gamma~::p\et(q\et r)\Fl\1\et r$.

\smallskip\noindent
En effet, supposons \ $\xi\force\C[p],A\to B$, $(\eta,q)\fforce A$ \ et \ $(\pi,r)\in\vv B\vv$.\\
On doit montrer \ $(\delta_{A\to B}\xi,p)\star(\eta,q)\ps(\pi,r)\in\bbbot$, soit
$(\delta_{A\to B}\xi\star\eta\ps\pi,p\et(q\et r))\in\bbbot$.\\
Soit donc $\tau\in\C[p\et(q\et r)]$~; on doit montrer \
$\delta_{A\to B}\xi\star\eta\ps\pi^\tau\in\bbot$.\\
On a \ $\alpha\tau\in\C[p],\beta\tau\in\C[q]$~; or, par hypothèse
de récurrence, on a \ $\delta'_A\eta\force\C[q]\to A$, donc \ $(\delta'_A\eta)(\beta)\tau\force A$ \ et \
$((\xi)(\alpha)\tau)(\delta'_A\eta)(\beta)\tau\force B$~;
d'où \ $\lbd d((\xi)(\alpha)\tau)(\delta'_A\eta)(\beta)\tau\force\C[\1]\to B$.\\
D'après l'hypothèse de récurrence, on a \
$((\delta_B)\lbd d((\xi)(\alpha)\tau)(\delta'_A\eta)(\beta)\tau,\1)\fforce B$, donc~:\\
$((\delta_B)\lbd d((\xi)(\alpha)\tau)(\delta'_A\eta)(\beta)\tau,\1)\star(\pi,r)\in\bbbot$, soit \
$((\delta_B)\lbd d((\xi)(\alpha)\tau)(\delta'_A\eta)(\beta)\tau\star\pi,\1\et r)\in\bbbot$.\\
Or, on a $\gamma\tau\in\C[\1\et r]$, donc \
$(\delta_B)\lbd d((\xi)(\alpha)\tau)(\delta'_A\eta)(\beta)\tau\star\pi^{\gamma\tau}\in\bbot$, d'où~:\\
$((\chi')(\delta_B)\lbd d((\xi)(\alpha)\tau)(\delta'_A\eta)(\beta)\tau)(\gamma)\tau\star\pi\in\bbot$. Par suite~:\\
$(\chi)\lbd z((\chi')(\delta_B)\lbd d((\xi)(\alpha)z)(\delta'_A\eta)(\beta)z)(\gamma)z\star\pi^\tau\in\bbot$ \
d'où \ $\delta_{A\to B}\xi\star\eta\ps\pi^\tau\in\bbot$.

\smallskip\noindent
On pose maintenant~:\\
$\delta'_{A\to B}=\lbd x\lbd y\lbd z((\delta'_B)(\ov{\alpha_0}x)(\delta_A)\lbd d\,z)(\alpha)y$ \
avec \ $\alpha::p\Fl p\et\1$.

\smallskip\noindent
Supposons \ $(\xi,p)\fforce A\to B$~; soient $\tau\in\C[p]$, $\eta\force A$ et $\pi\in\|B\|$. On doit montrer~:\\
$\delta'_{A\to B}\xi\star\tau\ps\eta\ps\pi\in\bbot$. On a $\lbd d\,\eta\force\C[\1]\to A$~; en appliquant l'hypothèse de récurrence, on a $((\delta_A)\lbd d\,\eta,\1)\fforce A$, donc \ $(\xi,p)((\delta_A)\lbd d\,\eta,\1)\fforce B$ \
soit \ $((\ov{\alpha_0}\xi)(\delta_A)\lbd d\,\eta,p\et\1)\fforce B$.\\
En appliquant de nouveau l'hypothèse de récurrence, on trouve~:\\
$(\delta'_B)(\ov{\alpha_0}\xi)(\delta_A)\lbd d\,\eta\force\C[p\et\1]\to B$.
Comme on a $\alpha\tau\in\C[p\et\1]$, on obtient~:\\
$(\delta'_B)(\ov{\alpha_0}\xi)(\delta_A)\lbd d\,\eta\star\alpha\tau\ps\pi\in\bbot$ et finalement \
$\delta'_{A\to B}\xi\star\tau\ps\eta\ps\pi\in\bbot$.

\smallskip\noindent
$\bullet$~~Si $F\equiv\R(\vec{q})\to B$, où $\R$ est un symbole de prédicat d'arité $k$ et
$\vec{p}\in P^k$, on pose~:\\
$\delta_{R\to B}=\lbd x\lbd y(\ov{\alpha})(\delta_B)\lbd z(x)zy$ avec \
$\alpha::p\et(\1\et r)\Fl p\et r$.\\
$\delta'_{R\to B}=\lbd x\lbd y\lbd z((\delta'_B)(\ov{\alpha}_0)xz)(\alpha')y$ \ avec \ $\alpha'::p\Fl p\et\1$.

\smallskip\noindent
Supposons $\xi\force\C[p],\R[\vec{q}]\to B$ et soient $\eta\in|\R[\vec{q}]|$, $(\pi,r)\in\vv B\vv$.
On doit montrer~:\\
$(\delta_{R\to B}\xi,p)\star(\eta,\1)\ps(\pi,r)\in\bbbot$, soit \ 
$(\delta_{R\to B}\xi\star\eta\ps\pi,p\et(\1\et r))\in\bbbot$. Soit donc \ $\tau\in\C[p\et(\1\et r)]$~;
on doit montrer $\delta_{R\to B}\xi\star\eta\ps\pi^\tau\in\bbot$. Or, on a \ $\lbd z(\xi)z\eta\force\C[p]\to B$,
donc \ $((\delta_B)\lbd z(\xi)z\eta,p)\fforce B$, par hypothèse de récurrence. Par suite, on a \
$((\delta_B)\lbd z(\xi)z\eta,p)\star(\pi,r)\in\bbbot$, soit~:\\
$((\delta_B)\lbd z(\xi)z\eta\star\pi,p\et r)\in\bbbot$. Mais on a \ $\alpha\tau\in\C[p\et r]$, donc
$(\delta_B)\lbd z(\xi)z\eta\star\pi^{\alpha\tau}\in\bbot$, donc \ 
$(\ov{\alpha})(\delta_B)\lbd z(\xi)z\eta\star\pi^\tau\in\bbot$, d'où
$\delta_{R\to B}\xi\star\eta\ps\pi^\tau\in\bbot$.

\smallskip\noindent
Supposons maintenant \ $(\xi,p)\fforce\R(\vec{q})\to B$~; soient \ $\tau\in\C[p]$, $\eta\in|\R[\vec{q}]|$
et $\pi\in\|B\|$. On doit montrer \ $\delta'_{R\to B}\xi\star\tau\ps\eta\ps\pi\in\bbot$. Or, on a \
$(\xi,p)(\eta,\1)\fforce B$, soit \ $((\ov{\alpha}_0)\xi\eta,p\et\1)\fforce B$, donc~:\\
$(\delta'_B)(\ov{\alpha}_0)\xi\eta\force\C[p\et\1]\to B$, par hypothèse de récurrence.\\
Or, on a \ $\alpha'\tau\in\C[p\et\1]$, donc \ $(\delta'_B)(\ov{\alpha}_0)\xi\eta\star\alpha'\tau\ps\pi\in\bbot$,
d'où le résultat.

\smallskip\noindent
$\bullet$~~Si $F\equiv(p_1=p_2\mapsto B)$, on pose \ $\delta_F=\delta_B$ \ et \ $\delta'_F=\delta'_B$.\\
En effet, supposons $\xi\force\C[p]\to(p_1=p_2\mapsto B)$ et $(\pi,q)\in\vv p_1=p_2\mapsto B\vv$. On doit
montrer $(\delta_B\xi,p)\star(\pi,q)\in\bbbot$. Comme $\vv p_1=p_2\mapsto B\vv\ne\vide$, on a $p_1=p_2$,
donc $(\pi,q)\in\vv B\vv$ et $\xi\force\C[p]\to B$. D'où le résultat, par hypothèse de récurrence.

\smallskip\noindent
Supposons maintenant \ $(\xi,p)\fforce p_1=p_2\mapsto B$, $\tau\force\C[p]$ et
$\pi\in\|p_1=p_2\mapsto B\|$. On doit montrer \ $\delta'_B\star\tau\ps\pi\in\bbot$. Comme
$\|p_1=p_2\mapsto B\|\ne\vide$, on a $p_1=p_2$, donc $\pi\in\|B\|$ et \
$(\xi,p)\fforce B$. D'où le résultat, par hypothèse de récurrence.

\smallskip\noindent
$\bullet$~~Si $F\equiv\pt x\,A$, on pose \ $\delta_F=\delta_A$ \ et \ $\delta'_F=\delta'_A$.

\smallskip\noindent
En effet, si $\xi\force\C[p]\to\pt x\,A$, on a $\xi\force\C[p]\to A[a/x]$ pour tout $a\in P$. Par
hypothèse de récurrence, on a $(\delta_A\xi,p)\fforce A[a/x]$~; donc \ $(\delta_A\xi,p)\fforce\pt x\,A$.

\smallskip\noindent
Si $(\xi,p)\fforce\pt x\,A$, on a \ $(\xi,p)\fforce A[a/x]$ pour tout $a\in P$. Par hypothèse de
récurrence, on a $\delta'_A\xi\force\C[p]\to A[a/x]$~; donc \ $\delta'_A\xi\force\C[p]\to\pt x\,A$.

\cqfd

\subsection*{L'idéal générique}\noindent
On définit un prédicat unaire ${\cal J}:P\to{\cal P}(\PPi)$ du modèle ${\cal N}$ (paramètre du
second ordre d'arité~1), en posant \ ${\cal J}(p)=\Pi\fois\{p\}$~; on l'appellera \emph{l'idéal générique}.\\
Le prédicat binaire ${\cal J}^+:P^2\to{\cal P}(\Pi)$ du modèle ${\cal M}$ qui lui correspond, est donc
tel que ${\cal J}^+(p,q)=\vide$ (resp. $\Pi$) si $p\ne q$ (resp. $p=q$). Autrement dit~:\\
\centerline{${\cal J}^+(p,q)$ est le prédicat $p\ne q$.}

\noindent
La formule $p\force{\cal J}(q)$ s'écrit $\pt r(\C[p\et r]\to{\cal J}^+(r,q)$. On a donc
$\|p\force{\cal J}(q)\|=\|\neg\C[p\et q]\|$. Autrement dit~:\\
\centerline{$p\force{\cal J}(q)$ est identique à $\neg\C[p\et q]$.}

\smallskip\noindent
{\bfseries Notations.}\\
$\bullet$~~On note $p\sqle q$ la formule $\pt r(\neg\C[q\et r]\to\neg\C[p\et r])$ \ et \
$p\sim q$ la formule $p\sqle q\land q\sqle p$, c'est-à-dire \ $\pt r(\neg\C[q\et r]\dbfl\neg\C[p\et r])$.\\
Dans la suite, on écrira souvent $F\to\C[p]$ au lieu de $\neg\C[p]\to\neg F$~;\\
$p\sqle q$ s'écrit alors $\pt r(\C[p\et r]\to\C[q\et r])$ et $p\sim q$ s'écrit
$\pt r(\C[p\et r]\dbfl\C[q\et r])$.\\
{\small{\bfseries Remarque.} On rappelle, en effet, que $\C[p]$ n'est pas une formule, mais une partie de $\Lbd$~; en fait, dans certains modèles de réalisabilité considérés plus loin, il existera  une formule $\CC[p]$ telle que $|\C[p]|=\{\tau\in\Lbd_c;$ $\tau\force\CC[p]\}$.
On pourra alors identifier $\C[p]$ à la formule $\CC[p]$.}\\
$\bullet$~~Si $F$ est une formule close, on écrira \ $\fforce F$ \ pour exprimer qu'il existe
une quasi-preuve $\theta$ telle que \ $(\theta,\1)\fforce F$. D'après la proposition~\ref{1_to_p}(i),
cela équivaut à dire qu'il existe une quasi-preuve \ $\theta$ telle que \ $(\theta,p)\fforce F$ pour
tout $p\in P$.

\begin{proposition}\label{xi_p_force_Jq}\ \\
i)~~$\xi\force\neg\C[p\et q]$ $\Fl$ $(\chi\xi,p)\fforce{\cal J}(q)$~;\\
\hspace*{0.8em}$(\xi,p)\fforce{\cal J}(q)$ $\Fl$ $\chi'\xi\force\neg\C[p\et q]$.\\
ii)~~$\xi\force\pt r(\C[p\et(\1\et r)],\C[q]\to\bot)$ $\Fl$ $(\chi\xi,p)\fforce\neg\C[q]$~;\\
\hspace*{1em}$(\xi,p)\fforce\neg\C[q]$ $\Fl$ $\chi'\xi\force\pt r(\C[p\et(\1\et r)],\C[q]\to\bot)$.\\
iii)~~Si $\xi\force\neg{\sf R}(a_1,\ldots,a_k)$ \ alors \ $(\xi,p)\fforce\neg{\sf R}(a_1,\ldots,a_k)$
pour tout $p$ (${\sf R}$ est un symbole de prédicat d'arité $k$).

\end{proposition}\noindent
i)~Si $\xi\force\neg\C[p\et q]$, alors $\xi\star\tau\ps\pi\in\bbot$ et donc
$\chi\xi\star\pi^\tau\in\bbot$ pour tout $\tau\in\C[p\et q]$. On a donc~:\\
$(\chi\xi\star\pi,p\et q)\in\bbbot$, soit $(\chi\xi,p)\star(\pi,q)\in\bbbot$ pour toute $\pi\in\Pi$,
c'est-à-dire \ $(\chi\xi,p)\fforce{\cal J}(q)$.

\smallskip\noindent
Si $(\xi,p)\fforce{\cal J}[q]$, on a $(\xi,p)\star(\pi,q)\in\bbbot$, donc $(\xi\star\pi,p\et q)\in\bbbot$
pour toute $\pi\in\Pi$. On a donc $\xi\star\pi^\tau\in\bbot$, soit
$\chi'\xi\star\tau\ps\pi\in\bbot$ pour tout $\tau\in\C[p\et q]$. Donc $\chi'\xi\force\neg\C[p\et q]$.

\smallskip\noindent
ii)~Si $\xi\force\pt r(\C[p\et(\1\et r)],\C[q]\to\bot)$, on a $\xi\star\upsilon\ps\tau\ps\pi\in\bbot$
pour $\upsilon\in\C[p\et(\1\et r)]$ et $\tau\in\C[q]$. Donc $\chi\xi\star\tau\ps\pi^\upsilon\in\bbot$,
d'où $(\chi\xi\star\tau\ps\pi,p\et(\1\et r))\in\bbbot$ soit $(\chi\xi,p)\star(\tau,\1)\ps(\pi,r)\in\bbot$.\\
Or $(\tau,\1)$ décrit $\C_{\cal N}[q]$, et donc $(\chi\xi,p)\fforce\C[q]\to\bot$.

\smallskip\noindent
Si $(\xi,p)\fforce\neg\C[q]$, on a $(\xi,p)\star(\tau,\1)\ps(\pi,r)\in\bbbot$, et donc
$(\xi\star\tau\ps\pi,p\et(\1\et r))\in\bbbot$ pour tout $\tau\in\C[q]$. On a donc \
$\xi\star\tau\ps\pi^\upsilon\in\bbot$ \ d'où \ $\chi'\xi\star\upsilon\ps\tau\ps\pi\in\bbot$
pour tout \ $\upsilon\in\C[p\et(\1\et r)]$.\\
On en déduit \ $\chi'\xi\force\pt r(\C[p\et(\1\et r)],\C[q]\to\bot)$.\\
iii)~~Soit $\tau\in|{\sf R}(a_1,\ldots,a_k)|$~; on a $\xi\star\tau\ps\pi\in\bbot$ pour toute $\pi\in\Pi$,
donc $(\xi\star\tau\ps\pi,a)\in\bbbot$ quel que soit $a\in P$, d'où $(\xi,p)\star(\tau,\1)\ps(\pi,q)\in\bbbot$.
 
\cqfd

\begin{theorem}[Propriétés élémentaires du générique]\label{elem_gen}\ \\
i) $(\ov{\alpha},\1)\fforce\neg{\cal J}(\1)$ \ avec \ $\alpha::\1\et(p\et q)\Fl p\et\1$.\\
ii) $(\theta,\1)\fforce\pt x(\neg\C[x]\to{\cal J}(x))$ \ où \
$\theta=\lbd x(\chi)\lbd y((\chi'x)(\beta)y)(\alpha)y$\\
avec \ $\alpha::\1\et(p\et q)\Fl q$ \ et \
$\beta::\1\et(p\et q)\Fl p\et(\1\et\1)$.\\
iii) $(\theta,\1)\fforce\pt x\pt y({\cal J}(x\et y),\neg{\cal J}(x)\to{\cal J}(y))$ \
où \ $\theta=\lbd x\lbd y(\ov{\alpha})(y)(\ov{\beta})x$\\
avec \ $\alpha::\1\et(p'\et(q'\et q))\Fl q'\et((q\et p')\et\1)$ \ et \
$\beta::(q\et p')\et p\Fl p'\et(p\et q)$.\\
iv) $(\theta,\1)\fforce\pt x(\pt y(\neg\C[x\et y]\to{\cal J}(y))\to\neg{\cal J}(x))$ \ où \
$\theta=\lbd x\lbd y(\ov{\gamma})(x)\lbd z(\chi'y)(\beta)z$, avec\\
$\beta::p\et q\Fl q\et p$ \ et \ $\gamma::\1\et(r\et(q\et r'))\Fl r\et(\1\et p)$.\\
v)~~$(\theta,\1)\fforce\pt x\pt y({\cal J}(x),y\sqle x\to{\cal J}(y))$\\
où \ $\theta=\lbd x\lbd y((\chi)\lbd z(((\chi')(\ov{\alpha}_0y)\lbd z'(\chi'x)(\beta)z')(\alpha)z)(\gamma)z$, \ avec\\
$\alpha::\1\et(p'\et(r\et q))\Fl(r\et\1)\et(\1\et\1)$~; \
$\alpha'::\1\et(p'\et(q'\et q))\Fl q\et p'$~; \ $\beta::p\et q\Fl q\et p$.
\end{theorem}\noindent
i) Soit $(\xi,p)\fforce{\cal J}(\1)$~; on doit montrer que \ $(\ov{\alpha},\1)\star(\xi,p)\ps(\pi,q)\in\bbbot$, c'est-à-dire~:\\
$(\ov{\alpha}\star\xi\ps\pi,\1\et(p\et q))\in\bbbot$. Mais, d'après la
proposition~\ref{gamma::tu}, on a~:\\
$(\ov{\alpha}\star\xi\ps\pi,\1\et(p\et q))\succ(\xi\star\pi,p\et\1)\equiv(\xi,p)\star(\pi,\1)$.\\
Or, on a $(\xi,p)\star(\pi,\1)\in\bbbot$ par hypothèse sur $(\xi,p)$.

\smallskip\noindent
ii)Soient $(\eta,p)\fforce\neg\C[q]$ et $(\pi,q)\in\vv{\cal J}(q)\vv$. On doit montrer que
$(\theta,\1)\star(\eta,p)\ps(\pi,q)\in\bbbot$, soit $(\theta\star\eta\ps\pi,\1\et(p\et q))\in\bbbot$.
Soit donc $\tau\in\C[\1\et(p\et q)]$~; on doit montrer que \ $\theta\star\eta\ps\pi^\tau\in\bbot$.\\
D'après la proposition~\ref{xi_p_force_Jq}, on a \ $\chi'\eta\force\C[p\et(\1\et\1)],\C[q]\to\bot$.\\
Or, on a $\beta\tau\in\C[p\et(\1\et\1)]$ et $\alpha\tau\in\C[q]$, donc 
$\chi'\eta\star\beta\tau\ps\alpha\tau\ps\pi\in\bbot$ \ d'où\\ 
$(\chi)\lbd y((\chi'\eta)(\beta)y)(\alpha)y\star\pi^\tau\in\bbot$ d'où
$\theta\star\eta\ps\pi^\tau\in\bbot$.

\smallskip\noindent
iii) Soient $(\xi,p')\fforce{\cal J}(p\et q)$, \ $(\eta,q')\fforce\neg{\cal J}(p)$ \ et \
$(\pi,q)\in\vv{\cal J}(q)\vv$. On doit montrer que~:\\
$(\theta,\1)\star(\xi,p')\ps(\eta,q')\ps(\pi,q)\in\bbbot$, soit \
$(\theta\star\xi\ps\eta\ps\pi,\1\et(p'\et(q'\et q)))\in\bbbot$.\\
D'après les propositions~\ref{1_to_p}(ii) et~\ref{gamma::tu}, on est ramené à montrer~:\\
$((\ov{\alpha})(\eta)(\ov{\beta})\xi\star\pi,\1\et(p'\et(q'\et q)))\in\bbbot$ \ puis \
$(\eta\star\ov{\beta}\xi\ps\pi,q'\et((q\et p')\et\1))\in\bbbot$, c'est-à-dire~:\\
$(\eta,q')\star(\ov{\beta}\xi,q\et p')\ps(\pi,\1)\in\bbbot$.\\
Par hypothèse sur $(\eta,q')$, il reste donc à montrer que \
$(\ov{\beta}\xi,q\et p')\fforce{\cal J}(p)$, c'est-à-dire~:\\
$(\ov{\beta}\xi,q\et p')\star(\varpi,p)\in\bbbot$, ou encore \
$(\ov{\beta}\xi\star\varpi,(q\et p')\et p)\in\bbbot$ pour toute $\varpi\in\Pi$.\\
Or, d'après la proposition~\ref{gamma::tu}, on a~:\\
$(\ov{\beta}\xi\star\varpi,(q\et p')\et p)\succ(\xi\star\varpi,p'\et(p\et q))
\equiv(\xi,p')\star(\varpi,p\et q)\in\bbbot$ par hypothèse sur $(\xi,p')$.

\smallskip\noindent
iv)~~Soient $(\xi,q)\fforce{\cal J}(p)$ et $(\eta,r)\fforce\pt q(\neg\C[p\et q]\to{\cal J}(q))$~;
on doit montrer que~:\\ $(\theta,\1)\star(\eta,r)\ps(\xi,q)\ps(\pi,r')\in\bbbot$, soit
$(\theta\star\eta\ps\xi\ps\pi,\1\et(r\et(q\et r')))\in\bbbot$.\\
D'après la proposition~\ref{xi_p_force_Jq}(i), on a $\chi'\xi\force\neg\C[q\et p]$. Soit $\tau\in\C[p\et q]$,
donc $\beta\tau\in\C[q\et p]$ d'où $\chi'\xi\star\beta\tau\ps\rho\in\bbot$ pour tout $\rho\in\Pi$.
On a donc $\lbd x(\chi'\xi)(\beta)x\star\tau\ps\rho\in\bbot$, donc\\
$\lbd z(\chi'\xi)(\beta)z\force\neg\C[p\et q]$. D'après la proposition~\ref{xi_p_force_Jq}(iii), on a \
$(\lbd z(\chi'\xi)(\beta)z,\1)\fforce\neg\C[p\et q]$.\\
Par hypothèse sur $(\eta,r)$, on a donc \ $(\eta,r)\star(\lbd z(\chi'\xi)(\beta)z,\1)\ps(\pi,q)\in\bbbot$, soit~:\\
$(\eta\star\lbd z(\chi'\xi)(\beta)z\ps\pi,r\et(\1\et q))\in\bbbot$, donc \
$((\ov{\gamma})(\eta)\lbd z(\chi'\xi)(\beta)z\star\pi,\1\et(r\et(q\et r')))\in\bbbot$\\
(proposition~\ref{gamma::tu}) et donc~ \
$(\theta\star\eta\ps\xi\ps\pi,\1\et(r\et(q\et r')))\in\bbbot$.

\smallskip\noindent
v) Soient $(\xi,p')\fforce{\cal J}(p)$ et $(\eta,r)\fforce q\sqle p$~; on doit montrer que~:\\
$(\theta,\1)\star(\xi,p')\ps(\eta,r)\ps(\pi,q)\in\bbbot$ pour toute $\pi\in\Pi$, soit
$(\theta\star\xi\ps\eta\ps\pi,\1\et(p'\et(r\et q)))\in\bbbot$.\\
D'après la proposition~\ref{xi_p_force_Jq}(i), on a $\chi'\xi\force\neg\C[p'\et p]$, donc
$\lbd z'(\chi'\xi)(\beta)z'\force\neg\C[p\et p']$~: en effet, si $\tau\in\C[p\et p']$ et $\rho\in\Pi$, on a
$\lbd z'(\chi'\xi)(\beta)z'\star\tau\ps\rho\succ(\chi'\xi)(\beta)\tau\star\rho\in\bbot$
puisque $\beta\tau\in\C[p'\et p]$.\\
D'après la proposition~\ref{xi_p_force_Jq}(iii), on a alors $(\lbd z'(\chi'\xi)(\beta)z',\1)\fforce\neg\C[p\et p']$.
Or, par hypothèse sur $(\eta,r)$, on a \ $(\eta,r)\fforce(\neg\C[p\et p']\to\neg\C[q\et p'])$. Il en résulte que~:\\
$(\eta,r)(\lbd z'(\chi'\xi)(\beta)z',\1)\fforce\neg\C[q\et p']$, soit \
($(\ov{\alpha}_0\eta)\lbd z'(\chi'\xi)(\beta)z',r\et\1)\fforce\neg\C[q\et p']$.\\
D'après la proposition~\ref{xi_p_force_Jq}(ii), on a \
$(\chi')(\ov{\alpha}_0\eta)\lbd z'(\chi'\xi)(\beta)z'\force\C[(r\et\1)\et(\1\et\1)],\C[q\et p']\to\bot$.\\
Soit $\tau\in\C[\1\et(p'\et(r\et q))]$, donc $\alpha\tau\in\C[(r\et\1)\et(\1\et\1)]$ et
$\alpha'\tau\in\C[q\et p']$. On a donc~:\\
$(((\chi')(\ov{\alpha}_0\eta)\lbd z'(\chi'\xi)(\beta)z')(\alpha)\tau)(\gamma)\tau\star\pi\in\bbot$, donc~:\\
$(\chi)\lbd z(((\chi')(\ov{\alpha}_0\eta)\lbd z'(\chi'\xi)(\beta)z')(\alpha)z)(\alpha')z\star\pi^\tau\in\bbot$.
Autrement dit~:\\
$((\chi)\lbd z(((\chi')(\ov{\alpha}_0\eta)\lbd z'(\chi'\xi)(\beta)z')(\alpha)z)(\alpha')z\star\pi,
\1\et(p'\et(r\et q)))\in\bbbot$\\
ou encore, d'après la proposition~\ref{1_to_p}(ii)~: \
$(\theta\star\xi\ps\eta\ps\pi,\1\et(p'\et(r\et q)))\in\bbbot$.

\cqfd

\begin{theorem}[Densité]\label{densite}\ \\
Pour toute fonction $\phi~:P\to P$, on a~:\\
$(\theta,\1)\fforce\pt x(\neg\C[x\et\phi(x)]\to{\cal J}(x)),\pt x\,{\cal J}(x\et\phi(x))\to\bot$\\
où $\theta=(\ov{\beta})\lbd x\lbd y(x)(\vartheta)y$, \
$\vartheta=(\chi)\lbd d\lbd x\lbd y(\chi'x)(\alpha)y$~;\\
avec \ $\alpha::q\et r\Fl q\et(q\et r)$~; \ $\beta::\1\et(p\et(q\et r))\Fl p\et(\1\et q)$.
\end{theorem}\noindent
Soient $(\xi,p)\fforce\pt x(\neg\C[x\et\phi(x)]\to{\cal J}(x))$,
$(\eta,q)\fforce\pt x\,{\cal J}(x\et\phi(x))$ et $(\pi,r)\in\PPi$.\\
On doit montrer que $(\theta\star\xi\ps\eta\ps\pi,\1\et(p\et(q\et r)))\in\bbbot$~;
soit donc \ $\tau_0\in\C[\1\et(p\et(q\et r))]$. On doit montrer \
$\theta\star\xi\ps\eta\ps\pi^{\tau_0}\in\bbot$.\\
On montre d'abord que $(\vartheta\eta,\1)\fforce\neg\C[q\et\phi(q)]$.\\
Soient donc $(\varpi,r')\in\PPi$ et $\tau\in\C[q\et\phi(q)]$~; on doit montrer \
$(\vartheta\eta,\1)\star(\tau,\1)\ps(\varpi,r')\in\bbbot$\\
soit \ $(\vartheta\eta\star\tau\ps\varpi,\1\et(\1\et r'))\in\bbbot$ ou encore \
$\vartheta\eta\star\tau\ps\varpi^{\tau'}\in\bbot$ \ pour tout $\tau'\in\C[\1\et(\1\et r')])$.\\
Or, $\vartheta\eta\star\tau\ps\varpi^{\tau'}\succ\eta\star\varpi^{\alpha\tau}$ \ et \
$\alpha\tau\in\C[q\et(q\et\phi(q))]$. Il suffit donc de montrer~:\\
$(\eta\star\varpi,q\et(q\et\phi(q)))\in\bbbot$ ou encore \
$(\eta,q)\star(\varpi,q\et\phi(q))\in\bbbot$.\\
Or, cela résulte de l'hypothèse sur $(\eta,q)$, qui implique \ $(\eta,q)\fforce{\cal J}(q\et\phi(q))$.

\smallskip\noindent
Par hypothèse sur $\xi$, on a $(\xi,p)\fforce\neg\C[q\et\phi(q)]\to{\cal J}(q)$. Il en résulte que~:\\
$(\xi,p)\star(\vartheta\eta,\1)\ps(\pi,q)\in\bbbot$, soit \
$(\xi\star\vartheta\eta\ps\pi,p\et(\1\et q))\in\bbbot$.\\
Or, on a $\tau_0\in\C[\1\et(p\et(q\et r))])$, donc $\beta\tau_0\in\C[p\et(\1\et q)]$.
Il en résulte que \ $\xi\star\vartheta\eta\ps\pi^{\beta\tau_0}\in\bbot$.\\
Cela donne le résultat voulu, puisque \
$\theta\star\xi\ps\eta\ps\pi^{\tau_0}\succ\xi\star\vartheta\eta\ps\pi^{\beta\tau_0}$.

\cqfd

\subsection*{Condition de chaîne dénombrable}\noindent
Dans cette section, on considère une algèbre de réalisabilité standard ${\cal A}$ et un
${\cal A}$-modèle ${\cal M}$. On suppose que l'ensemble $P$ (domaine de variation des variables
d'individu) est de cardinal $\ge2^{\aleph_0}$.
On fixe une surjection \ $\varepsilon:P\to{\cal P}(\Pi)^\ennl$ et on définit un prédicat binaire
du modèle ${\cal M}$, noté aussi $\varepsilon$, en posant~:\\
$\|n\eps p\|=\varepsilon(p)(n)$ si $n\in\ennl$~; \ $\|n\eps p\|=\vide$ si $n\notin\ennl$\\
(on utilise, pour le prédicat $\varepsilon$, la notation \ $n\eps p$ au lieu de
$\varepsilon(n,p)$).\\
Le prédicat $\varepsilon$ permet donc d'associer, à chaque individu, un ensemble d'entiers qui
sont ses \emph{éléments}. La proposition~\ref{RPN} montre que l'axiome suivant est réalisé~:

\smallskip\noindent
\emph{Pour tout ensemble, il existe un individu qui a les mêmes éléments entiers}.

\smallskip\noindent
Cet axiome sera appelé \emph{axiome de représentation des prédicats sur $\NN$} et noté RPN.

\begin{proposition}[RPN]\label{RPN}\ \\
$\lbd x(x)\ul{0}\,\ul{0}\force\pt X\ex x\pt n\inde(Xn\dbfl n\eps x)$.
\end{proposition}\noindent
Cette formule s'écrit \
$\pt X(\pt x[\pt n($ent$(n),Xn\to n\eps x),\pt n($ent$(n),n\eps x\to Xn)\to\bot]\to\bot)$.\\
On considère donc un paramètre ${\cal X}:P\to{\cal P}(\Pi)$ d'arité $1$ et un terme $\xi\in\Lbd$
tel que~:

\smallskip\noindent
$\xi\force\pt x[\pt n($ent$(n),{\cal X}n\to n\eps x),\pt n($ent$(n),n\eps x\to{\cal X}n)\to\bot]$.

\smallskip\noindent
On doit montrer que \ $\lbd x(x)\ul{0}\,\ul{0}\star\xi\ps\pi\in\bbot$, ou encore
$\xi\star\ul{0}\ps\ul{0}\ps\pi\in\bbot$ pour toute pile $\pi\in\Pi$.\\
Par définition de $\varepsilon$, il existe $p_0\in P$ tel que l'on ait ${\cal X}n=\|n\eps p_0\|$
pour tout entier $n$. Or, on a~:\\
$\xi\force\pt n($ent$(n),{\cal X}n\to n\eps p_0),\pt n($ent$(n),n\eps p_0\to{\cal X}n)\to\bot$.\\
Il suffit donc de montrer que \ $\ul{0}\force\pt n($ent$(n),{\cal X}n\to n\eps p_0)$ \ et \
$\ul{0}\force\pt n($ent$(n),n\eps p_0\to{\cal X}n)$.\\
On rappelle que le prédicat ent$(x)$ est défini par~:

\smallskip\noindent
|ent$(n)|=\{\ul{n}\}$ \ si $n\in\ennl$ \ et \ |ent$(n)|=\vide$ si $n\notin\ennl$.

\smallskip\noindent
On doit donc montrer~:\\
$\ul{0}\star\ul{n}\ps\eta\ps\rho\in\bbot$ \ pour tout $n\in\ennl$, $\eta\force{\cal X}(n)$ \ et \
$\rho\in\|n\eps p_0\|$~;\\
$\ul{0}\star\ul{n}\ps\eta'\ps\rho'\in\bbot$ \ pour tout $n\in\ennl$, $\eta'\force n\eps p_0$ \ et \
$\rho'\in{\cal X}(n)$.\\
Or ceci s'écrit \ $\eta\star\rho\in\bbot$ \ et \ $\eta'\star\rho'\in\bbot$, ce qui est trivialement
vérifié, puisque ${\cal X}n=\|n\eps p_0\|$.

\cqfd

\smallskip\noindent
On suppose maintenant que $\{\C,\et,\1\}$ est une structure de forcing dans ${\cal M}$.
On définit alors également le symbole $\varepsilon$ dans le ${\cal B}$-modèle ${\cal N}$ en posant~:\\
$\vv n\eps p\vv=\|n\eps p\|\fois\{\1\}$ pour $n,p\in P$. Autrement dit\\
$\vv n\eps p\vv=\{(\pi,\1);\;\pi\in\eps(p)(n)\}$ si $n\in\NN$~; $\vv n\eps p\vv=\vide$ si $n\notin\NN$.

\begin{proposition}\label{n_eps_p}
Le prédicat \ $\varepsilon^+(q,n,p)$ est \ $q=\1\mapsto n\eps p$.\\
La formule \ $q\forcec n\eps p$ \ est \ $\C[q\et\1]\to n\eps p$.
\end{proposition}\noindent
Immédiat, par définition de $\vv n\eps p\vv$.

\cqfd

\begin{proposition}\label{n_eps_p_b}\ \\
i)~~$\xi\force(\C[p]\to n\eps q)$ \ $\Fl$ \ $(\delta\xi,p)\fforce n\eps q$ \
où \ $\delta=\lbd x(\chi)\lbd y(x)(\alpha)y$ \ et \ $\alpha::p\et\1$ $\Fl$ $p$.\\
ii)~~$(\xi,p)\fforce n\eps q$ \ $\Fl$ \ $\delta'\xi\force(\C[p]\to n\eps q)$ \
où \ $\delta'=\lbd x\lbd y(\chi'x)(\alpha')y$ \ et \ $\alpha'::p$ $\Fl$ $p\et\1$.
\end{proposition}\noindent
On a \ $(\xi,p)\fforce n\eps p$ $\Dbfl$ $(\xi,p)\star(\pi,\1)\in\bbbot$ \ pour toute $\pi\in\|n\eps p\|$,
ou encore~:\\
$(\xi,p)\fforce n\eps p$ \ $\Dbfl$ \ $\xi\star\pi^\tau\in\bbot$ \ pour tout $\tau\in\C[p\et\1]$ \ et \
$\pi\in\|n\eps p\|$.\\
i)~Supposons \ $\xi\force(\C[p]\to n\eps q)$, $\tau\in\C[p\et\1]$ \ et \ $\pi\in\|n\eps p\|$. On a alors~:\\
$\delta\xi\star\pi^\tau\succ\xi\star\alpha\tau\ps\pi\in\bbot$, puisque $\alpha\tau\in\C[p]$.\\
ii)~Supposons $(\xi,p)\force n\eps q$, $\tau\in\C[p]$ \ et \ $\pi\in\|n\eps p\|$. On a alors~:\\
$\delta'\xi\star\tau\ps\pi\succ\xi\star\pi^{\alpha'\tau}\in\bbot$, puisque \ $\alpha'\tau\in\C[p\et\1]$.

\cqfd

\smallskip\noindent
La notion de \emph{formule du premier ordre} a été définie plus haut (voir théorème~\ref{premier_ordre}).
On étend cette définition en y ajoutant la clause suivante~:

\smallskip\noindent
$\bullet$~~$t\eps u$ est du premier ordre, quels que soient les termes $t,u$.

\smallskip\noindent
La proposition~\ref{n_eps_p_b} montre que le théorème~\ref{premier_ordre} reste valable pour cette notion
étendue.

\smallskip\noindent
On dira que la structure de forcing $\{\C,\et,\1\}$ satisfait la
\emph{condition de chaîne dénombrable} (en abrégé \emph{c.c.d.}) s'il existe une quasi-preuve
{\sf ccd} telle que~:

\smallskip\noindent
${\sf ccd}\force\pt X[
\pt n\inde\ex p\,X(n,p),\pt n\inde\pt p\pt q(X(n,p),X(n,q)\to p=q),\\
\hspace*{2.5em}\pt n\inde\pt p\pt q(X(n,p),X(sn,q)\to q\sqle p)\to\\
\hspace*{2.5em}\ex p'\{\pt n\inde\pt p(X(n,p)\to p'\sqle p),
(\pt n\inde\pt p(X(n,p)\to\C[p])\to\C[p'])\}]$.

\smallskip\noindent
{\small Le sens intuitif de cette formule est~:\\
Si $X(n,p)$ est une suite décroissante de conditions, alors il existe une condition $p'$ qui
les minore toutes~; de plus, si toutes ces conditions sont non triviales, alors $p'$ est non triviale.}

\smallskip\noindent
On se propose, dans cette section de montrer le~:

\begin{theorem}[Conservation des réels]\label{cons_reels}\ \\
Si la c.c.d. est vérifiée, il existe une quasi-preuve \ {\sf crl} \ telle que~:\\
({\sf crl}$,\1)\fforce\pt X\ex x\pt n\inde(Xn\dbfl n\eps x)$.
\end{theorem}\noindent
Cela signifie que l'axiome RPN, qui est réalisé dans le ${\cal A}$-modèle ${\cal M}$ (voir proposition~\ref{RPN}) l'est aussi dans le ${\cal B}$-modèle générique ${\cal N}$.

\smallskip\noindent
{\bfseries Notation.}\\
La formule $\pt q(\C[p\et q],q\forcec Xn\to p\forcec Xn)$  se lit
``~$p$ décide $Xn$~'', et est notée \ $p\forcec\pm Xn$.\\
Elle s'écrit aussi \ $\pt q\pt r(\C[p\et q],q\forcec Xn,\C[p\et r]\to X^+(r,n))$.\\
Si ${\cal X}:P\to{\cal P}(\Pi\fois P)$ un prédicat unaire du ${\cal B}$-modèle ${\cal N}$,
et ${\cal X}^+:P^2\to{\cal P}(\Pi)$ est le prédicat binaire correspondant du
${\cal A}$-modèle standard ${\cal M}$, la formule
$\pt q(\C[p\et q],q\forcec{\cal X}n\to p\forcec{\cal X}n)$ est donc notée
aussi \ $p\forcec\pm{\cal X}n$.

\begin{theorem}\label{pX_decide}
Si la c.c.d. est vérifiée, il existe une quasi-preuve \ {\sf dec} \ telle que~:\\
{\sf dec}$\force\pt X\pt p_0\ex p'\{(\C[p_0]\to\C[p']),p'\sqle p_0,\pt n\inde(p'\forcec\pm Xn)\}$.
\end{theorem}\noindent
On montre d'abord comment le théorème~\ref{cons_reels} se déduit de ce théorème~\ref{pX_decide}.\\
D'après le théorème~\ref{forcec-fforce}, il suffit de trouver une quasi-preuve {\sf crl0} telle que~:\\
{\sf crl0}$\force\1\forcec\pt X\ex x\pt n\inde(Xn\dbfl n\eps x)$\\
ou encore, puisque \ $\1\forcec\neg A\equiv\pt p_0((p_0\forcec A),\C[\1\et p_0]\to\bot)$~:\\
{\sf crl0}$\force\pt X\pt p_0[(p_0\forcec\pt q\{\pt n\inde(Xn\dbfl n\eps q)\to\bot\}),
\C[\1\et p_0]\to\bot]$.\\
D'après le théorème~\ref{pX_decide}, il suffit de trouver une quasi-preuve {\sf crl1} telle que~:\\
{\sf crl1}$\force\pt X\pt p_0\pt p'\{(\C[p_0]\to\C[p']),p'\sqle p_0,
\pt n\inde(p'\forcec\pm Xn),\\
\hspace*{\fill}(p_0\forcec\pt q(\pt n\inde(Xn\dbfl n\eps q)\to\bot)),\C[\1\et p_0]\to\bot\}$.\\
Il suffit de trouver une quasi-preuve {\sf crl2} telle que~:\\
{\sf crl2}$\force\pt X\pt p_0\pt p'\{(p_0\forcec\pt q(\pt n\inde(Xn\dbfl n\eps q)\to\bot)),p'\sqle p_0,
\pt n\inde(p'\forcec\pm Xn),\C[p']\to\bot\}$.\\
On prendra alors {\sf crl1}$=\lbd x\lbd y\lbd z\lbd u\lbd v((x)(${\sf crl2}$)uyz)(\delta)v$
avec \ $\delta::\1\et p\Fl p$~;\\
(rappelons que la formule $\C[p_0]\to\C[p']$ s'écrit, en fait, $\neg\C[p']\to\neg\C[p_0]$).

\smallskip\noindent
On fixe \ ${\cal X}^+:P^2\to{\cal P}(\Pi)$, $p_0,p'\in P$,
$\xi\force(p_0\forcec\pt q(\pt n\inde({\cal X}n\dbfl n\eps q)\to\bot))$, $\eta\force p'\sqle p_0$,
$\zeta\force\pt n\inde(p'\forcec\pm{\cal X}n)$ et $\tau\in\C[p']$. On doit avoir \
({\sf crl2})$\xi\eta\zeta\tau\force\bot$.\\
On choisit $q_0\in P$ tel que l'on ait $\|n\eps q_0\|=\|p'\forcec{\cal X}n\|$ pour tout $n\in\NN$,
ce qui est possible, par définition de $\varepsilon$.
On a trivialement \
$\xi\force(p_0\forcec(\pt n\inde(n\eps q_0\to{\cal X}n),\pt n\inde({\cal X}n\to n\eps q_0)\to\bot))$.\\
Or, la formule \
$p_0\forcec(\pt n\inde(n\eps q_0\to{\cal X}n),\,\pt n\inde({\cal X}n\to n\eps q_0)\to\bot)$ s'écrit~:\\
$\pt r\pt r'(r\forcec\pt n\inde(n\eps q_0\to{\cal X}n),\;r'\forcec\pt n\inde({\cal X}n\to n\eps q_0),\;
\C[(p_0\et r)\et r']\to\bot)$.\\
En remplaçant $r$ et $r'$ par $p'$, on obtient donc~:\\
$\xi\force(p'\forcec\pt n\inde(n\eps q_0\to{\cal X}n),\;p'\forcec\pt n\inde({\cal X}n\to n\eps q_0),\;
\C[(p_0\et p')\et p']\to\bot)$.\\
De \ $\tau\in\C[p']$ et $\eta\force\pt r(\neg\C[p_0\et r]\to\neg\C[p'\et r])$, on déduit~:\\
$\lbd h((\eta)\lbd x(h)(\beta)x)(\alpha)\tau\force\neg\neg\C[(p_0\et p')\et p']$\\
où $\alpha,\beta$ sont des $\C$-expressions telles que $\alpha:p\Fl p\et p$~; \
$\beta::p\et q\Fl(p\et q)\et q$.\\
On a donc~:\\
(1)~~$\lbd y\lbd z((\eta)\lbd x(\xi yz)(\beta)x)(\alpha)\tau\force
(p'\forcec\pt n\inde(n\eps q_0\to{\cal X}n)),(p'\forcec\pt n\inde({\cal X}n\to n\eps q_0))\to\bot$.

\smallskip\noindent
$\bullet$~~La formule \ $p'\forcec\pt n\inde(n\eps q_0\to{\cal X}n)$ s'écrit \
$\pt n\inde\pt r(r\forcec n\eps q_0\to p'\et r\forcec{\cal X}n)$.\\
Mais \ $r\forcec n\eps q_0\equiv\C[r\et\1]\to n\eps q_0$ (proposition~\ref{n_eps_p}) \
$\equiv\C[r\et\1]\to p'\forcec{\cal X}(n)$ \ par définition de~$q_0$.
Donc \ $p'\forcec\pt n\inde(n\eps q_0\to{\cal X}n)\equiv
\pt n\inde\pt r((\C[r\et\1]\to p'\forcec{\cal X}(n))\to p'\et r\forcec{\cal X}n)\equiv$\\
$\pt n\inde\pt r\pt q'[\pt q(\C[r\et\1],\C[p'\et q]\to{\cal X}^+(q,n)),\C[(p'\et r)\et q']\to{\cal X}^+(q',n)]$.\\
On a donc~:\\
(2)~~$\lbd d\lbd x\lbd y((x)(\alpha')y)(\beta')y\force(p'\forcec\pt n\inde(n\eps q_0\to{\cal X}n))$\\
avec \ $\alpha'::(p\et r)\et q\Fl r\et\1$ \ et \ $\beta'::(p\et r)\et q\Fl p\et q$.

\smallskip\noindent
$\bullet$~~La formule \ $p'\forcec\pt n\inde({\cal X}n\to n\eps q_0)$ s'écrit \
$\pt n\inde\pt r(r\forcec{\cal X}n\to p'\et r\forcec n\eps q_0)$, ou encore~:\\
$\pt n\inde\pt r(r\forcec{\cal X}n,\C[(p'\et r)\et\1]\to n\eps q_0)$, c'est-à-dire, par définition de $q_0$~:\\
$\pt n\inde\pt r(r\forcec{\cal X}n,\C[(p'\et r)\et\1]\to p'\forcec{\cal X}n)$. \ Or, on a~:\\
$\zeta\force\pt n\inde(p'\forcec\pm{\cal X}n)$, autrement dit \
$\zeta\force\pt n\inde\pt r(r\forcec{\cal X}n,\C[p'\et r]\to p'\forcec{\cal X}n)$. Par suite~:\\
(3)~~$\lbd n\lbd x\lbd y(\zeta nx)(\alpha'')y\force p'\forcec\pt n\inde({\cal X}n\to n\eps q_0)$ \
avec \ $\alpha''::(p\et r)\et\1\Fl p\et r$.

\smallskip\noindent
Il résulte de (1,2,3) que~:\\
$((\lbd y\lbd z((\eta)\lbd x(\xi yz)(\beta)x)(\alpha)\tau)\;\lbd d\lbd x\lbd y((x)(\alpha')y)(\beta')y)
\;\lbd n\lbd x\lbd y(\zeta nx)(\alpha'')y\force\bot$.\\
On peut donc poser \ \ {\sf crl2} $=\\
\lbd x_0\lbd y_0\lbd z_0\lbd u((\lbd y\lbd z((y_0)\lbd x(x_0yz)(\beta)x)(\alpha)u)\;
\lbd d\lbd x\lbd y((x)(\alpha')y)(\beta')y)\;\lbd n\lbd x\lbd y(z_0nx)(\alpha'')y$.

\cqfd

\smallskip\noindent
Le reste de cette section est consacré à la preuve du théorème~\ref{pX_decide}.

\subsubsection*{Définition d'une suite par récurrence}\noindent
On se donne une suite finie de formules $\vec{F}(n,p,p')$ avec paramètres et $p_0\in P$.
On se donne aussi une quasi-preuve \ {\sf dse} \ telle que \
{\sf dse}$\force\pt n\pt p\ex p'\,\vec{F}(n,p,p')$.

\smallskip\noindent
{\small{\bfseries Remarque.} Dans l'application qu'on a en vue, la suite $\vec{F}$ est composée de trois formules.}

\smallskip\noindent
D'après le théorème~\ref{ACI}(ii) (axiome du choix pour les individus), il existe une fonction\\
$f:P^3\to P$ telle que~:\\
$\vsig\force\pt n\pt p(\pt k\inde(\vec{F}(n,p,f(n,p,k))\to\bot)\to\pt p'(\vec{F}(n,p,p')\to\bot))$.

\smallskip\noindent
Il en résulte que \ $\lbd x(${\sf dse}$)(\vsig)x\force\pt n\pt p(\pt k\inde(\vec{F}(n,p,f(n,p,k))\to\bot)\to\bot)$.

\smallskip\noindent
On définit une fonction notée $(m\ppt n)$, de $P^2$ dans $P$, en posant, pour $m,n\in P$~:\\
$(m\ppt n)=1$ si $m,n\in\NN$ et $m<n$~; $(m\ppt n)=0$ sinon.

\smallskip\noindent
La relation $(m\ppt n)=1$ est évidemment bien fondée sur $P$.\\
D'après le théorème~\ref{bien_fonde}(ii), on a donc~:\\
$\Y\force\pt k(\pt l($ent$(l),\vec{F}(n,p,f(n,p,l))\to(l\ppt k)\ne1),$ent$(k),\vec{F}(n,p,f(n,p,k))\to\bot)\\
\hspace*{\fill}\to\pt k($ent$(k),\vec{F}(n,p,f(n,p,k))\to\bot)$.\\
En posant $\Y'=\lbd x(\Y)\lbd y\lbd z(x)zy$, on a donc~:\\
$\Y'\force\pt k\inde\{\pt l\inde(\vec{F}[n,p,f(n,p,l)]\to(l\ppt k)\ne1),\vec{F}[n,p,f(n,p,k)]\to\bot\}\\
\hspace*{\fill}\to\pt k\inde(\vec{F}[n,p,f(n,p,k)]\to\bot)$.\\
On a donc~:\\
$\lbd x(${\sf dse}$)(\vsig)(\Y')x\force\pt k\inde\{\pt l\inde(\vec{F}[n,p,f(n,p,l)]\to(l\ppt k)\ne1),\vec{F}[n,p,f(n,p,k)]\to\bot\}\to\bot$.

\smallskip\noindent
On définit la formule \ $G(n,p,k)\equiv\pt l\inde(\vec{F}(n,p,f(n,p,l))\to(l\ppt k)\ne1)$ et la suite de
formules \ $\vec{H}(n,p,k)\equiv\{G(n,p,k),\vec{F}(n,p,f(n,p,k))\}$. On a donc montré~:

\begin{lemma}\label{H_existe}
{\sf dse0}$\force\pt n\pt p\ex k\inde\{\vec{H}(n,p,k)\}$, avec {\sf dse0} $=\lbd x(${\sf dse}$)(\vsig)(\Y')x$.
\end{lemma}

\begin{lemma}\label{H_unicite}
Soit \ \cp \ une quasi-preuve telle que, quels que soient $m,n\in\NN$, on ait~:\\
\cp$\star\ul{m}\ps\ul{n}\ps\xi\ps\eta\ps\zeta\ps\pi\succ\xi\star\pi$ (resp. \ $\eta\star\pi$, \ $\zeta\star\pi$)
si \ $m<n$ \ (resp. \ $n<m$, $m=n$). \ Alors~:\\
i)~\cp$\force\pt m\inde\pt n\inde((m\ppt n)\ne1,(n\ppt m)\ne1,m\ne n\to\bot)$.\\
ii)~{\sf dse1}$\force\pt n\pt p\pt k\inde\pt{k'}\,\inde(\vec{H}(n,p,k),\vec{H}(n,p,k'),k\ne k'\to\bot)$\\
avec \ {\sf dse1}$\,=\lbd k\lbd k'\lbd x\lbd\vec{y}\lbd x'\lbd\vec{y}'(($\cp$\,k'k)(x)k'\vec{y}')(x')k\vec{y}$,
où \ $\vec{y},\vec{y}'$ sont deux suites de variables distinctes de même longueur que la suite $\vec{F}$.
\end{lemma}\noindent
i)~Trivial.\\
ii)~Soient \ $\xi\force G(n,p,k)$, $\vec{\eta}\force\vec{F}(n,p,f(n,p,k))$, \
$\xi'\force G(n,p,k')$, $\vec{\eta}'\force\vec{F}(n,p,f(n,p,k'))$\\
et \ $\zeta\force k\ne k'$.
On doit montrer \
$\cp\star\ul{k}'\ps\ul{k}\ps(\xi)\ul{k}'\vec{\eta}'\ps(\xi')\ul{k}\vec{\eta}\ps\zeta\ps\pi\in\bbot$.\\
Si $k=k'$, on est ramené à \ $\zeta\star\pi\in\bbot$~; c'est vrai parce qu'on a alors $\zeta\force\bot$.\\
Si $k'<k$, on est ramené à montrer \ $\xi\star\ul{k}'\ps\vec{\eta}'\ps\pi\in\bbot$. Cela résulte
immédiatement de~:\\
$\xi\force\pt{k'}\,\inde(\vec{F}(n,p,f(n,p,k'))\to (k'\ppt k)\ne1)$ \ et \ donc \
$\xi\force$ent$(k'),\vec{F}(n,p,f(n,p,k'))\to\bot$,\\
puisque $k'<k$.

\cqfd

\smallskip\noindent
On définit maintenant le prédicat~:\\
$\Phi(x,y)\equiv\pt X(\pt n\pt p\pt k\inde(\vec{H}(n,p,k),X(n,p)\to X(sn,f(n,p,k))),X(0,p_0)\to X(x,y))$\\
et on montre que $\Phi(x,y)$ est une suite de conditions (relation fonctionnelle sur
$\NN$) et quel\-ques autres propriétés de $\Phi$.

\begin{lemma}\label{rec_Phi}\ \\
i)~$\lbd x\lbd y\,y\force\Phi(0,p_0)$.\\
ii)~$\lbd x(x)II\force\pt y(\Phi(0,y)\to y=p_0)$.\\
iii)~~{\sf rec}$\force\pt x\pt y\pt k\inde(\vec{H}(x,y,k),\Phi(x,y)\to\Phi(sx,f(x,y,k)))$\\
où \ {\sf rec} $=\lbd k\lbd x\lbd\vec{y}\lbd x'\lbd z\lbd u(zkx\vec{y})(x')zu$\\
$\vec{y}$ étant une suite de variables distinctes de même longueur que $\vec{F}$.
\end{lemma}\noindent
i)~Trivial.

\smallskip\noindent
ii)~On définit le prédicat binaire ${\cal X}:P^2\to{\cal P}(\Pi)$
en posant \ ${\cal X}(0,q)=\|q=p_0\|$ \ et \ ${\cal X}(p,q)=\vide$ pour $p\ne0$. On remplace $X$
par ${\cal X}$ dans la définition de $\Phi(0,y)$. Comme on a $sn\ne0$ pour tout $n\in P$, on obtient \
$\|\Phi(0,y)\|\supset\|\top,p_0=p_0\to y=p_0\|$~; d'où le résultat.

\smallskip\noindent
iii)~Soient $\xi\force G(x,y,k)$, $\vec{\eta}\force\vec{F}(x,y,f(x,y,k))$, \ $\xi'\force\Phi(x,y)$,\\
$\zeta\force\pt n\pt p\pt k\inde(\vec{H}(n,p,k),X(n,p)\to X(sn,f(n,p,k)))$,\\
$\upsilon\force X(0,p_0)$ \ et \ $\pi\in\|X(sx,f(x,y,k))\|$.\\
Alors \ $\xi'\zeta\upsilon\force X(x,y)$, donc \
$\zeta\star\ul{k}\ps\xi\ps\vec{\eta}\ps\xi'\zeta\upsilon\ps\pi\in\bbot$ \
soit \ ({\sf rec}$)\ul{k}\xi\vec{\eta}\xi'\zeta\upsilon\star\pi\in\bbot$.

\cqfd

\begin{lemma}\label{ccd1}
{\sf ccd1}$\force\pt n\inde\ex p\,\Phi(n,p)$ \
où \ {\sf ccd1}$\,=\lbd n((n)\lbd x\lbd y(x)\lbd z(${\sf cd1}$)zy)\lbd x(x)\lbd x\lbd y\,y$\\
avec {\sf cd1}$\,=\lbd x\lbd y(${\sf dse0}$)\lbd l\lbd\vec{z}(y)(${\sf rec}$)l\vec{z}x$~;\\
$\vec{z}$ est une suite de variables distinctes de même longueur que $\vec{H}$.
\end{lemma}\noindent
Preuve par récurrence sur $n$~; on a $\lbd x\lbd y\,y\force\Phi(0,p_0)$, donc
$\lbd x(x)\lbd x\lbd y\,y\force\ex y\,\Phi(0,y)$.\\
On montre maintenant \ {\sf cd1}$\force\Phi(x,y)\to\ex y\Phi(sx,y)$.\\
On considère donc $\xi\force\Phi(x,y)$, \ $\eta\force\pt y(\Phi(sx,y)\to\bot)$.\\
On a \ {\sf rec}$\force\pt l\inde(\vec{H}(x,y,l),\Phi(x,y)\to\Phi(sx,f(x,y,l)))$
(lemme~\ref{rec_Phi}iii),\\
$\eta\force(\Phi(sx,f(x,y,l))\to\bot)$, et donc~:\\
$\lbd l\lbd\vec{z}(\eta)(${\sf rec}$)l\vec{z}\xi\force
\force\pt l\inde(\vec{H}(x,y,l)\to\bot)$, où $\vec{z}$ est de même longueur que $\vec{H}$.\\
Or, on a \ {\sf dse0}$\force\ex k\inde\{\vec{H}(x,y,k)\}$ (lemme~\ref{H_existe})~; donc~:\\
({\sf dse0})$\lbd l\lbd\vec{z}(\eta)(${\sf rec}$)l\vec{z}\xi\force\bot$, soit
({\sf cd1})$\xi\eta\force\bot$.

\smallskip\noindent
On a donc montré \ {\sf cd1}$\force\pt y(\Phi(x,y)\to\ex y\Phi(sx,y))$,
d'où il résulte que~:\\
$\lbd x\lbd y(x)\lbd z(${\sf cd1}$)zy\force\ex y\Phi(x,y)\to\ex y\Phi(sx,y)$.

\cqfd

\begin{lemma}\label{ccd2}
Il existe une quasi-preuve \ {\sf ccd2} telle que~:\\
{\sf ccd2}$\force\pt n\inde\pt p\pt q(\Phi(n,p),\Phi(n,q)\to p=q)$.
\end{lemma}\noindent
On fait la preuve détaillée par récurrence sur $n$. Elle permet d'écrire explicitement la
quasi-preuve \ {\sf ccd2}.

\smallskip\noindent
Pour $n=0$, le lemme~\ref{rec_Phi}(ii) donne le résultat~: \ $\Phi(0,p),\Phi(0,q)\to p=q$.\\
On fixe $m$ et on suppose $\pt p\pt q(\Phi(m,p),\Phi(m,q)\to p=q)$.\\
On définit le prédicat binaire~:\\
$\Psi(n,q)\equiv\pt p\pt k\inde(n=sm,\vec{H}(m,p,k),\Phi(m,p)\to q=f(m,p,k))$.

\smallskip\noindent
On montre \ $\force\pt p\pt k\inde(\vec{H}(n,p,k),\Phi(n,p)\to\Psi(sn,f(n,p,k)))$, c'est-à-dire~:\\
$\force\pt p\pt q\pt k\inde\pt l\inde\{\vec{H}(n,p,k),\Phi(n,p),sn=sm,\vec{H}(m,q,l),\Phi(m,q)\to f(n,p,k)=f(m,q,l)\}$.\\
Or on a $\|sn=sm\|=\|n=m\|$, $\Phi(m,p),\Phi(m,q)\to p=q$ par hypothèse de récurrence~; $\vec{H}(m,p,k),\vec{H}(m,p,l)\to k=l$ (lemme~\ref{H_unicite}(ii)), d'où $f(n,p,k)=f(m,q,l)$.

\smallskip\noindent
En posant $\Psi'(x,y)\equiv\Phi(x,y)\land\Psi(x,y)$, on a~:\\
$\force\pt p\pt k\inde(\vec{H}(n,p,k),\Psi'(n,p)\to\Psi'(sn,f(n,p,k)))$~; on a aussi $\force\Psi'(0,p_0)$. Cela montre que $\force(\Phi(x,y)\to\Psi'(x,y))$ en faisant $X\equiv\Psi'$ dans la définition de~$\Phi$.\\
On a donc \ $\force\Phi(sm,q)\to\pt p\pt k\inde(\vec{H}(m,p,k),\Phi(m,p)\to q=f(m,p,k))$.
D'où~:\\ $\force\Phi(sm,q),\Phi(sm,q')\to
\pt p\pt k\inde(\vec{H}(m,p,k),\Phi(m,p)\to(q=f(m,p,k))\land(q'=f(m,p,k)))$\\
et donc \
$\force\Phi(sm,q),\Phi(sm,q')\to\pt p\pt k\inde(\vec{H}(m,p,k),\Phi(m,p)\to q=q')$.\\
On obtient donc \ $\force\Phi(sm,q),\Phi(sm,q')\to q=q'$ puisqu'on a \
{\sf ccd1}$\force\ex p\,\Phi(m,p)$ \ (lemme~\ref{ccd1}) et \
{\sf dse0}$\force\pt p\ex k\inde\{\vec{H}(m,p,k)\}$ (lemme~\ref{H_existe}).

\cqfd

\subsubsection*{Fin de la preuve du théorème~\ref{pX_decide}}\noindent
Pour montrer le théorème~\ref{pX_decide}, on fixe $p_0\in P$ et un prédicat binaire
${\cal X}:P^2\to{\cal P}(\Pi)$.\\
Il s'agit de trouver une quasi-preuve \ {\sf dec} \ telle que~:\\
{\sf dec}$\force\ex p'\{(\C[p_0]\to\C[p']),p'\sqle p_0,\pt n\inde(p'\forcec\pm{\cal X}n)\}$.

\smallskip\noindent
On applique les résultats précédents, en prenant pour $\vec{F}(n,p,p')$ la suite des trois
formules suivantes~: \ $\{(\C[p]\to\C[p']),\,(p'\sqle p),\,p'\forcec\pm{\cal X}n\}$.\\
Le lemme~\ref{densite_decide} ci-dessous donne une quasi-preuve \ {\sf dse} \ telle que
{\sf dse}$\force\pt n\pt p\ex p'\{\vec{F}(n,p,p')\}$.

\begin{lemma}\label{densite_decide}
{\sf dse}$\force\pt p\ex p'\{\vec{F}(n,p,p')\}$\\
où \ {\sf dse}$\,=
\lbd a(\lbd h(aII)\lbd x\lbd y\,h)\lbd z(\ccc)\lbd k((a\lbd x\,xz)\beta')\lbd x\lbd y(k)(y)(\alpha)x$\\
avec \ $\beta'=\lbd x\lbd y(x)(\beta)y$, \ $\alpha::(p\et q)\et r\Fl r\et q$ \ et \
$\beta::(p\et q)\et r\Fl p\et r$.
\end{lemma}\noindent
La formule considérée s'écrit \
$\pt p'[(\C[p]\to\C[p']),p'\sqle p\,,(p'\forcec\pm{\cal X}n)\to\bot]\to\bot$.\\
Soit donc $\xi\force\pt p'[(\C[p]\to\C[p']),p'\sqle p\,,(p'\forcec\pm{\cal X}n)\to\bot]$.
On doit montrer \ ({\sf dse}$)\xi\force\bot$.

\smallskip\noindent
$\bullet$~~On montre \ $\lbd h(\xi II)\lbd x\lbd y\,h\force\neg(p\forcec{\cal X}n)$~:\\
Soit $\zeta\force(p\forcec{\cal X}n)$~; on a donc \ $\lbd x\lbd y\,\zeta\force(p\forcec\pm{\cal X}n)$~; en effet~:\\
$p\forcec\pm{\cal X}n\equiv\pt q(\C[p\et q],q\forcec{\cal X}n\to p\forcec{\cal X}n)$.\\
Or, on a $\xi\force\;(\C[p]\to\C[p]),p\sqle p\,,(p\forcec\pm{\cal X}n)\to\bot$~; on a
$I\force\C[p]\to\C[p]$ et $I\force p\sqle p$\\
(puisque $p'\sqle p\equiv\pt q(\neg\C[p\et q]\to\neg\C[p'\et q])$).
Donc $(\xi II)\lbd x\lbd y\,\zeta\force\bot$, d'où le résultat.

\smallskip\noindent
$\bullet$~~On montre maintenant \
$\lbd z(\ccc)\lbd k((\xi\lbd x\,xz)\beta')\lbd x\lbd y(k)(y)(\alpha)x\force(p\forcec{\cal X}n)$.\\
Soient donc \ $\tau\in\C[p\et q]$ et $\pi\in{\cal X}^+(q,n)$. On doit montrer~:\\
$((\xi\lbd x\,x\tau)\beta')\lbd x\lbd y(\kk_\pi)(y)(\alpha)x\star\pi\in\bbot$. Or, on a \
$\lbd x\,x\tau\force\neg\neg\C[p\et q]$,\\
$\beta'\force p\et q\sqle p$ (lemme~\ref{p_et_q_le_p}) et
$\xi\force(\neg\C[p\et q]\to\neg\C[p]),p\et q\sqle p\,,(p\et q\forcec\pm{\cal X}n)\to\bot$~; donc~:\\
$(\xi\lbd x\,x\tau)\beta'\force((p\et q\forcec\pm{\cal X}n)\to\bot)$. Il suffit donc de montrer~:\\
$\lbd x\lbd y(\kk_\pi)(y)(\alpha)x\force(p\et q\forcec\pm{\cal X}n)$, c'est-à-dire~:\\
$\lbd x\lbd y(\kk_\pi)(y)(\alpha)x\force\pt r(\C[(p\et q)\et r],\,r\forcec{\cal X}n\to p\et q\forcec{\cal X}n)$.
On montre, en fait~:\\
$\lbd x\lbd y(\kk_\pi)(y)(\alpha)x\force\pt r(\C[(p\et q)\et r],\,r\forcec{\cal X}n\to\bot)$.\\
Soient donc \ $\upsilon\in\C[(p\et q)\et r]$ \ et \ $\eta\force(r\forcec{\cal X}n)$. On doit montrer~:\\
$(\kk_\pi)(\eta)(\alpha)\upsilon\star\rho\in\bbot$ pour tout $\rho\in\Pi$, soit \
$(\eta)(\alpha)\upsilon\star\pi\in\bbot$. Or, on a  \ $(\alpha)\upsilon\in\C[r\et q]$,\\
donc \ $(\eta)(\alpha)\upsilon\force{\cal X}^+(q,n)$, d'où le résultat, puisque $\pi\in{\cal X}^+(q,n)$.

\smallskip\noindent
$\bullet$~~Il en résulte que \
$(\lbd h(\xi II)\lbd x\lbd y\,h)\lbd z(\ccc)\lbd k((\xi\lbd x\,xz)\beta')\lbd x\lbd y(k)(y)(\alpha)x\force\bot$\\
soit \ ({\sf dse})$\xi\force\bot$, ce qui termine la preuve.

\cqfd

\begin{lemma}\label{p_et_q_le_p}
Soit $\beta::(p\et q)\et r\Fl p\et r$. Alors $\lbd x\lbd y(x)(\beta)y\force\pt p\pt q((p\et q)\sqle p)$.
\end{lemma}\noindent
Cette formule s'écrit $\pt p\pt q\pt r(\neg\C[p\et r],\C[(p\et q)\et r]\to\bot)$.\\
Soient donc $\xi\force\neg\C[p\et r],\tau\in\C[(p\et q)\et r]$, d'où $\beta\tau\in\C[p\et r]$ \ et \
$(\xi)(\beta)\tau\force\bot$. On a donc bien \
$\lbd x\lbd y(x)(\beta)y\star\xi\ps\tau\ps\pi\in\bbot$ pour tout $\pi\in\Pi$.

\cqfd

\smallskip\noindent
On se propose d'appliquer la condition de chaîne dénombrable au prédicat binaire $\Phi(x,y)$.
Les lemmes~\ref{ccd1} et~\ref{ccd2} montrent que les deux premières hypothèses de la c.c.d.
sont réalisées par {\sf ccd1} et {\sf ccd2}. La troisième est donnée par le lemme~\ref{ccd3}
ci-dessous.

\begin{lemma}\label{ccd3}
Il existe deux quasi-preuves \ {\sf ccd3} \ et \ {\sf for}  telles que~:\\
i)~~{\sf ccd3}$\force\pt n\inde\pt p\pt q(\Phi(n,p),\Phi(sn,q)\to q\sqle p)$.\\
ii)~~{\sf for}$\force\pt n\inde\pt q(\Phi(sn,q)\to q\forcec\pm{\cal X}n)$.
\end{lemma}\noindent
D'après le lemme~\ref{rec_Phi}(iii), on a~:\\
{\sf rec}$\force\pt k\inde(\vec{H}(n,p,k),\Phi(n,p)\to\Phi(sn,f(n,p,k)))$.
En utilisant {\sf ccd2} (lemme~\ref{ccd2}), on a~:\\
$\force\pt k\inde(\vec{H}(n,p,k),\Phi(n,p),\Phi(sn,q)\to q=f(n,p,k))$.\\
Or, $\vec{H}(n,p,k)$ est une suite de quatre formules dont les deux dernières sont~:\\
$f(n,p,k)\sqle p$ \ et \ $f(n,p,k)\forcec\pm{\cal X}n$.\\
i)~On en déduit d'abord \ $\force\pt k\inde(\vec{H}(n,p,k),\Phi(n,p),\Phi(sn,q)\to q\sqle p)$.\\
D'où le résultat, puisqu'on a \
{\sf dse0}$\force\ex k\inde\{\vec{H}(n,p,k)\}$ (lemme~\ref{H_existe}).\\
ii)~On en déduit aussi \ $\force\pt k\inde(\vec{H}(n,p,k),\Phi(n,p),\Phi(sn,q)\to q\forcec\pm{\cal X}n)$.\\
On obtient donc \ $\force\pt n\inde\pt q(\Phi(sn,q)\to q\forcec\pm{\cal X}n)$ puisqu'on a~:\\
{\sf ccd1}$\force\pt n\inde\ex p\,\Phi(n,p)$ (lemme~\ref{ccd1}) et \
{\sf dse0}$\force\pt n\pt p\ex k\inde\{\vec{H}(n,p,k)\}$ (lemme~\ref{H_existe}).

\cqfd

\smallskip\noindent
On peut maintenant appliquer la c.c.d. au prédicat $\Phi(x,y)$, ce qui donne une quasi-preuve
{\sf ccd0} telle que \ {\sf ccd0}$\force\ex p'\{\vec{\Omega}(n,p,p')\}$ \ avec~:\\
$\vec{\Omega}(n,p,p')\equiv\{\pt n\inde\pt p(\Phi(n,p)\to p'\sqle p),\;
\pt n\inde\pt p(\Phi(n,p),\neg\C[p]\to\bot),\neg\C[p']\to\bot\}$.

\smallskip\noindent
Pour terminer la preuve du théorème~\ref{pX_decide}, il suffit donc de trouver des quasi-preuves\\
{\sf dec0,dec1,dec2} \ telles que~:

\smallskip\noindent
{\sf dec0}$\force\pt p'(\vec{\Omega}(n,p,p'),\neg\C[p_0],\C[p']\to\bot)$~;\\
{\sf dec1}$\force\pt p'(\vec{\Omega}(n,p,p')\to p'\sqle p_0)$~;\\
{\sf dec2}$\force\pt p'(\vec{\Omega}(n,p,p')\to\pt n\inde(p'\forcec\pm{\cal X}n))$.

\smallskip\noindent
Soient donc \ $\omega_0,\omega_1\in\Lbd$ \ tels que~:\\
$\omega_0\force\pt n\inde\pt p(\Phi(n,p)\to p'\sqle p)$ \ et \
$\omega_1\force\pt n\inde\pt p(\Phi(n,p),\neg\C[p]\to\bot),\neg\C[p']\to\bot$

\smallskip\noindent
En appliquant le lemme~\ref{rec_Phi}(i) avec $n=0,p=p_0$, on obtient \ $(\omega_0)\lbd x\lbd y\,y\force p'\sqle p_0$.\\ On peut donc prendre \ {\sf dec1} $=\lbd a\lbd b(a)\lbd x\lbd y\,y$.

\begin{lemma}\label{ccd4}
{\sf ccd4}$\force(\C[p_0]\to\pt n\inde\pt p(\Phi(n,p),\neg\C[p]\to\bot))$\\
où \ {\sf ccd4}$\,=
\lbd a\lbd b\lbd c((b\lbd x_0\lbd x_1\lbd x_2\lbd x_3\lbd x\lbd y(x)(x_1)y)\lbd x\,xa)c$.
\end{lemma}\noindent
Soient $\tau\in\C[p_0]$, $\xi\force\Phi(n,p)$ et $\eta\force\neg\C[p]$.\\
En faisant $X(x,y)\equiv\neg\neg\C[y]$ dans la définition de $\Phi$, on a~:\\
$\xi\force\pt n'\pt p'\pt k\inde
(G[n',p',k],\vec{F}[n',p',f(n',p',k)],\neg\neg\C[p']\to\neg\neg\C[f(n',p',k)]),\\
\hspace*{\fill}\neg\neg\C[p_0],\neg\C[p]\to\bot$.\\
On a \ $\lbd x(x)\tau\force\neg\neg\C[p_0]$.\\
Par ailleurs, puisque \ $\vec{F}[n',p',q]\equiv
\{(\neg\C[q]\to\neg\C[p']),(q\sqle p'),q\forcec\pm{\cal X}n\}$, on a facilement~:\\
\nopagebreak
$\lbd x_0\lbd x_1\lbd x_2\lbd x_3\lbd x\lbd y(x)(x_1)y\force\\
\hspace*{\fill}\pt n'\pt p'\pt k\inde(G[n',p',k],\vec{F}[n',p',f(n',p',k)],
\neg\neg\C[p']\to\neg\neg\C[f(n',p',k)])$.\\
Il en résulte que \ $((\xi\lbd x_0\lbd x_1\lbd x_2\lbd x_3\lbd x\lbd y(x)(x_1)y)\lbd x(x)\tau)\eta\force\bot$,
soit \ ({\sf ccd4})$\tau\xi\eta\force\bot$.

\cqfd

\smallskip\noindent
Du lemme~\ref{ccd4}, on déduit immédiatement \
$\lbd x(\omega_1)(${\sf ccd4}$)x\force\C[p_0],\neg\C[p']\to\bot$.\\
On peut donc poser \ {\sf dec0} $=\lbd a\lbd b\lbd x(b)(${\sf ccd4}$)x$.

\begin{lemma}\label{force_inf}\ \\
i)~~{\sf lef0}$\force\pt p\pt q(p\forcec{\cal X}n,\,q\sqle p\to q\forcec{\cal X}n)$ \ avec \
{\sf lef0}$\,=\lbd x\lbd y\lbd z(\ccc)\lbd k((y)\lbd u(k)(x)u)z$.\\
ii)~~{\sf lef1}$\force\pt p\pt q(p\forcec\pm{\cal X}n,\,q\sqle p\to q\forcec\pm{\cal X}n)$ \ avec\\
{\sf lef1}$\,=\lbd x\lbd y\lbd z\lbd u((${\sf lef0}$)(\ccc)\lbd h((y)\lbd v(h)(x)vu)z$.
\end{lemma}\noindent
i) Immédiat en explicitant les formules~:\\
$p\forcec{\cal X}n\equiv\pt r(\C[p\et r]\to{\cal X}^+(r,n))$~;\\
$q\sqle p\equiv\pt r(\neg\C[p\et r]\to\neg\C[q\et r])$~;\\
$q\forcec{\cal X}n\equiv\pt r(\C[q\et r]\to{\cal X}^+(r,n))$.\\
On déclare \ $x:p\forcec{\cal X}n$, \ $y:q\sqle p$, \ $z:\C[q\et r]$, \ $k:\neg{\cal X}^+n$.\\
ii)~~On écrit les formules~:\\
$p\forcec\pm{\cal X}n\equiv\pt r(\C[p\et r],r\forcec{\cal X}n\to p\forcec{\cal X}n)$~;\\
$q\sqle p\equiv\pt r(\neg\C[p\et r]\to\neg\C[q\et r])$~;\\
$q\forcec\pm{\cal X}n\equiv\pt r(\C[q\et r],r\forcec{\cal X}n\to q\forcec{\cal X}n)$.\\
On déclare \ $x:p\forcec\pm{\cal X}n$, \ $y:q\sqle p$, \ $z:\C[q\et r]$, \ $u:r\forcec{\cal X}n$, \
$v:\C[p\et r]$, \ $h:\neg(p\force{\cal X}n)$.

\cqfd

\smallskip\noindent
En utilisant les lemmes~\ref{ccd3}(ii) et~\ref{force_inf} ainsi que \
$\omega_0\force\pt n\inde\pt p(\Phi(n,p)\to p'\sqle p)$, on obtient~:\\
$\lbd n\lbd x((${\sf lef1}$)(${\sf for}$)nx)(\omega_0)nx
\force\pt n\inde\pt q(\Phi(sn,q)\to p'\forcec\pm{\cal X}n)$.\\
Or, on a \ {\sf ccd1}$\force\pt n\inde\ex p\,\Phi(n,p)$ (lemme~\ref{ccd1})~; on en déduit~:\\
$\lbd n(\ccc)\lbd k((${\sf ccd1}$)(s)n)\lbd x(k)((${\sf lef1}$)(${\sf for}$)nx)(\omega_0)nx
\force\pt n\inde(p_{\cal X}\forcec\pm{\cal X}n)$.\\
On peut donc poser \
{\sf dec2} $=\lbd a\lbd b\lbd n(\ccc)\lbd k((${\sf ccd1}$)(s)n)\lbd x(k)((${\sf lef1}$)(${\sf for}$)nx)(a)nx$.

\smallskip\noindent
Cela termine la preuve du théorème~\ref{pX_decide}.

\cqfd

\section*{L'axiome d'ultrafiltre sur $\NN$}\noindent
On considère une algèbre de réalisabilité standard ${\cal A}$ et un ${\cal A}$-modèle ${\cal M}$ dans
lequel l'en\-semble d'individus (qui est aussi l'ensemble des conditions) est $P={\cal P}(\Pi)^{\NN}$.\\
La relation binaire $\eps$ est définie par $\|n\eps p\|=p(n)$ si $n\in\NN$~; sinon, $\|n\eps p\|=\vide$.\\
$\1$ est défini par $\1(n)=\vide$ pour tout $n\in\NN$~;\\
$\et$ est définie par $\|n\eps(p\et q)\|=\|n\eps p\land n\eps q\|$ pour tout $n\in\NN$.

\subsubsection*{L'axiome de représentation des prédicats sur $\NN$ (RPN)}\noindent
On définit la fonction récursive d'arité $k$, notée $(n_1,\ldots,n_k)$ (codage des $k$-uplets)~:\\
$(n_1,n_2)=n_1+(n_1+n_2)(n_1+n_2+1)/2$~; $(n_1,\ldots,n_{k+1})=((n_1,\ldots,n_k),n_{k+1})$.

\begin{proposition}\label{rep_pred}
$\force\pt X\ex x\pt y_1\indi\ldots\pt y_k\indi((y_1,\ldots,y_k)\eps x\dbfl X(y_1,\ldots,y_k))$ où $X$ est
une variable de prédicat d'arité $k$.
\end{proposition}\noindent
Soit \ ${\cal X}:P^k\to{\cal P}(\Pi)$ un prédicat d'arité $k$. On définit $a\in P$ en posant~:\\
$a(n)={\cal X}(n_1,\ldots,n_k)$ pour $n\in\NN$, $n=(n_1,\ldots,n_k)$. On a alors immédiatement~:\\
$I\force\pt y_1\inde\ldots\pt y_k\inde((y_1,\ldots,y_k)\eps a\to{\cal X}(y_1,\ldots,y_k))$ \ et\\
$I\force\pt y_1\inde\ldots\pt y_k\inde({\cal X}(y_1,\ldots,y_k)\to(y_1,\ldots,y_k)\eps a)$.\\
On en déduit~:\\
$\lbd x(x)I\force\pt X\ex x\pt y_1\inde\ldots\pt y_k\inde
((y_1,\ldots,y_k)\eps x\to X(y_1,\ldots,y_k))$ \ et\\
$\lbd x(x)I\force\pt X\ex x\pt y_1\inde\ldots\pt y_k\inde
(X(y_1,\ldots,y_k)\to(y_1,\ldots,y_k)\eps x)$.

\smallskip\noindent
Il suffit alors d'appliquer le théorème~\ref{mm1}.

\cqfd

\subsubsection*{Le schéma de compréhension pour $\NN$ (SCN)}\noindent
Soit \ $F[y,x_1,\ldots,x_k]$ \ une formule dont les variables libres sont parmi $y,x_1,\ldots,x_k$.
On définit une fonction d'arité $k$, soit \ $g_F:P^k\to P$, autrement dit \
$g_F:P^k\fois\NN\to{\cal P}(\Pi)$ \ en posant \
$g_F(p_1,\ldots,p_k)(n)=\|F[n,p_1,\ldots,p_k]\|$ pour tout $n\in\NN$.

\begin{proposition}\label{comp_ent}
On a \ $\force\pt x_1\ldots\pt x_k\pt y\indi(y\eps g_F(x_1,\ldots,x_k)\dbfl F[y,x_1,\ldots,x_k])$ pour toute formule \ $F[y,x_1,\ldots,x_k]$.
\end{proposition}\noindent
En effet, on a trivialement~:\\
$I\force \pt x_1\ldots\pt x_k\pt y\inde(y\eps g_F(x_1,\ldots,x_k)\to F[y,x_1,\ldots,x_k])$ \ et\\
$I\force \pt x_1\ldots\pt x_k\pt y\inde(F[y,x_1,\ldots,x_k]\to y\eps g_F(x_1,\ldots,x_k))$.

\smallskip\noindent
Il suffit alors d'appliquer le théorème~\ref{mm1}.

\cqfd

\smallskip\noindent
{\small{\bfseries Remarque.} Le symbole de fonction binaire $\et$ est obtenu en appliquant SCN à la formule
$y\eps x_1\land y\eps x_2$.} 

\subsubsection*{Le modèle générique}\noindent
On désigne par $\C[x]$ la formule $\pt m\indi\ex n\indi(m+n)\eps x$, qui exprime que l'ensemble d'en\-tiers~$x$
est infini. Le prédicat $\C$ est défini par cette formule~: pour tout $p\in P$, $|\C[p]|$ est, par
définition, l'ensemble $\{\tau\in\Lbd;\;\tau\force\C[p]\}$.\\
Il en résulte que la condition \ $\gamma::t(p_1,\ldots,p_n)$ $\Fl$ $u(p_1,\ldots,p_n)$ s'écrit~:\\
$\lbd x\,\gamma x\force\pt p_1\ldots\pt p_n(\C[t(p_1,\ldots,p_n)]\to\C[u(p_1,\ldots,p_n)])$.

\smallskip\noindent
Pour terminer la définition de l'algèbre ${\cal B}$ (et du ${\cal B}$-modèle ${\cal N}$), il reste donc à trouver
des quasi-preuves $\alpha_0,\alpha_1,\alpha_2,\beta_0,\beta_1,\beta_2$ telles que~:

\smallskip\noindent
$\alpha_0\force\pt p\pt q\pt r(\C[(p\et q)\et r]\to\C[p\et(q\et r)])$~; \ $\alpha_1\force\pt p(\C[p]\to\C[p\et\1])$~;\\
$\alpha_2\force\pt p\pt q(\C[p\et q]\to\C[q])$~; \ $\beta_0\force\pt p(\C[p]\to\C[p\et p])$~; \
$\beta_1\force\pt p\pt q(\C[p\et q]\to\C[q\et p])$~;\\
$\beta_2\force\pt p\pt q\pt r\pt s(\C[((p\et q)\et r)\et s]\to\C[(p\et(q\et r))\et s])$.

\smallskip\noindent
Or, on a facilement, en déduction naturelle~:\\
$\vdash\theta:\pt n(n\eps x\to n\eps x')\to(\C[x]\to\C[x'])$ \ avec \
$\theta=\lbd f\lbd u\lbd m\lbd h(um)\lbd n\lbd x(hn)(f)x$.\\
D'après le théorème~\ref{adequat} (lemme d'adéquation), on peut donc poser $\alpha_i=\theta\alpha^{*}_i$
et $\beta_i=\theta\beta_i^*$, pour des quasi-preuves $\alpha_i^*,\beta_i^*(0\le i\le 2)$ telles que~:\\
$\vdash\alpha_0^*:\pt X\pt Y\pt Z\{(X\land Y)\land Z\to X\land(Y\land Z)\}$~; \
$\vdash\alpha_1^*:\pt X\{X\to X\land\top\}$~;\\
$\vdash\alpha_2^*:\pt X\pt Y\{X\land Y\to Y\}$~; \ $\vdash\beta_0^*:\pt X\{X\to X\land X\}$~; \
$\vdash\beta_1^*:\pt X\pt Y\{X\land Y\to Y\land X\}$~;\\
$\vdash\beta_2^*:\pt X\pt Y\pt Z\pt U\{((X\land Y)\land Z)\land U\to(X\land(Y\land Z))\land U\}$.

\subsection*{La condition de chaîne dénombrable}\noindent
On montre, dans cette section le~:

\begin{theorem}\label{ccd_ver1}\ \\
La structure de forcing $\{\C,\et,\1\}$ satisfait la condition de chaîne dénombrable dans ${\cal M}$.
\end{theorem}\noindent
Il s'agit de trouver une quasi-preuve \ {\sf ccd} \ telle que~:\\
{\sf ccd}$\force\pt X\ex x\{\pt n\inde\ex p\,X(n,p),\pt n\inde\pt p\pt q(X(n,p), X(n,q)\to p=q),\\
\hspace*{4.2em}\pt n\inde\pt p\pt q(X(n,p),X(sn,q)\to q\sqle p)\to\\
\hspace*{4.2em}\pt n\inde\pt p(X(n,p)\to x\sqle p)
\land(\pt n\inde\pt p(X(n,p)\to\C[p])\to\C[x])\}$\\
où $p\sqle q$ est la formule $\pt r(\C[p\et r]\to\C[q\et r])$.

\smallskip\noindent
D'après le théorème~\ref{mm1}, cela revient à trouver une quasi-preuve \ {\sf ccd'} \ telle que~:

\smallskip\noindent
{\sf ccd'}$\force\pt X\ex x\{\pt n\indi\ex p\,X(n,p),\pt n\indi\pt p\pt q(X(n,p), X(n,q)\to p=q),\\
\hspace*{4.2em}\pt n\indi\pt p\pt q(X(n,p),X(sn,q)\to q\sqle p)\to\\
\hspace*{4.2em}\pt n\indi\pt p(X(n,p)\to x\sqle p)
\land(\pt n\indi\pt p(X(n,p)\to\C[p])\to\C[x])\}$.

\smallskip\noindent
D'après le théorème~\ref{adequat} (lemme d'adéquation), nous pouvons utiliser la méthode suivante
pour montrer \ $\force F$~:\\
Montrer \ $\force A_1,\ldots,\force A_k$, puis\\
montrer \ $A_1,\ldots,A_k\vdash F$ \ au moyen des règles de la déduction naturelle classique
du second ordre (qui contient le schéma de compréhension), et des axiomes suivants qui sont réalisés par des quasi-preuves dans le ${\cal A}$-modèle ${\cal M}$~:

\smallskip\noindent
$\bullet$~~$t\ne u$ pour tous les termes clos $t,u$  qui ont des valeurs distinctes dans ${\cal M}$.\\
$\bullet$~~$\pt x_1\indi\ldots\pt x_k\indi(t(x_1,\ldots,x_k)=u(x_1,\ldots,x_k))$
pour toutes les équations entre termes qui sont vraies dans $\NN$.\\
$\bullet$~~Le schéma de fondation (SCF\/, voir théorème~\ref{bien_fonde}ii) qui est constitué des formules~:\\
$\pt X_1\ldots\pt X_k\{\pt x\indi[\pt y\indi(X_1y,\ldots,X_ky\to f(y,x)\ne1),X_1x,\ldots,X_kx\to\bot]\\
\hspace*{\fill}\to\pt x\indi(X_1x,\ldots,X_kx\to\bot)\}$\\
où $f:P^2\to P$ est telle que la relation $f(y,x)=1$ soit bien fondée sur $\NN$.\\
$\bullet$~~Le schéma d'axiome du choix pour les individus (ACI, voir théorème~\ref{ACI}) qui est constitué des
formules \ \ $\pt\vec{x}(\pt y\indi F(\vec{x},f_F(\vec{x},y))\to\pt y\,F(\vec{x},y))$~;\\
$\vec{x}=(x_1,\ldots,x_k)$ est une suite finie de variables, $\pt\vec{x}\pt y\indi F$ une formule close
quelconque, et $f_F$ un symbole de fonction d'arité $k+1$.\\
$\bullet$~~L'axiome de représentation des prédicats sur $\NN$ (RPN, voir proposition~\ref{rep_pred})
qui est constitué des formules \ \ $\pt X\ex x\pt\vec{y}\indi((y_1,\ldots,y_k)\eps x\dbfl X\vec{y})$~;\\
$\vec{y}=(y_1,\ldots,y_k)$ est une suite de $k$ variables et $X$ est une variable de prédicat d'arité $k$.\\
$\bullet$~~Le schéma de compréhension pour les entiers (SCN, voir proposition~\ref{comp_ent}),
qui est constitué des formules \ \ $\pt\vec{x}\pt y\indi(y\eps g_F(\vec{x})\dbfl F[y,\vec{x}])$~;\\
$\vec{x}=(x_1,\ldots,x_k)$ est une suite de $k$ variables, $\pt\vec{x}\pt y\indi F$ est une formule close quelconque,
et $g_F$ est un symbole de fonction d'arité $k$.

\begin{lemma}
$\vdash\pt p\pt q(p\sqle q\dbfl\ex m\indi\pt n\indi(n+m\eps p\to n+m\eps q))$.
\end{lemma}\noindent
On applique le SCN à la formule $F[y,x]\equiv y\neps x$~; on obtient donc~:

\smallskip\noindent
\centerline{$\vdash\pt x\pt y\indi(y\eps\neg x\dbfl y\neps x)$}

\noindent
en utilisant la notation $\neg x$ pour $g_F(x)$.

\smallskip\noindent
On a \ $p\sqle q\equiv\pt r(\C[p\et r]\to \C[q\et r])$ \ et \ donc \
$p\sqle q\;\vdash\C[p\et\neg q]\to \C[q\et\neg q]$.\\
Or, on a \ $\C[q\et\neg q]\vdash\pt m\indi\ex n\indi(m+n\eps q\land m+n\neps q)\;\vdash\bot$, \ d'où~:\\
$p\sqle q\;\vdash\neg\C[p\et\neg q]$, \ soit \
$\vdash p\sqle q\to\ex m\indi\pt n\indi\neg(m+n\eps p\land\neg(m+n\eps q))$.

\smallskip\noindent
Inversement, des hypothèses~:\\
$\pt n{'}\,\indi(m'+n'\eps p\to m'+n'\eps q),\pt m\indi\ex n\indi(m+n\eps p\land m+n\eps r)$, \ on déduit~:\\
$\pt m\indi\ex n\indi((m'+m)+n\eps p\land(m'+m)+n\eps r)$, \ puis~:\\
$\pt m\indi\ex n\indi(m+(m'+n)\eps q\land m+(m'+n)\eps r)$ \ puis~:\\
$\pt m\indi\ex n\indi(m+n\eps q\land m+n\eps r)$. \ Par suite~:\\
$\pt n{'}\,\indi(m'+n'\eps p\to m'+n'\eps q)\;\vdash\;\C[p\et r]\to\C[q\et r]$ \ et donc~:\\
$\ex m'\pt n{'}\,\indi(m'+n'\eps p\to m'+n'\eps q)\;\vdash\;\C[p\et r]\to\C[q\et r]$.

\cqfd

\smallskip\noindent
En appliquant RPN et le schéma de compréhension, on obtient~:\\
$\force\pt X\ex h\,D(h,X)$ \ avec \
$D(h,X)\equiv\pt k\indi\pt n\indi((k,n)\eps h\dbfl\pt q\pt i\indi(i\le n,X(i,q)\to k\eps q))$.\\
Le sens intuitif de $D(h,X)$ est~: ``$h$ est l'individu associé à la suite de conditions $X$
rendue décroissante''.

\smallskip\noindent
On applique SCN à la formule $F(k,n,h)\equiv(k,n)\eps h$. On obtient donc~:\\
$\vdash\pt n\pt h\pt k\indi\pt n(k\eps g_F(n,h)\dbfl(k,n)\eps h)$.\\
On utilisera la notation $h_n$ pour $g_F(n,h)$. On a donc~:

\smallskip\noindent
\centerline{$\vdash\pt n\pt h\pt k\indi(k\eps h_n\dbfl (k,n)\eps h)$.}

\smallskip\noindent
et par suite~:\\
\centerline{$D(h,X)\vdash
\pt k\indi\pt n\indi(k\eps h_n\dbfl\pt q\pt i\indi(i\le n,X(i,q)\to k\eps q))$}

\smallskip\noindent
On pose \ $\Phi(k,h,n)\equiv\ex i\indi\{\pt j\indi(j+n\eps h_n\to(j<i)\ne1),\,i+n\eps h_n,\,k=i+n\}$.\\
Le sens intuitif de $\Phi(k,h,n)$ est~: ``~$k$ est le premier élément de $h_n$ qui est $\ge n$~''.\\
On applique SCN à la formule $F(k,h)\equiv\ex n\indi\,\Phi(k,h,n)$. On obtient donc~:\\
$\vdash\pt h\pt k\indi(k\eps g_F(h)\dbfl\ex n\indi\,\Phi(k,h,n))$.\\
On utilisera la notation \ $\inf(h)$ pour $g_F(h)$. On a donc~:

\smallskip\noindent
\centerline{$\vdash\pt h\pt k\indi(k\eps\inf(h)\dbfl\ex n\indi\,\Phi(k,h,n))$.}

\smallskip\noindent
Les hypothèses de la c.c.d. sont~:

\smallskip\noindent
$H_0[X]\equiv\pt n\indi\ex p\,X(n,p)$~;\\
$H_1[X]\equiv\pt n\indi\pt p\pt q(X(n,p), X(n,q)\to p=q)$~;\\
$H_2[X]\equiv\pt n\indi\pt p\pt q(X(n,p),X(sn,q)\to q\sqle p)$~;\\
$H_3[X]\equiv\pt n\indi\pt p(X(n,p)\to\C[p])$.

\smallskip\noindent
On pose \ $\vec{H}[X]\equiv\{H_0[X],H_1[X],H_2[X],H_3[X]\}$ \ et \
$\vec{H}_*\![X]=\{H_0[X],H_1[X],H_2[X]\}$.

\smallskip\noindent
Il suffit donc de montrer~:\\
$D(h,X),\vec{H}_*\![X]\,\vdash\pt n\indi\pt p(X(n,p)\to\inf(h)\sqle p)$ \ et\\
$D(h,X),\vec{H}[X]\,\vdash\,\C[\inf(h)]$.

\smallskip\noindent
{\bfseries Notation.} La formule \ $\pt n\indi(n\eps p\to n\eps q)$ \ est notée \ $p\subseteq q$.

\begin{lemma}\label{DhH1}
$D(h,X)\vdash\pt m\indi\pt n\indi(h_{n+m}\subseteq h_n)$.
\end{lemma}\noindent
Cette formule s'écrit \ $\pt m\indi\pt n\indi\pt k\indi(k\eps h_{n+m}\to k\eps h_n)$. Or, on a~:

\smallskip\noindent
$D(h,X)\vdash\pt m\indi\pt n\indi\pt k\indi(k\eps h_{n+m}\to\pt q\pt i\indi(i\le n+m,X(i,q)\to
k\eps q))$~;\\
$\vdash\pt m\indi\pt n\indi\pt k\indi[\pt q\pt i\indi(i\le n+m,X(i,q)\to k\eps q)\to
\pt q\pt i\indi(i\le n,X(i,q)\to k\eps q)]$~:\\
$D(h,X)\vdash\pt m\indi\pt n\indi\pt k\indi(\pt q\pt i\indi(i\le n,X(i,q)\to k\eps q)\to k\eps h_n)$.

\cqfd

\begin{lemma}\label{DhH2_0}
$D(h,X),H_0[X],H_1[X]\vdash\pt n\indi\pt k\indi\pt p(X(sn,p),\,k\eps p,\,k\eps h_n\to k\eps h_{sn})$.
\end{lemma}\noindent
On a \ $D(h,X),$ int$(k),$ int$(n)\vdash\pt p\pt i\indi(i\le sn,X(i,p)\to k\eps p)\to k\eps h_{sn}$.\\
Or, on a \ \ int$(n),$ int$(i),\,i\le sn\,\vdash\,i\le n\lor i=sn$, et donc~:\\
\nopagebreak
int$(n),\,\pt p\pt i\indi(i\le n,X(i,p)\to k\eps p),\,\pt p(X(sn,p)\to k\eps p)\;\vdash\;
\pt p\pt i\indi(i\le sn,X(i,p)\to k\eps p)$.\\
Par suite, on a~:\\
$D(h,X),$ int$(k),$ int$(n)\vdash
\pt p\pt i\indi(i\le n,X(i,p)\to k\eps p),\pt p(X(sn,p)\to k\eps p)\to k\eps h_{sn}$, \ soit~:\\
$D(h,X),$ int$(k),$ int$(n)\vdash k\eps h_n,\pt p(X(sn,p)\to k\eps p)\to k\eps h_{sn}$. Par suite~:\\
$D(h,X),$ int$(k),$ int$(n),H_0[X],H_1[X]\vdash\pt p(k\eps h_n,X(sn,p),k\eps p\to k\eps h_{sn})$.

\cqfd

\begin{lemma}\label{DhH2}
$D(h,X),\vec{H}_*\![X]\vdash\pt n\indi\pt p(X(n,p)\to p\sqle h_n)$.
\end{lemma}\noindent
Preuve par récurrence sur $n$. On doit montrer~:\\
$D(h,X),\vec{H}_*\![X],$ int$(n)\,\vdash\,
\pt p\ex m\indi\pt l\indi(X(n,p),l+m\eps p\to l+m\eps h_n)$.\\
Pour $n=0$, on a \ $D(h,X)\vdash\pt k\indi(\pt q(X(0,q)\to k\eps q)\to k\eps h_0)$.
Il suffit donc de montrer~:\\
$D(h,X),\vec{H}_*\![X]\,\vdash\,
\pt p\ex m\indi\pt l\indi\pt q(X(0,p),l+m\eps p,X(0,q)\to l+m\eps q)$,\\
ce qui découle, en fait, de $H_1[X]$, à savoir $X(0,p),X(0,q)\to p=q$.\\
L'hypothèse de récurrence est $\pt p(X(n,p)\to p\sqle h_n)$~;\\
$H_2[X]$ est $\pt p\pt q(X(n,p),X(sn,q)\to q\sqle p)$~; $H_0[X]$ est $\ex p\,X(n,p)$.\\
Par ailleurs, on a facilement \ $q\sqle p,p\sqle r\vdash q\sqle r$. On en déduit donc~:\\
$\pt p(X(sn,p)\to p\sqle h_n)$, soit \ $\pt p\ex m\indi\pt l\indi(X(sn,p),l+m\eps p\to l+m\eps h_n)$.\\
Or, on a \ $D(h,X),H_0[X],H_1[X]\,\vdash X(sn,p),\,l+m\eps p,\,l+m\eps h_n\to l+m\eps h_{sn}$
par le lemme~\ref{DhH2_0}.\\
On a donc  \ $\pt p\ex m\indi\pt l\indi(X(sn,p),l+m\eps p\to l+m\eps h_{sn})$ \ c'est-à-dire~:\\
$\pt p(X(sn,p)\to p\sqle h_{sn})$, ce qui est le résultat voulu.

\cqfd

\begin{lemma}\label{DhH2_1}
$D(h,X),\vec{H}(X)\vdash\pt n\indi\C[h_n]$.
\end{lemma}\noindent
On a \ $\pt n\indi\pt p(X(n,p)\to\C[p])$ \ d'après $H_3$. Par ailleurs, on a facilement~:\\
$\vdash\pt p\pt q(\C[p],p\sqle q\to\C[q])$. En appliquant le lemme~\ref{DhH2}, on obtient donc~:\\
$D(h,X),\vec{H}(X)\vdash\pt n\indi\pt p(X(n,p)\to\C[h_n])$. D'où le résultat, en appliquant $H_0[X]$.

\cqfd

\begin{lemma}\label{DhH3}
$D(h,X),\vec{H}[X]\vdash\pt n\indi\ex k\indi\Phi(k,h,n)$.
\end{lemma}\noindent
D'après le schéma de fondation (SCF), on a~:\\
$\vdash\,\pt i\indi\{\pt j\indi(j+n\eps h_n\to(j\ppt i)\ne1),i+n\eps h_n\to\bot\}
\to\pt i\indi(i+n\eps h_n\to\bot)$.\\
Or, on a \ $D(h,X),\vec{H}[X]\vdash\pt n\indi\C[h_n]$ (lemme~\ref{DhH2_1}), donc \
$D(h,X),\vec{H}[X]\vdash\pt n\indi\ex i\indi i+n\eps h_n$.
On en déduit \ $D(h,X),\vec{H}[X]\vdash\pt n\indi\ex i\indi
\{\pt j\indi(j+n\eps h_n\to(j\ppt i)\ne1),i+n\eps h_n\}$.

\cqfd

\begin{lemma}\label{DhH4}
$D(h,X),\vec{H}[X]\;\vdash\;\C[\inf(h)]$.
\end{lemma}\noindent
On a \ $\C[\inf(h)]\equiv\pt m\indi\ex i\indi(i+m\eps\inf(h))$.\\
Or, par définition du symbole de fonction $\inf$, on a \
$\vdash\pt h\pt k\indi(k\eps\inf(h)\dbfl\ex n\indi\Phi(k,h,n))$.\\
Donc \ $\vdash\C[\inf(h)]\dbfl\pt m\indi\ex i\indi\ex n\indi\Phi(i+m,h,n)$.\\
Par définition de $\Phi$, on a trivialement \
$\vdash\pt n\indi\pt k\indi(\Phi(k,h,n)\to\ex i\indi(k=i+n))$.\\
Par ailleurs, on a \ $D(h,X),\vec{H}[X]\;\vdash\;\pt n\indi\ex k\indi\Phi(k,h,n)$ (lemme~\ref{DhH3}).\\
Donc \ $D(h,X),\vec{H}[X]\;\vdash\;\pt n\indi\ex i\indi\,\Phi(i+n,h,n)$, d'où \
$D(h,X),\vec{H}[X]\;\vdash\;\C[\inf(h)]$.

\cqfd

\begin{lemma}\label{DhH5}\ \\
$D(h,X),\vec{H}_*\![X]\vdash\pt h\pt k\indi\pt k{'}\indi
\pt n\indi\pt n{'}\indi(\Phi(k,h,n),\Phi(k',h,n'),k'>k\to n'>n)$.
\end{lemma}\noindent
On a \ $\Phi(k,h,n)\equiv\ex i\indi\vec{\Psi}(k,h,n,i)$, avec~:\\
$\vec{\Psi}(k,h,n,i)\equiv\{\pt j\indi(j+n\eps h_n\to(j\ppt i)\ne1),\; i+n\eps h_n,\; k=i+n\}$.\\
On doit donc montrer~:\\
$D(h,X),\vec{H}_*\![X],$ int$(k),$ int$(k'),$ int$(n),$ int$(n'),$ int$(i),$ int$(i')\,\vdash\,
\vec{\Xi}(h,k,n,i,k',n',i')\to\bot$\\
avec \ $\vec{\Xi}(h,k,n,i,k',n',i')\equiv
\{\vec{\Psi}(k,h,n,i),\,\vec{\Psi}(k',h,n',i'),\,k'>k,\,n'\le n\}$ \ c'est-à-dire~:\\
$\vec{\Xi}(h,k,n,i,k',n',i')\equiv\\
\{\pt j\indi(j+n\eps h_n\to(j\ppt i)\ne1),\; i+n\eps h_n,\; k=i+n,\\
\pt j{'}\,\indi(j'+n'\eps h_{n'}\to(j'\ppt i')\ne1),\; i'+n'\eps h_{n'},\; k'=i'+n',\\
k'>k,\,n'\le n\}$.\\
De \ $n'\le n$ \ et \ $k=i+n$, on déduit \ $n'\le k$, donc \ $k=j'+n'$.\\
De \ $k'>k$, on déduit \ $i'+n'>k$, et donc \ $j'<i'$.\\
On a donc \ $j'+n'\neps h_{n'}$, soit \ $k\neps h_{n'}$. Or, de \ $n'\le n$, on déduit \ $h_n\subseteq h_{n'}$
(lemme~\ref{DhH1}), donc \ $k\neps h_n$, ce qui contredit \ $i+n\eps h_n,\; k=i+n$.

\cqfd

\smallskip\noindent
Par définition de $\Phi$, on a trivialement \
$\vdash\pt n\indi\pt k\indi(\Phi(k,h,n)\to k\eps h_n)$.

\smallskip\noindent
D'après les lemmes~\ref{DhH1} et~\ref{DhH5}, on en déduit~:\\
$D(h,X),\vec{H}_*\![X]\vdash\pt h\pt k\indi\pt k{'}\indi
\pt n\indi\pt n{'}\indi(\Phi(k,h,n),\Phi(k',h,n'),k'>k\to k'\eps h_n)$.\\
Le lemme~\ref{DhH3} donne \ $\pt n\indi\ex k\indi\Phi(k,h,n)$. On en déduit~:\\
$D(h,X),\vec{H}_*\![X]\vdash\pt n\indi\ex k\indi\pt n{'}\,\indi\pt k{'}\,\indi
(\Phi(k',h,n'),k'>k\to k'\eps h_n)$,\\
et donc \ $D(h,X),\vec{H}_*\![X]\vdash\pt n\indi(\inf(h)\sqle h_n)$.

\smallskip\noindent
Or, on a trivialement \ $D(h,X)\vdash\pt n\indi\pt k\indi\pt p(k\eps h_n,X(n,p)\to k\eps p)$. D'où, finalement~:\\
$D(h,X),\vec{H}_*\![X]\vdash\pt n\indi\pt p(X(n,p)\to\inf(h)\sqle p)$.

\smallskip\noindent
On a ainsi obtenu la quasi-preuve {\sf ccd'} cherchée, ce qui termine la preuve du théorème~\ref{ccd_ver1}.

\cqfd

\subsection*{L'ultrafiltre}\noindent
Dans le modèle ${\cal N}$, nous avons défini \emph{l'idéal générique} ${\cal J}$, qui est un
prédicat unaire, en posant~: \ ${\cal J}(p)=\Pi\fois\{p\}$ \ pour tout $p\in P$.

\smallskip\noindent
D'après le théorème~\ref{elem_gen}, on a~:

\smallskip\noindent
i)~$\fforce\neg{\cal J}(\1)$\\
ii)~$\fforce\pt x(\neg\C[x]\to{\cal J}(x))$\\
iii)$\fforce\pt x\pt y({\cal J}(x\et y)\to{\cal J}(x)\lor{\cal J}(y))$\\
iv)~$\fforce\pt x(\pt y(\neg\C[x\et y]\to{\cal J}(y))\to\neg{\cal J}(x))$\\
v)~$\fforce\pt x\pt y({\cal J}(x),y\sqle x\to{\cal J}(y))$

\smallskip\noindent
D'après le théorème~\ref{premier_ordre}, on a \ $\force F$ $\Dbfl$ $\fforce F$ \ pour toute
formule close $F$ du premier ordre.

\smallskip\noindent
{\small{\bfseries Remarque.} Une formule ``du premier ordre'' comporte des quantificateurs
sur les individus qui, à l'aide du symbole $\eps$, représentent les parties de $\NN$.C'est donc
une formule du second ordre du point de vue de l'arithmétique. Mais elle ne comporte pas de
quantificateur sur les ensembles d'individus.}

\smallskip\noindent
D'après les théorèmes~\ref{mm1} et~\ref{entMN}, on peut utiliser, dans $F$, le quantificateur
$\pt x\indi$, puisque le quantificateur $\pt x\inde$ est du premier ordre.

\smallskip\noindent
On a donc~:

\smallskip\noindent
vi)~$\fforce\C[x]\dbfl\pt m\indi\ex n\indi(m+n\eps x)$\\
vii)~$\fforce y\sqle x\dbfl\ex m\indi\pt n\indi(m+n\eps y\to m+n\eps x)$\\
viii)~$\fforce\pt n\indi n\eps\1$~; $\fforce\pt x\pt y\pt n\indi(n\eps x\et y\dbfl n\eps x\land n\eps y)$

\smallskip\noindent
puisque les formules considérées sont du premier ordre.
Les propriétés (i) à (viii) montrent que, dans le ${\cal B}$-modèle ${\cal N}$, la formule suivante est
réa\-lisée~:\\
${\cal J}$ est un idéal maximal non trivial sur l'algèbre de Boole des parties de $\NN$ \emph{qui sont
représentées par des individus.}

\smallskip\noindent
Or, d'après les théorèmes~\ref{cons_reels} et~\ref{ccd_ver1}, la formule suivante
est réalisée dans ${\cal N}$~:\\
\emph{Toute partie de $\NN$ est représentée par un individu.}

\smallskip\noindent
La formule suivante est donc réalisée dans ${\cal N}$~:\\
\emph{${\cal J}$ est un idéal maximal non trivial sur l'algèbre de Boole des parties de $\NN$.}

\subsubsection*{Programmes obtenus à partir de preuves}\noindent
Soit $F$ une formule de \emph{l'arithmétique du second ordre}, c'est-à-dire une formule du
se\-cond ordre dont tous les quantificateurs d'individu sont restreints à $\NN$ et tous les
quantificateurs du second ordre sont restreints à ${\cal P}(\NN)$.\\
On lui associe une formule  $F^\dag$ \emph{du premier ordre}, définie par récurrence sur $F$~:

\smallskip\noindent
$\bullet$~~Si $F$ est $t=u$, $F^\dag\equiv F$.\\
$\bullet$~~Si $F$ est $Xt$, $F^\dag$ est $t\eps X^-$, où $X^-$ est une variable \emph{d'individu} associée à la
variable de prédicat unaire $X$.\\
$\bullet$~~Si $F$ est $A\to B$, $F^\dag$ est $A^\dag\to B^\dag$.\\
$\bullet$~~Si $F$ est $\pt x\,A$, $F^\dag$ est $\pt x\indi\,A^\dag$.\\
$\bullet$~~Si $F$ est $\pt X\,A$, $F^\dag$ est $\pt X^-\,A^\dag$.

\smallskip\noindent
On note que, si $F$ est une formule de \emph{l'arithmétique du premier ordre}, alors $F^\dag$ est simplement
la restriction $F^{\mbox{\small int}}$ de $F$ au prédicat int$(x)$. 

\smallskip\noindent
Soit $F$ une formule close de l'arithmétique du second ordre et considérons une preuve de~$F$ à l'aide de
l'axiome du choix dépendant ACD et l'axiome AU de l'ultrafiltre sur $\NN$, énoncé
sous la forme~: ``${\cal J}$ est un idéal maximal non trivial sur ${\cal P}(\NN)$''.\\
On en déduit immédiatement une preuve de $F^\dag$ en ajoutant l'axiome RPN de représentation des prédicats
sur $\NN$~: \ $\pt X\ex x\pt y(y\eps x\dbfl Xy)$. On obtient donc~:\\
$x:$ AU, $y:$ RPN, $z:$ ACD$^\dag\vdash t[x,y,z]:F^\dag$.\\
On a donc \ $\vdash u:\,$AU, RPN $\to G$ \ avec $u=\lbd x\lbd y\lbd z\,t[x,y,z]$ et $G\equiv$ ACD$^\dag\to F^\dag$.\\
$G$ est donc une \emph{formule du premier ordre}.\\
Dans la section précédente, on a obtenu des quasi-preuve \ $\theta,\theta'$ \ telles que \
$(\theta,\1)\fforce AU$ \ et \ $(\theta',\1)\fforce$ RPN (théorèmes~\ref{cons_reels} et~\ref{ccd_ver1}).\\
Le théorème~\ref{adequat_B} (lemme d'adéquation) donne donc
$(u^*,\1_u)(\theta,\1)(\theta',\1)\fforce G$, c'est-à-dire~:\\
$(v,(\1_u\et\1)\et\1)\fforce G$ \ avec $v=((\ov{\alpha}_0)(\ov{\alpha}_0)u^*\theta)\theta'$.\\
D'après le théorème~\ref{premier_ordre}, on a donc \
$\delta'_Gv\force\C[(\1_u\et\1)\et\1]\to G$, c'est-à-dire~:\\
$\delta'_Gv\force\C[(\1_u\et\1)\et\1],$ ACD$^\dag\to F$.\\
L'axiome ACD$^\dag$ est conséquence de ACI (axiome du choix pour les individus). D'après le
théorème~\ref{ACI}, on a donc une quasi-preuve \ $\eta_0\force$ ACD$^\dag$.\\
Par ailleurs, on a évidemment une quasi-preuve $\xi_0\force\C[(\1_u\et\1)\et\1]$.\\
On a donc finalement $\delta'_Gv\xi_0\eta_0\force F$.\\
On peut alors appliquer au programme \ $\zeta=\delta'_Gv\xi_0\eta_0$ \ tous les résultats obtenus dans le cadre de la réalisabilité usuelle. Le cas où $F$ est une formule arithmétique (resp. $\Pi_1^1$) est étudié dans~\cite{krivine3}
(resp.~\cite{krivine4}).\\
Pour prendre deux exemples très simples~:

\smallskip\noindent
Si $F\equiv\pt X(X1,X0\to X1)$, on a \ $\zeta\star\kappa\ps\kappa'\ps\pi\succ\kappa\star\pi$
quels que soient les termes $\kappa,\kappa'\in\Lbd$ et la pile~$\pi\in\Pi$.

\smallskip\noindent
Si $F\equiv\pt m\indi\ex n\indi(\phi(m,n)=0)$, où $\phi$ est un symbole de fonction, alors pour tout $m\in\NN$, \
il exis\-te $n\in\NN$ tel que \ $\phi(m,n)=0$ \ et \
$\zeta\star\ul{m}\ps T\kappa\ps\pi\succ\kappa\star\ul{n}\ps\pi'$.\\
$T$ est la quasi-preuve de mise en mémoire des entiers donnée au théorème~\ref{mm1}(i).\\
$\pi,\kappa$ sont quelconques~; en prenant pour $\kappa$ une constante, on obtient donc un programme de calcul
de $n$ en fonction de $m$.

\section*{Bon ordre sur $\mathbb{R}$}\noindent
Le ${\cal A}$-modèle ${\cal M}$ est le même que dans la section précédente~: l'ensemble d'individus est
$P={\cal P}(\Pi)^{\NN}$. Rappelons qu'un élément de $P$ est appelé tantôt \emph{individu}, tantôt
\emph{condition}, suivant le contexte.

\smallskip\noindent
On pose $(m,n)=m+(m+n)(m+n+1)/2$ (bijection de $\NN^2$ sur $\NN$). On définit une fonction binaire
$\phi:P^2\to P$ en posant~:\\
$\phi(n,p)(i)=p(i,n)$ si $n\in\NN$~; $\phi(n,p)$ est arbitraire (par exemple $0$) si $n\notin\NN$.

\smallskip\noindent
{\bfseries Notation.} Dans la suite, on écrira $p_n$ au lieu de $\phi(n,p)$. La donnée d'un individu $p$
est donc équivalente à celle d'une suite d'individus $p_n (n\in\NN$). Si $i,n\in\NN$, on a
$\|(i,n)\eps p\|=\|i\eps p_n\|$.

\smallskip\noindent
On fixe un bon ordre strict \ $\trl$ \ sur $P={\cal P}(\Pi)^{\NN}$, isomorphe au cardinal $2^{\aleph_0}$~:
tout segment initial propre de $\trl$ est donc de cardinal $<2^{\aleph_0}$. On définit une fonction binaire,
notée \ $(p\trl q)$ \ en posant \ $(p\trl q)=1$ si $p\trl q$~; \ $(p\trl q)=0$ sinon.\\
Comme la relation $(p\trl q)=1$ est bien fondée sur $P$, on a (théorème~\ref{bien_fonde})~:\\
$\Y\force\pt X[\pt x(\pt y((y\trl x)=1\mapsto Xy)\to Xx)\to\pt x\,Xx]$\\
dans le ${\cal A}$-modèle ${\cal M}$, mais aussi dans tout ${\cal B}$-modèle ${\cal N}$.\\
On écrira, en abrégé, $y\trl x$ pour $(y\trl x)=1$.\\
Dans ${\cal M}$ et ${\cal N}$, la relation $\trl$ est donc bien fondée mais, en général, pas totale.\\
C'est une relation d'ordre strict, dans ces deux modèles, car on a immédiatement, dans le
modèle~${\cal M}$~: \ $I\force\pt x((x\trl x)\ne1)$~; \
$I\force\pt x\pt y\pt z((x\trl y)=1\mapsto((y\trl z)=1\mapsto(x\trl z)=1))$.\\
Comme il s'agit de formules du premier ordre, d'après le théorème~\ref{premier_ordre}, on a aussi,
dans le modèle~${\cal N}$~: \ $\fforce\pt x((x\trl x)\ne1)$~; \
$\fforce\pt x\pt y\pt z((x\trl y)=1\mapsto((y\trl z)=1\mapsto(x\trl z)=1))$.

\smallskip\noindent
Une condition $p\in P$ est aussi une suite d'individus $p_k$. On va la considérer intuitivement comme
``~l'ensemble des individus $p_{k+1}$ pour $k\eps p_0$~''~; cela pour définir la condition $\1$, la
formule $\C[p]$ qui exprime que $p$ est une condition non triviale, et l'opération binaire $\et$.

\smallskip\noindent
$\1$ est l'ensemble vide, autrement dit \ $i\eps\1_0$ (c'est-à-dire $(i,0)\eps\1$) doit être faux. On pose donc~:\\
$\1(n)=\Pi$ pour tout $n\in\NN$.

\smallskip\noindent
Une condition est non triviale si l'ensemble d'individus qui lui est associé est totalement ordonné par \ $\trl$.
On pose donc~:\\
$\C[p]\equiv\pt i\inde\pt j\inde(i\eps p_0,j\eps p_0\to E[p_{i+1},p_{j+1}])$ \ avec~:\\
$E[x,y]\equiv(x=y\lor x\trl y\lor y\trl x)$ \ c'est-à-dire \
$E[x,y]\equiv(x\ne y,(x\trl y)\ne1,(y\trl x)\ne1\to\bot)$.

\smallskip\noindent
L'ensemble associé à $p\et q$ est la réunion des ensembles associés à $p$ et à $q$~; on pose donc~:\\
$p\et q=r$ \ où $r_0$ est défini par~: \ $\|2i\eps r_0\|=\|i\eps p_0\|$~; $\|2i+1\eps r_0\|=\|i\eps q_0\|$~;\\
$r_{j+1}$ est défini par~: \ $r_{2i+1}=p_{i+1}$~; \ $r_{2i+2}=q_{i+1}$.

\smallskip\noindent
La notation \ $p\subset q$ \ signifie que l'ensemble associé à $q$ contient celui associé à $p$.\\
On pose donc~:\\
$p\subset q\equiv\pt i\inde(i\eps p_0\to\ex j\inde\{j\eps q_0,p_{i+1}=q_{j+1}\})$.

\begin{lemma}\label{subset_trans}\ \\
i)~~$\theta\force\pt p\pt q\pt r(p\subset q,q\subset r\to p\subset r)$ \ avec \
$\theta=\lbd f\lbd g\lbd i\lbd x\lbd h(fix)\lbd j\lbd y(g)jyh$.\\
ii)~~$\theta'\force\pt p\pt q\pt r(p\subset q\to p\et r\subset q\et r)$ \ avec \
$\theta'=\lbd f\lbd i\lbd y\lbd u((ei)(u)iy)(((f)(d_2)iy)\lbd j(u)(d_0)j$\\
où $d_0,d_1,d_2,e$ sont des quasi-preuves représentant respectivement les fonctions récursives~:\\
$n\mapsto 2n$, $n\mapsto 2n+1$, $n\mapsto[n/2]$, $n\mapsto$ parité de $n$ ($e$ est à valeurs booléennes).
\end{lemma}\noindent
i)~~On suppose~:\\
$f\force\pt i($ent$(i),i\eps p_0,\pt j($ent$(j),j\eps q_0\to p_{i+1}\ne q_{j+1})\to\bot)$~;\\
$g\force\pt j($ent$(j),j\eps q_0,\pt k($ent$(k),k\eps r_0\to q_{j+1}\ne r_{k+1})\to\bot)$~;\\
$x\force i\eps p_0$~; $h\force\pt k($ent$(k),k\eps r_0\to p_{i+1}\ne r_{k+1})$~; et on a \ $\ul{i}\in|$ent$(i)|$.\\
D'où \ $f\ul{i}x\force\pt j($ent$(j),j\eps q_0\to p_{i+1}\ne q_{j+1})\to\bot$.\\
Supposons \ $y\force j\eps q_0$ \ et soit \ $\ul{j}\in|$ent$(j)|$.

\smallskip\noindent
Si $p_{i+1}=q_{j+1}$, alors \ $g\ul{j}yh\force\bot$~; \ donc  \ $g\ul{j}yh\force p_{i+1}\ne q_{j+1}$.
On a montré~:\\
$\lbd j\lbd y(g)jyh\force\pt j($ent$(j),j\eps q_0\to p_{i+1}\ne q_{j+1})$. Donc \
$(f\ul{i}x)\lbd j\lbd y(g)jyh\force\bot$.

\smallskip\noindent
ii)~~On suppose~:\\
$f\force\pt i($ent$(i),i\eps p_0,\pt j($ent$(j),j\eps q_0\to p_{i+1}\ne q_{j+1})\to\bot)$~;\\
$y\force i'\eps(p\et r)_0$~; \
$u\force\pt j'($ent$(j'),j'\eps(q\et r)_0\to(p\et r)_{i'+1}\ne(q\et r)_{j'+1})$.\\
Si on remplace $j'$ par \ $2j''$, puis par \ $2j''+1$, on obtient, d'après la définition de \ $\et$~:

\smallskip\noindent
(1) \ \ $(u)(d_0)\ul{j}''\force j''\eps q_0\to(p\et r)_{i'+1}\ne q_{j''+1}$~;\\
(2) \ \ $(u)(d_1)\ul{j}''\force j''\eps r_0\to(p\et r)_{i'+1}\ne r_{j''+1}$.

\smallskip\noindent
Il y a alors deux cas~:

\smallskip\noindent
$\bullet$~~Si $i'=2i''$, on a \ $y\force i''\eps p_0$ \ et, d'après~(1), \
$(u)(d_0)\ul{j}''\force j''\eps q_0\to p_{i''+1}\ne q_{j''+1}$. \ Donc~:\\
$\lbd j(u)(d_0)j\force\pt j($ent$(j),j\eps q_0\to p_{i''+1}\ne q_{j+1})$ et, par suite~:\\
\centerline{$(((f)(d_2)\ul{i}')y)\lbd j(u)(d_0)j\force\bot$.}

\smallskip\noindent
$\bullet$~~Si $i'=2i''+1$, on a \ $y\force i''\eps r_0$ \ et, d'après~(2), \
$(u)(d_1)\ul{j}''\force j''\eps r_0\to r_{i''+1}\ne r_{j''+1}$.\\
En faisant $j''=i''$, on obtient \ $(u)(d_1)\ul{i}''\force i''\eps r_0\to\bot$ \ et donc~:\\
\centerline{$(u)\ul{i}'y\force\bot$.}

\smallskip\noindent
On a donc, dans les deux cas~: \ \
$((e\ul{i}')(u)\ul{i}'y)(((f)(d_2)\ul{i}')y)\lbd j(u)(d_0)j\force\bot$.

\cqfd

\begin{lemma}\label{p_sub_q_Cq_Cp}\ \\
i)~~$\theta\force\pt p\pt q(p\subset q,\C[q]\to\C[p])$ \ avec\\
$\theta=\lbd f\lbd g\lbd i\lbd i'\lbd x\lbd x'\lbd u\lbd v\lbd w(fi'x')\lbd j'\lbd y'(fix)\lbd j\lbd y(g)jj'yy'uvw$.\\
ii)~~$\force\pt p\pt q\pt r(p\subset q,\C[q\et r]\to\C[p\et r])$ \ autrement dit \
$\force\pt p\pt q(p\subset q\to q\sqle p)$.
\end{lemma}\noindent
i)~~Soient \ $f\force p\subset q,g\force\C[q]$, \ c'est-à-dire~:\\
$f\force\pt i($ent$(i),i\eps p_0,\pt j($ent$(j),j\eps q_0\to p_{i+1}\ne q_{j+1})\to\bot)$~;\\
$g\force\pt j\pt j'($ent$(j),$ ent$(j'),j\eps q_0,j'\eps q_0\to E[q_{j+1},q_{j'+1}])$ \ avec~:\\
$E[x,y]\equiv(x\ne y,(x\trl y)\ne1,(y\trl x)\ne1\to\bot)$.\\
Soient\ $x\force  i\eps p_0,x'\force i'\eps p_0,u\force p_{i+1}\ne p_{i'+1},
v\force(p_{i+1}\trl p_{i'+1})\ne1,w\force(p_{i'+1}\trl p_{i+1})\ne1$.\\
Soient \ $y\force j\eps q_0$, $y'\force j'\eps q_0$.\\
On a \ $g\ul{j}\,\ul{j}'yy'\force E[q_{j+1},q_{j'+1}]$~; si $p_{i+1}=q_{j+1}$ et $p_{i'+1}=q_{j'+1}$,
alors \ $g\ul{j}\,\ul{j}'yy'\force E[p_{i+1},p_{i'+1}]$, donc \ $g\ul{j}\,\ul{j}'yy'uvw\force\bot$.\\
On a donc \ $\lbd j\lbd y(g)jj'yy'uvw\force$ent$(j),j\eps q_0\to\bot$ si \
$p_{i+1}=q_{j+1}$ et $p_{i'+1}=q_{j'+1}$.\\
Par suite, \
$\lbd j\lbd y(g)jj'yy'uvw\force\pt j($ent$(j),j\eps q_0\to p_{i+1}\ne q_{j+1})$ \
si \ $p_{i'+1}=q_{j'+1}$, d'où~:\\
$(f\ul{i}x)\lbd j\lbd y(g)jj'yy'uvw\force\bot$ \ si \ $p_{i'+1}=q_{j'+1}$, d'où~:\\
$\lbd j'\lbd y'(f\ul{i}x)\lbd j\lbd y(g)jj'yy'uvw\force
\pt j'($ent$(j'),j'\eps q_0\to p_{i'+1}\ne q_{j'+1})$. Donc~:\\
$(f\ul{i}'x')\lbd j'\lbd y'(f\ul{i}x)\lbd j\lbd y(g)jj'yy'uvw\force\bot$.

\smallskip\noindent
ii)~~Se déduit immédiatement de (i) et \ $\force\pt p\pt q\pt r(p\subset q\to p\et r\subset q\et r)$ \
(lemme~\ref{subset_trans}).

\cqfd

\smallskip\noindent
Le lemme suivant montre qu'on peut construire l'algèbre ${\cal B}$ et le ${\cal B}$-modèle ${\cal N}$.

\begin{lemma}\label{six_qp}
Il existe six quasi-preuves $\alpha_0,\alpha_1,\alpha_2,\beta_0,\beta_1,\beta_2$ telles que~:\\
$\alpha_0\force\pt p\pt q\pt r(\C[(p\et q)\et r]\to\C[p\et(q\et r)])$~; \
$\alpha_1\force\pt p(\C[p]\to\C[p\et\1])$~;\\
$\alpha_2\force\pt p\pt q(\C[p\et q]\to\C[q])$~; \
$\beta_0\force\pt p(\C[p]\to\C[p\et p])$~; \
$\beta_1\force\pt p\pt q(\C[p\et q]\to\C[q\et p])$~;\\
$\beta_2\force\pt p\pt q\pt r\pt s(\C[((p\et q)\et r)\et s]\to\C[(p\et(q\et r))\et s])$.
\end{lemma}\noindent
On montre seulement le premier cas. D'après le lemme~\ref{p_sub_q_Cq_Cp}(i), il suffit de trouver
une quasi-preuve \ $\theta\force\pt p\pt q\pt r(p\et(q\et r)\subset (p\et q)\et r)$.
On suppose donc~:\\
$y\force i\eps(p\et(q\et r))_0$~;
$u\force\pt j($ent$(j),j\eps((p\et q)\et r)_0\to(p\et(q\et r))_{i+1}\ne((p\et q)\et r)_{j+1})$.\\
Il y a trois cas~:\\
$\bullet$~~$i=2i'$~; on a alors \ $y\force i'\eps p_0$. On fait $j=2i=4i'$, donc
$u\force$ ent$(2i),i'\eps p_0\to p_{i'+1}\ne p_{i'+1}$.
On a donc~: \ \ $(u)(d_0)\ul{i}y\force\bot$.\\
$\bullet$~~$i=4i'+1$~; on a alors \ $y\force i'\eps q_0$. On fait $j=i+2=4i'+3$, donc~:\\
$u\force$ ent$(i+2),i'\eps q_0\to q_{i'+1}\ne q_{i'+1}$. On a donc~: \ \ $((u)(\sig)^2\ul{i})y\force\bot$.\\
$\bullet$~~$i=4i'+3$~; on a alors \ $y\force i'\eps r_0$. On fait $j=i-3=4i'$, donc~:\\
$u\force$ ent$(i-3),i'\eps r_0\to r_{i'+1}\ne r_{i'+1}$. On a donc~: \ \ $((u)(\p)^3\ul{i})y\force\bot$\\
($\p$ est le programme pour le prédécesseur).\\
On pose donc \ $\theta=\lbd i\lbd y\lbd u(((e_4i)(u)(d_0)iy)((u)(\sig)^2i)y)((u)(\p)^3i)y$, \ où $e_4$ est
défini par sa règle d'exécution~: \ $e_4\star\ul{i}\ps\xi\ps\eta\ps\zeta\ps\pi\succ\xi\ps\pi$
(resp. $\eta\ps\pi$, $\zeta\ps\pi$) si $i=4i'$ (resp. $4i'+1,4i'+3$).

\cqfd

\smallskip\noindent
On montre maintenant le~:

\begin{theorem}\label{ccd_ver2}\ \\
La structure de forcing $\{\C,\et,\1\}$ satisfait la condition de chaîne dénombrable dans ${\cal M}$.
\end{theorem}\noindent
Les hypothèses de la c.c.d. sont~:

\smallskip\noindent
$H_0\equiv\pt n\ex p\,{\cal X}(n,p)$~;\\
$H_1\equiv\pt n\inde\pt p\pt q\{{\cal X}(n,p),{\cal X}(n,q)\to p=q\}$~;\\
$H_2\equiv\pt n\inde\pt p\pt q({\cal X}(n,p),{\cal X}(sn,q)\to q\sqle p)$~;\\
$H_3\equiv\pt n\inde\pt p({\cal X}(n,p)\to\C[p])$.

\smallskip\noindent
Par ailleurs, d'après le théorème~\ref{ACI}, on a une fonction binaire $f:P^2\to P$ telle que~:\\
$\vsig\force\pt n\inde(\ex p\,{\cal X}(n,p)\to\ex k\inde{\cal X}(n,f(n,k)))$.\\
D'après $H_0$, on peut donc aussi utiliser la formule~:

\smallskip\noindent
$H'_0\equiv\pt n\inde\ex k\inde\,{\cal X}(n,f(n,k))$.

\smallskip\noindent
On pose \ $\vec{H}=\{H_0,H'_0,H_1,H_2,H_3\}$ \ et \ $\vec{H}_*=\{H_0,H'_0,H_1,H_2\}$.

\begin{lemma}\label{Cp_et_q}
$\vec{H}\vdash\pt p\pt q\pt m\inde\pt n\inde({\cal X}(m,p),{\cal X}(n,q)\to\C[p\et q])$.
\end{lemma}\noindent
On montre \ $\pt m\indi\pt n\indi({\cal X}(m,p),{\cal X}(m+n,q)\to q\sqle p)$ \ par récurrence sur $n$.\\
Pour $n=0$, cela découle de \ $H_1,H_3$. Pour le pas de la récurrence, on utilise $H_2$.

\smallskip\noindent
On a donc \ $\pt p\pt q\pt m\inde\pt n\inde({\cal X}(m,p),{\cal X}(n,q)\to p\sqle q\lor q\sqle p)$.\\
De $p\sqle q$, on déduit \ $\C[p\et p]\to\C[q\et p]$, \ d'où le résultat d'après $H_3$ \ et \
$\C[p]\to\C[p\et p]$.

\cqfd

\smallskip\noindent
On définit la limite cherchée $h$ en définissant $h_0$ et $h_{m+1}$ pour chaque $m\in\NN$.\\
Pour  $m=(i,n,k)$ (\,c'est-à-dire \ $(i,(n,k))$\,), on pose \
$\|m\eps h_0\|=\|{\cal X}(n,f(n,k))\land i\eps(f(n,k))_0\|$~;\\
puis \ $h_{m+1}=(f(n,k))_{i+1}$.\\
Intuitivement, ${\cal X}$ définit une suite d'ensembles dénombrables, et $h$ est la réunion de
ces ensembles.

\smallskip\noindent
$\bullet$~~Preuve de \ $\vec{H}_*\vdash{\cal X}(n,p)\to h\sqle p$.\\
D'après le lemme~\ref{p_sub_q_Cq_Cp}(ii), il suffit de montrer \
${\cal X}(n,p)\to p\subset h$, soit~:\\
${\cal X}(n,p),i\eps p_0,\pt m\inde(m\eps h_0,\to h_{m+1}\ne p_{i+1})\to\bot$, pour $n,i\in\NN$.\\
On fixe $k\in\NN$ \ et on pose \ $m=(i,n,k)$. D'après la définition de $h$, il suffit de montrer~:\\
${\cal X}(n,p),i\eps p_0,\pt k\inde({\cal X}(n,f(n,k)),i\eps(f(n,k))_0,
\to (f(n,k))_{i+1}\ne p_{i+1})\to\bot$.\\
Or, de \ $H_1,{\cal X}(n,p),{\cal X}(n,f(n,k))$, \ on déduit \ $f(n,k)=p$ \ et donc~:\\
$(f(n,k))_0=p_0$ \ et $(f(n,k))_{i+1}=p_{i+1}$. On est donc ramené à montrer~:\\
${\cal X}(n,p),i\eps p_0,\pt k\inde({\cal X}(n,f(n,k)),i\eps p_0\to p_{i+1}\ne p_{i+1})\to\bot$.\\
Or, cette formule se déduit immédiatement de $H'_0$.

\smallskip\noindent
$\bullet$~~Preuve de \ $\vec{H}\vdash\C[h]$.\\
On doit montrer $\C[h]$, soit \ $m\eps h_0,m'\eps h_0\to E[h_{m+1},h_{m'+1}]$. \ Or, on a~:

\smallskip\noindent
$m=(i,n,k)$~; \ $\|m\eps h_0\|=\|{\cal X}(n,f(n,k))\land i\eps(f(n,k))_0\|$~; \
$h_{m+1}=(f(n,k))_{i+1}$~;\\
$m'=(i',n',k')$~; \ $\|m'\eps h_0\|=\|{\cal X}(n',f(n',k'))\land i'\eps(f(n',k'))_0\|$~; \
$h_{m'+1}=(f(n',k'))_{i'+1}$.

\smallskip\noindent
De \ ${\cal X}(n,f(n,k)),{\cal X}(n',f(n',k'))$, on déduit \ $\C[u]$ avec $u=f(n,k)\et f(n',k')$
(lemme~\ref{Cp_et_q}).\\
On a donc~:\\
$\|i\eps(f(n,k))_0\|=\|2i\eps u\|$~; \ $\|i'\eps(f(n',k'))_0\|=\|2i'+1\eps u\|$~; \
$h_{m+1}=u_{2i+1}$~; \ $h_{m'+1}=u_{2i'+2}$.\\
De $\C[u]$, on déduit \ $E[u_{2i+1},u_{2i'+2}]$, c'est-à-dire \ $E[h_{m+1},h_{m'+1}]$.

\smallskip\noindent
Cela termine la preuve du théorème~\ref{ccd_ver2}.

\cqfd

\subsubsection*{Le bon ordre sur ${\cal P}(\NN)$}\noindent
Dans le modèle ${\cal N}$, on définit le prédicat unaire \
${\cal G}(x)\equiv\ex p\ex i\inde\{\neg{\cal J}(p),i\eps p_0,x=p_{i+1}\}$.

\begin{lemma}\label{G_total}
$\fforce{\cal G}(x),{\cal G}(y)\to E[x,y]$.
\end{lemma}\noindent
On doit montrer \
$\fforce\neg{\cal J}(p),\neg{\cal J}(q),i\eps p_0,x=p_{i+1},j\eps q_0,y=q_{j+1}\to E[x,y]$, soit~:\\
$\fforce\neg{\cal J}(p),\neg{\cal J}(q),i\eps p_0,j\eps q_0\to E[p_{i+1},q_{j+1}]$.\\
D'après le théorème~\ref{elem_gen}(ii) et (iii), on a \ $\fforce\neg{\cal J}(p),\neg{\cal J}(q)\to\C[p\et q]$.\\
Il suffit donc de montrer \ $\fforce\C[p\et q],i\eps p_0,j\eps q_0\to E[p_{i+1},q_{j+1}]$.\\
On montre ci-dessous qu'on a \
$I\force\C[p\et q],i\eps p_0,j\eps q_0\to E[p_{i+1},q_{j+1}]$.
Comme c'est une formule du premier ordre, cela donne le résultat voulu, d'après le théo\-rème~\ref{premier_ordre}.\\
On a, en effet~: \ $p_{i+1}=(p\et q)_{2i+1}$~; $q_{j+1}=(p\et q)_{2j+2}$~;\\
$\|i\eps p_0\|=\|2i\eps(p\et q)_0\|$~; \ $\|j\eps q_0\|=\|2j+1\eps(p\et q)_0\|$.\\
On est donc ramené à montrer~:\\
$I\force\C[p\et q],2i\eps(p\et q)_0,2j+1\eps(p\et q)_0\to E[(p\et q)_{2i+1},(p\et q)_{2j+2}]$\\
ce qui est évident, par définition de $\C[p\et q]$.

\cqfd

\smallskip\noindent
Le lemme~\ref{G_total} montre que $\trl$ est une relation \emph{totale} sur ${\cal G}$. Mais, par ailleurs,
 dans ${\cal N}$, $\trl$ est une relation bien fondée. On a donc~:

\smallskip\noindent
\centerline{$\fforce\,{\cal G}$ est \emph{bien ordonné} par $\trl$.}

\smallskip\noindent
On définit maintenant deux fonctions sur $P$~:

\smallskip\noindent
$\bullet$~~une fonction unaire \ $\delta:P\to P$ \ en posant \
$\|i\eps\delta(p)_0\|=\|i+1\eps p_0\|$~; \ $\delta(p)_{i+1}=p_{i+2}$.\\
$\bullet$~~une fonction binaire \ $\phi:P^2\to P$ \ en posant \
$\|0\eps\phi(p,q)_0\|=\vide$~; $\|i+1\eps\phi(p,q)_0\|=\|i\eps p_0\|$~;\\
$\phi(p,q)_1=q$~; \ $\phi(p,q)_{i+2}=p_{i+1}$ pour tout $i\in\NN$.\\
On a donc \ $\delta(\phi(p,q))=p$ \ et \ $\phi(p,q)_1=q$ \  quels que soient $p,q\in P$ \ et par suite~:\\
$I\force\pt p\pt q(\delta(\phi(p,q))=p)$~; \ $\mathbf{I}\fforce\pt p\pt q(\delta(\phi(p,q))=p)$~;\\
$I\force\pt p\pt q(\phi(p,q)_1=q)$~; \ $\mathbf{I}\fforce\pt p\pt q(\phi(p,q)_1=q)$.

\smallskip\noindent
Intuitivement, $\delta(p)$ définit l'ensemble obtenu en ôtant $p_1$ de l'ensemble associé à $p$~;\\ $\phi(p,q)$ définit l'ensemble obtenu en ajoutant $q$ à l'ensemble associé à $p$.

\begin{lemma}\label{ajout}
Si $p,q\in P$, il existe $q'\in P$ tel que \ $\delta(q')=q$ \ et \ $p_i\trl q'$ pour tout $i\in\NN$.
\end{lemma}\noindent
Pour chaque $a\in P$, on a $\delta(\phi(q,a))=q$. Mais l'application \ $a\mapsto\phi(q,a)$ \ est
évidemment injective, puisque \ $\phi(q,a)_1=a$. Donc l'ensemble $\{\phi(q,a);\;a\in P\}$ est de cardinal $2^{\aleph_0}$. Or, par hypothèse sur $\trl$, tout segment initial propre de $P$, pour le bon ordre $\trl$,
est de cardinal $<2^{\aleph_0}$. Il existe donc $a_0\in P$ tel que $p_i\trl\phi(q,a_0)$ \ pour tout $i\in\NN$. Il suffit alors de poser $q'=\phi(q,a_0)$.

\cqfd

\smallskip\noindent
On peut donc définir une fonction binaire \ $\psi:P^2\to P$ telle que l'on ait~:\\
$\delta(\psi(p,q))=q$ \ et \ $(p_i\trl\psi(p,q))=1$ \ quels que soient $p,q\in P$ et $i\in\NN$.
On a donc~:

\smallskip\noindent
$I\force\pt p\pt q(\delta(\psi(p,q))=q)$~; \ $\mathbf{I}\fforce\pt p\pt q(\delta(\psi(p,q))=q)$.\\
$KI\force\pt p\pt q\pt i\inde(p_i\trl\psi(p,q))$~; \
$\mathbf{KI}\fforce\pt p\pt q\pt i\inde(p_i\trl\psi(p,q))$.

\begin{lemma}\label{G_atteint}
On a \ $\fforce\pt q\ex x\{{\cal G}(x),\delta(x)=q\}$.
\end{lemma}\noindent
Ceci s'écrit \ $\fforce\pt q[\pt x\pt p\pt i\inde(\delta(x)=q,\,i\eps p_0,\,x=p_{i+1}
\to{\cal J}(p))\to\bot]$ \
ou encore~:\\
$\fforce\pt q[\pt p\pt i\inde(i\eps p_0,\,\delta(p_{i+1})=q\to{\cal J}(p))\to\bot]$.\\
En faisant $i=0$, il suffit de montrer~:\\
(1)\hspace{5em}$\fforce\pt q[\pt p(0\eps p_0,\delta(p_1)=q\to{\cal J}(p))\to\bot]$.

\smallskip\noindent
En remplaçant $p$ par $\phi(p,\psi(p,q))$ dans (1), on voit qu'il suffit de montrer~:\\
\centerline{$\fforce\pt q\neg\pt p\,{\cal J}(\phi(p,\psi(p,q)))$.}

\begin{lemma}\label{CpC_phi_ap}
$\force\pt p\pt q(\C[p]\to\C[\phi(p,\psi(p,q))])$.
\end{lemma}\noindent
On a \ $\C[r]\equiv\pt i\inde\pt j\inde(i\eps r_0,j\eps r_0\to E[r_{i+1},r_{j+1}])$. Donc, pour
montrer \ $\force\C[p]\to\C[r]$, il suffit de montrer~:\\
(1) \ \ $\force\C[p]\to\pt i\inde\pt j\inde(i+1\eps r_0,j+1\eps r_0\to E[r_{i+2},r_{j+2}])$ \ et\\
(2) \ \ $\force\C[p]\to\pt j\inde(0\eps r_0,j+1\eps r_0\to E[r_1,r_{j+2}])$.\\
On applique cette remarque en posant \ $r=\phi(p,\psi(p,q))$. Alors \ (1) s'écrit \
$\force\C[p]\to\C[p]$ \ puisque \ $\|i+1\eps r_0\|=\|i\eps p_0\|$ et $r_{i+2}=p_{i+1}$ \
et de même pour $j$.\\
Il suffit donc de montrer (2), c'est-à-dire~:\\
$\force\C[p]\to\pt j\inde(0\eps\phi(p,\psi(p,q))_0,j+1\eps\phi(p,\psi(p,q))_0
\to E[\phi(p,\psi(p,q))_1,\phi(p,\psi(p,q))_{j+2}])$.\\
Or, on a \ $I\force\pt p\pt q(0\eps\phi(p,q)_0)$~; \
$I\force\pt p\pt q(j\eps p_0\to j+1\eps\phi(p,\psi(p,q))_0)$~;\\
$I\force\pt p\pt q(\phi(p,\psi(p,q))_1=\psi(p,q))$~; \
$I\force\pt p\pt q(\phi(p,\psi(p,q))_{j+2}=p_{j+1})$.\\
Il suffit donc de montrer~:\\
$\force\C[p]\to\pt j\inde(j\eps p_0\to E[\psi(p,q),p_{j+1}])$\\
ce qui est trivial, puisqu'on a \ $KI\force\pt p\pt q\pt j\inde(p_{j+1}\trl\psi(p,q))$.

\cqfd

\begin{lemma}\label{p_sub_phi_pq}
$\lbd i\lbd x\lbd y((y)(\sig)i)x\force\pt p\pt q(p\subset\phi(p,q))$.
\end{lemma}\noindent
Ceci s'écrit~:\\
$\lbd i\lbd x\lbd y((y)(\sig)i)x\force\pt i($ent$(i),i\eps p_0,\pt j($ent$(j),j\eps\phi(p,q)_0
\to\phi(p,q)_{j+1}\ne p_{i+1})\to\bot)$\\
ce qui est immédiat, en faisant $j=i+1$.

\cqfd

\smallskip\noindent
On a \ $\force p\subset\phi(p,\psi(p,q))$ (lemme~\ref{p_sub_phi_pq}), d'où on déduit \ $\force\phi(p,\psi(p,q))\sqle p$ (lemme~\ref{p_sub_q_Cq_Cp}ii), \ et donc \
$\force\C[\phi(p,\psi(p,q))]\to\C[p\et\phi(p,\psi(p,q))]$. D'après le lemme~\ref{CpC_phi_ap},
on a donc~:\\
$\force\pt p\pt q(\C[p]\to\C[p\et\phi(p,\psi(p,q))])$. Comme il s'agit d'une formule du premier ordre,
on a, d'après le théorème~\ref{premier_ordre}~: \
$\fforce\pt p\pt q(\C[p]\to\C[p\et\phi(p,\psi(p,q))])$\\
et donc, d'après le théorème~\ref{elem_gen}(ii)~: \
$\fforce\pt p\pt q(\neg\C[p\et\phi(p,\psi(p,q))]\to{\cal J}(p))$.\\
On applique alors le théorème~\ref{densite} qui donne~: \
$\fforce\pt q\neg\pt p\,{\cal J}(\phi(p,\psi(p,q)))$\\
ce qui est le résultat cherché.

\cqfd

\begin{theorem}\label{bon_ordre}
Les formules suivantes sont réalisées dans ${\cal N}$~:\\
i)~~Il existe un bon ordre sur l'ensemble des individus.\\
ii)~~Il existe un bon ordre sur l'ensemble des parties de \ $\NN$.
\end{theorem}\noindent
i)~~Le lemme~\ref{G_atteint} montre que, dans ${\cal N}$, la fonction $\delta$ est une surjection
de ${\cal G}$ sur l'ensemble $P$ des individus. Or, on a vu que la formule~: \
``~${\cal G}$ est bien ordonné par $\trl$~'' \ est réalisée dans ${\cal N}$.\\
ii)~~D'après les théorèmes~\ref{cons_reels} et~\ref{ccd_ver2}, la formule suivante est réalisée dans
${\cal N}$~: \ ``~Toute partie de~$\NN$ est représentée par un individu~''.
D'où le résultat, d'après (i).

\cqfd

\smallskip\noindent
Le théorème~\ref{bon_ordre}(ii) permet de transformer en programme n'importe quelle preuve d'une formule
de l'arithmé\-tique du second ordre, utilisant l'existence d'un bon ordre sur $\mathbb{R}$. La méthode
est la même que celle exposée ci-dessus pour l'axiome de l'ultrafiltre.

\end{document}